\documentclass[aps,prd,onecolumn, showpacs,nofootinbib,amsmath,groupedaddress]{revtex4}
\usepackage{epsfig}
\usepackage{graphicx}
\usepackage{amsfonts}
\usepackage{latexsym}
\usepackage{amsmath,amssymb}
\usepackage{verbatim}
\usepackage{mathtools}
\usepackage{wrapfig}
\usepackage{setspace}
\usepackage{slashed}
\usepackage[all]{xypic}
\usepackage{tikz}
\usepackage[hypertex]{hyperref}

\begin{document}
\title{Non-equilibrium fluctuation-dissipation relation from holography}

\author{Ayan Mukhopadhyay}
\affiliation{LPTHE,
UPMC -- Paris 6; CNRS UMR
7589, 
Tour 13-14, 4$^{\grave{e}me}$ \'etage, Boite 126, 4 Place
Jussieu, 75252 Paris Cedex 05, France 
}

\date{\today}
\begin{abstract}
{\noindent We derive a non-equilibrium fluctuation-dissipation relation for bosonic correlation functions from holography in the classical gravity approximation at strong coupling. This generalizes the familiar thermal fluctuation-dissipation relation in absence of external sources. This also holds universally for any non-equilibrium state which can be obtained from a stable thermal equilibrium state in perturbative derivative (hydrodynamic) and amplitude (non-hydrodynamic) expansions. Therefore, this can provide a strong experimental test for the applicability of the holographic framework. We discuss how it can be tested in heavy ion collisions. We also make a conjecture regarding multi-point holographic non-equilibrium Green's functions.}

\end{abstract}

\maketitle
\tableofcontents
\section{Introduction}

Nature challenges us to understand phenomena in real time. This is very difficult to do in the microscopic formulation of the laws of nature given in the framework of quantum field theory. 

Though quantum field theory is experimentally successful in understanding microscopic processes when perturbation theory works, it has very limited success in describing macroscopic phenomena in real time. The best we can typically do is to calculate rate and cross-sections using perturbation theory, and use the results as inputs for kinetic or phenomenological equations. As for example, for dilute quantum gases we can use the Boltzmann equation where the two body scattering cross-section is an input. However, for few systems we know how to formulate systematic  corrections to the Boltzmann equations, which takes into account the uncertainty principles of quantum dynamics.

The main difficulty is that time-dependent perturbation theory gives us behavior in time only in a Taylor series. This fails to give a uniform approximation in time away from the initial phase of evolution, and describe decoherence, thermalization and hydrodynamics in a unified framework. Even when the coupling constants are small, we need an essentially non-perturbative approach to understand the origin of irreversibility from the microscopic quantum field theory.

In recent decades, there has been impressive progress in non-equilibrium quantum field theory, which attempts to bridge this gap using the two-particle irreducible action formalism\footnote{This started with the closed-time-path formalism of Schwinger and Keldysh \cite{Schwinger}. The next advancement came with the two-particle irreducible (2PI) action formalism due to Cornwall, Jackiw and Tomboulis \cite{Cornwall}; and incorporation of the Schwinger-Keldysh closed time contour in it by Calzetta and Hu \cite{Calzetta}. This will be briefly reviewed in section II of this paper. For a review of recent progress of systematic approximations of the two-particle irreducible action and it's applications please see \cite{review1, review2, Rammer}. }. Still a lot of progress is needed to develop approximation schemes which takes into account unitarity and conservation laws in a controlled manner. As these methods are non-perturbative, some progress is also needed in understanding renormalizability in these formalisms.

Holography in the form of gauge/gravity duality is a new tool which gives a non-perturbative reformulation of quantum field theory \cite{adscft}. This is particularly tractable when the quantum field theory is a strongly coupled gauge theory and the rank of the gauge group is large. In such a case, gauge/gravity duality maps a quantum field theory to a familar classical theory of gravity in one higher dimension. At present this is the best non-perturbative reformulation of quantum gauge theories at hand, which can describe phenomena in real time. 

In this paper, we will try to unearth a special feature of holographic duality at strong coupling and large rank of the gauge group (i.e. large $N$), which gives a genuinely non-equilibrium result and is also a model-independent (i.e. universal) feature \footnote{At this stage we can make a comarison to the famous universal holographic result at strong coupling and large $N$ that the viscosity to entropy density ratio $\eta/s$ equals $1/4\pi$ \cite{etabys}. However, this is not a genuinely non-equilibrium result by our definition, because it can be computed via holographic prescritions from thermal correlators, and applying Kubo formula \cite{fluidgravity1}.}. By a genuinely non-equilibrium result, we imply one which cannot be obtained readily from equilibrium correlation functions. We also try to focus on a feature which gives fundamental insight on thermalization and nature of quantum kinetics. This leads us to study non-equilibrium fluctuation-dissipation relation as described below.

It is known from non-equilibrium quantum field theory, that in order to get quantum formulation of kinetic equations we need the non-equilibrium spectral and statistical functions. We define them below.

The spectral function can be understood as an off-shell generalization of density of states. Here we will define it for a bosonic operator $O(\mathbf{x},t)$. It can be obtained as a Wigner transform of the commutator, i.e. Fourier transform of the ralative coordinate as below :
\begin{equation}\label{specb}
\mathcal{A}(\omega, \mathbf{k}, \mathbf{x}, t) =\int d^3 r \, dt_{r} \, e^{i(\omega t_r - \mathbf{k}\cdot \mathbf{r})} \Big{\langle}\Big[O\Big(\mathbf{x} + \frac{\mathbf{r}}{2}, t + \frac{t_r}{2}\Big), O\Big(\mathbf{x} - \frac{\mathbf{r}}{2}, t - \frac{t_r}{2}\Big)\Big] \Big{\rangle}.
\end{equation}
The expectation value value above is taken in a non-equilibrium state. In case of fermionic fields we need to take the anti-commutator above.

It can be proved (see appendix A) that the spectral function is proportional to the imaginary part of the Wigner transformed retarded Green's function as below :
\begin{equation}
\mathcal{A}(\omega, \mathbf{k}, \mathbf{x}, t) = - 2 \, \text{Im}\, G_R(\omega, \mathbf{k}, \mathbf{x}, t).
\end{equation}
Therefore the spectral function is real. In case of fermionic Hermitean operators it is also non-negative.

Physically $\mathcal{A}(\omega, \mathbf{k}, \mathbf{x}, t)$ encodes the probability that a quasiparticle with momentum $\mathbf{k}$ at a given point in space-time given by $\mathbf{x}$ and $t$ will have energy $\omega$. We note that the uncertainty principle forces us to consider the quasi-particle off-shell. At strong coupling we expect broadening of peaks and typically no sharply defined quasi-particles.

At equiibrium, due to translational invariance, the spectral function $\mathcal{A}(\omega, \mathbf{k}, \mathbf{x}, t)$ do not depend on the center-of-mass coordinates $\mathbf{x}$ and $t$. It is measurable by angle-resolved photoemission spectroscopy (ARPES) of electrons. Away from equilibrium, time resolved version of the same (tr-ARPES) is necessary (see for instance \cite{trarpes}).

The statistical function can be understood as an off-shell generalization of quasi-particle distribution function in phase-space. It is the Wigner transform of the anti-commutator for bosonic fields (commutator for fermionic fields) as below :
\begin{equation}\label{statb}
G_\mathcal{K}(\omega, \mathbf{k}, \mathbf{x}, t) =-\frac{i}{2} \int d^3 r \, dt_{r} \, e^{i(\omega t_r - \mathbf{k}\cdot \mathbf{r})} \Big{\langle}\Big{\{ }O\Big(\mathbf{x} + \frac{\mathbf{r}}{2}, t + \frac{t_r}{2}\Big), O\Big(\mathbf{x} - \frac{\mathbf{r}}{2}, t - \frac{t_r}{2}\Big)\Big{\} } \Big{\rangle} .
\end{equation}
We will show in the next section that the statistical function is related to the imaginary part of the Wigner-transformed Feynman propagator as follows :
\begin{equation}
G_{\mathcal{K}}(\omega, \mathbf{k}, \mathbf{x}, t) = i \text{Im}\, G_F(\omega, \mathbf{k}, \mathbf{x}, t).
\end{equation}
Thus the statistical function is purely imaginary. In literature, this is often defined without a $-i$ factor in front as in (\ref{statb}). In that case it is real, and when integrated over $\omega$ at fixed $\mathbf{k}$, $\mathbf{x}$ and $t$, it gives the time-dependent phase-space distribution of quasiparticles in the semiclassical limit at weak coupling. 
Here we will keep the $-i$ factor - in this case it is also known as the Keldysh propagator. 

The statistical function/ Keldysh propagator can be indirectly measured by following space-time evolution of expectation values of operators, like conserved currents and the energy-momentum tensor. We note that both the non-equilibrium statistical and spectral functions need regularization for removing ultraviolet divergences, which are however expected to be independent of the state. What is relevant for physical observations is their dependence on the non-equilibrium hydrodynamic and non-hydrodynamic variables characterizing the state. 

At equilibrium, the statistical function is again independent of $\mathbf{x}$ and $t$ due to translational invariance. Also it is determined from the spectral function using analyticity. The relation between the spectral function and the statistical function at equilibrium is known as the fluctuation-dissipation relation, which at thermal equilibrium is as follows in the bosonic case:
\begin{equation}\label{therm}
G_{\mathcal{K}}(\omega, \mathbf{k})= -i \Big( n_{\text{BE}}(\omega) + \frac{1}{2}\Big) \mathcal{A}(\omega, \mathbf{k}).
\end{equation}
It is called the fluctuation-dissipation relation because it implies the fluctuation-dissipation theorems \cite{Rammer}, like that relating the thermal electrical noise (current fluctuations) to the electric resistance.

Away from equilibrium, analyticity is insufficient to relate the spectral and statistical functions. They evolve in a coupled way given by the so-called Kadanoff-Baym equations which can be derived from the two-particle irreducible action (see section II). In the semi-classical limit at weak coupling, the Kadanoff-Baym equations reduce to Boltzmann equation with quantum corrections. Formally these equations relating off-shell quantities are valid at strong coupling also. Even at weak coupling however, non-perturbative methods are needed to make complete sense of these equations and for making systematic expansions as mentioned earlier. In general we do not expect any generalization of the flucutation-dissipation relation away from equilibrium which can be stated in a universal way independent of the details of the non-equilibrium state and the nature of external perturbations. 

Numerical simulations of Kadanoff-Baym equations based on two-particle irreducible (2PI) effective action at weak coupling indeed show apparent thermalization - at long time the spectral and statistical functions satisfy (\ref{therm}). Such simulations, as for instance, in $O(N)$ scalar field theories at large $N$ and weak coupling \cite{ON}, indeed show thermalization; but far away from equilibrium, there seems to be no simple fluctuation-dissipation relation between spectral and statistical functions.

\textbf{Summary of results :} In this paper we study the non-equilibrium spectral and statistical functions in states which are perturbatively connected to thermal equilibrium with external sources absent. These are non-equilibrium states undergoing strongly coupled hydrodynamic and non-hydrodynamic relaxation.

At strong coupling we expect a few operators to suffice in describing typical states. This is so because most of the operators except the relevant order parameters, conserved currents and the energy-momentum tensor will be expected to have large anomalous dimensions. This will be more generic in non-supersymmetric theories \cite{hardwall, noneqspec} where we do not have chiral primary operators \footnote{The chiral pimary operators are special because they are typically protected from receiving large quantum corrections to their anomalous dimensions}, unless there are emergent symmetries at large $N$ (as in \cite{nonsusyhol}). 

Thus we can describe the non-equilibrium states holographically at strong coupling and large $N$ using solutions using Einstein's gravity coupled to a few gauge fields (dual to conserved currents) and scalar fields (dual to relevant order parameters). Here we will further specialize to states where conserved currents and order parameters vanish in the course of evolution. Thus we will be able to describe 
these states using the hydrodynamic and non-hydrodynamic quasinormal modes of pure gravity along with their non-linear evolution.

The holographic prescription for obtaining the spectral function in these class of states systematically in the derivative (hydrodynamic) and amplitude (non-hydrodynamic) expansions has been obtained before. In this paper we will generalize this methodology to obtain the statistical function in the same expansions systematically. We will find that the non-equilibrium statistical function, like the non-equilibrium spectral  function individually carries detailed imprint of the non-equilibrium state, particularly detailed information about the dispersion relation of the relaxation modes and their non-linear interactions. However, there is a simple generalized non-equilibrium fluctuation-dissipation relation between the spectral and statistical functions, again set exactly by the final temperature of thermal equilibrium. 

This non-equilibrium fluctuation-dissipation relation being independent of the non-equilibrium state concerned is universal and is as follows :
\begin{equation}
G_{\mathcal{K}}(\omega, \mathbf{k}, \mathbf{x}, t)= -i \Big( n_{\text{BE}}(\omega) + \frac{1}{2}\Big) \mathcal{A}(\omega, \mathbf{k}, \mathbf{x}, t),
\end{equation}
where $ n_{\text{BE}}(\omega) $ is set by the final equilibrium temperature. We note the definition of both, the spectral and statistical functions involve Wigner transform, so this relation is non-local both in space and time. However, we recover the equilibrium fluctuation-dissipation relation at long times when the spectral and statistical functions are independent of $\mathbf{x}$ and $t$. Also this relation is strictly true in absence of external sources, so conservation of energy implies that the final equilibrium temperature is set by the total energy of the initial state. We elaborate later in section V that the relation is not in contradiction with locality and causality of the underlying field theory. 

Furthermore, at strong coupling we expect the Kadanoff-Baym equations to have strong quantum memory, thus a Boltzmann like limit where the spectral and statistical functions will behave quasi-locally is likely to be absent. Therefore, the non-equilibrium fluctuation-dissipation relation need not be specified by local data. What is surprising is that the only data required is the total conserved energy of the state in the large $N$ and strong coupling limit.

This relation has been obtained in the probe-brane limit in \cite{Baier}, \cite{Julian}. 

\textbf{Methodology in brief and organization of this paper :} We describe briefly the methodology we will follow here along with the organization of this paper. 

We start from non-equilibrium field theory in section II. We review briefly the Schwinger-Keldysh formalism. We then obtain new parametrizations of the non-equilibrium fluctuation-dissipation relation which will be necessary for having consistent derivative/amplitude expansions of the spectral and statistical functions. We show that the parameters must satisfy additional field-theoretic constraints. Based on this non-equilibrium fluctuation dissipation relation, we find that any non-equilibrium Green's function can be written as an appropriate weighted sum of the non-equilibrium retarded and advanced Green's function. The weights are also subject to field-theoretic constraints. Additionally we obtain new formal understanding of some important features of the effective action for non-equilibrium Green's functions.

In section III, we review the construction of gravity duals of non-equilibrium states and introduce the derivative/amplitude expansions. We then study the dynamics of the scalar field dual to the bosonic order parameter in the same expansions. We find the unique boundary conditions for the non-equilibrium modes consistent with the derivative/amplitude expansions which give regularity at the horizon. The only new element here will be study of conjugation properties of the boundary conditions necessary for later analysis of field-theoretic constraints.

In section IV, we map boundary conditions of the non-equilibrium modes to the parametrization of non-equilibrium Green's functions as an appropriate sum of the non-equilibrium retarded and advanced Green's functions, obtained in section II.

This map is possible because of our previous work in which the boundary conditions for the non-equilibrium modes which give solutions regular at the horizon have been identified, and have been utilized to obtain the causal response function which gives the holographic non-equilibrium retarded Green's function \cite{noneqspec}. This builds on the holographic prescription for thermal retarded Green's function proposed by Son and Starinets \cite{incoming}. We show the advanced response function obtained from the regular non-equilibrium solution gives the holographic non-equilibrium advanced Green's function which passes field-theoretic consistency tests.

We then show that the sum of causal and advanced gravitational response functions evaluated using appropriate boundary conditions map to the parametrization of the non-equilibrium Feynman propagator required for it to have consistent derivative/amplitude expansions. The boundary conditions which depart from regularity are responsible for the non-equilibrium shifts in the weights of the non-equilibrium retarded and advanced Green's function in the weighted sum. We then study reality constraints on the parameters obtained from field-theory and show that the only boundary conditions which are consistent are those which give regularity at the horizon. Thus we find that there are no non-equilibrium shifts in the weights of the non-equilibrium retarded and advanced Green's function in the weighted sum which gives the non-equilibrium Feynman propagator.

We also give intuitive aguments why regularity at the horizon in combination with linear response theory should determine all non-equilibrium propagators holographically.

In section V, we obtain the non-equilibrium statistical function from the non-equilibrium Feynman propagator, and hence the non-equilibrium fluctuation-dissipation relation.
We argue why it's form is consistent with the locality and causality of the underlying field theory. We also give a conjecture on higher point non-equilibrium correlation functions in absence of external sources.

In section VI, we show though our holographic prescriptions generalize to some extent in higher derivative gravity, we cannot expect our results to hold beyond strong coupling even when the classical gravity approximation is valid. The main reason for this will be that the bulk scalar will not be minimally coupled to gravity generically.

In section VII, we conclude with a summary of the key results and a discussion on how the holographic non-equilibrium fluctuation-dissipation result can be tested in heavy ion collisions particularly.

The appendices give relevant supporting material.

Before proceeding further, we would like to mention that an interesting formalism has been proposed by van Rees and Skenderis in \cite{Skenderis} for doing real-time holography by construction of a gravitational analogue of the Schwinger-Keldysh closed time contour. Though this formalism seems ideally suited for doing thermal field theory in real time, it is very difficult to adapt this to non-equilibrium states. The main difficulty lies in construction of smooth Euclidean caps for non-equilibrium geometries, which is necessary in this formalism to set non-trivial initial boundary conditions, i.e. to represent creation of non-trivial states holographically.

Furthermore, there is an interesting approach \cite{Herzog} due to Son and Herzog for obtaining the thermal Schwinger-Keldysh propagators. It is difficult to generalize this to non-equilibrium context as well as this will involve construction of Penrose diagrams of non-equilibrium geometries. Indeed it is not known how to do this for non-equilibrium geometries in perturbation theory.

However, elements of these approaches may be concretely realized in special non-equilibrium geometries. It will be also necessary to understand how to construct the non-equilibrium spectral and statistical functions holographically in presence of external sources using systematic expansions. We leave such investigations to future work. 

\textbf{Notations :} We will denote the position and momentum four-vectors without bold fonts, as for instance by $x$ and $k$ respectively. The spatial position and momentum vectors will be denoted in bold fonts, as for instance by $\mathbf{x}$ and $\mathbf{k}$ respectively. 

\section{Aspects of non-equilibrium quantum field theory}

In this section, we will first briefly review the Schwinger-Keldysh formalism of non-equilibrium field theory. In particular we will define the effective action which gives the equation of motion of the expectation value of operators and their Green's functions. Then we will show formally this effective action contains no new information from the usual effective action, which is the Legendre transform of the generating functional of connected vacuum correlation functions. Finally we will establish some new relations between non-equilibrium Green's functions of bosonic Hermitean local operators.

\subsection{The non-equilibrium 
propagators from an effective action}

The basic tool of non-equilibrium field theory is construction of an effective action which is a functional of the operator and correlation functions.  The evolution of the expectation value of the operator and the Green's functions in all states, including those which are away from equilibrium, can be obtained by extremizing the effective action  $\Gamma[\mathcal{O}(x), G(x,y)]$. By constuction, this effective action does not depend on any equilibrium or non-equilibrium variables like temperature, velocity or shear-stress tensor. These non-equilibrium variables parametrize the solutions, which can be obtained from definite initial conditions. \footnote{Typically we need infinite variables to parametrize non-equilibrium states, hence non-equilibrium expectation values of operators and their correlation functions. However, we will see in the next section, we may expect that at least in field theories admitting holographic gravity duals, there are non-equilibrium states which can be parametrized by hydrodynamic variables and shear-stress tensor alone.} 

The effective action $\Gamma[\mathcal{O}(x), G(x,y)]$ can be defined as the double Legendre transform of a generalization of the generating functional of vacuum correlation functions - $Z[J(x), K(x,y)]$. This generalized partition function includes not only a source $J(x)$, but also an arbitrary bi-local interaction term $K(x,y)$, and is defined as below :
\begin{eqnarray}
Z[J, K]&=& e^{i W[J, K]}\nonumber\\
\phantom{G} &=& \int \mathcal{D}\Phi_s \exp \Bigg[i\Bigg(S[\Phi] + \int d^{4}x \,J(x)O(x) +\frac{1}{2}\int d^4 xd^4 y \,O(x)K(x,y)O(y)\Bigg)\Bigg]. 
\end{eqnarray}
Above $\Phi$ collectively denotes all the elementary fields. The operator $O$ which couples to $J$ locally and $K$ bi-locally is a polynomial of these elementary fields $\Phi$ and their derivatives.

We then define the expectation value of the operator $\mathcal{O}(x)$ and it's Green's function $G(x,y)$ through :
\begin{eqnarray}\label{fd}
\frac{\delta W [J, K]}{\delta J(x)} &=& \mathcal{O} (x), \nonumber\\
\frac{\delta W [J, K]}{\delta K(x,y)} &=& \frac{1}{2}\Bigg(\mathcal{O}(x)\mathcal{O} (y)+ G(x,y)\Bigg).
\end{eqnarray}
Eliminating $J(x)$ and $K(x,y)$ in favor of $\mathcal{O}(x)$ and $G(x,y)$, we can now do double Legendre transform to define the effective action as below :
\begin{eqnarray}\label{fd1}
\Gamma[\mathcal{O}(x), G(x,y)] &=& W[J, K] -  \int d^{4}x\, J(x)\mathcal{O}(x) -\frac{1}{2}\int d^4 xd^4y \ K(x,y)\Bigg(\mathcal{O}(x)\mathcal{O}(y)+G(x,y)\Bigg). 
\end{eqnarray}
Clearly,
\begin{eqnarray}\label{extrem}
\frac{\delta\Gamma[\mathcal{O}, G]}{\delta \mathcal{O}(x)} &=& -J (x) - \int d^4y \ K(x,y) \mathcal{O}(y), \nonumber\\
\frac{\delta\Gamma[O, G]}{\delta G(x,y)} &=& - \frac{1}{2} K(x,y).
\end{eqnarray}
Therefore, in absence of sources $J(x)$ and $K(x,y)$, extremizing the generalized effective action $\Gamma[\mathcal{O}(x), G(x,y)]$ gives the dynamics of both the expectation value of the operator and their Green's functions.

There is one important point in the above construction which we mention now. As stated in the introduction, for non-equilibrium states we need to know separately the spectral function (the commutator/anti-commutator) and the statistical function (the anti-commutator/commutator) for both bosonic/fermionic operators. 

\begin{figure}[h!]
\includegraphics[width = 2.5in]{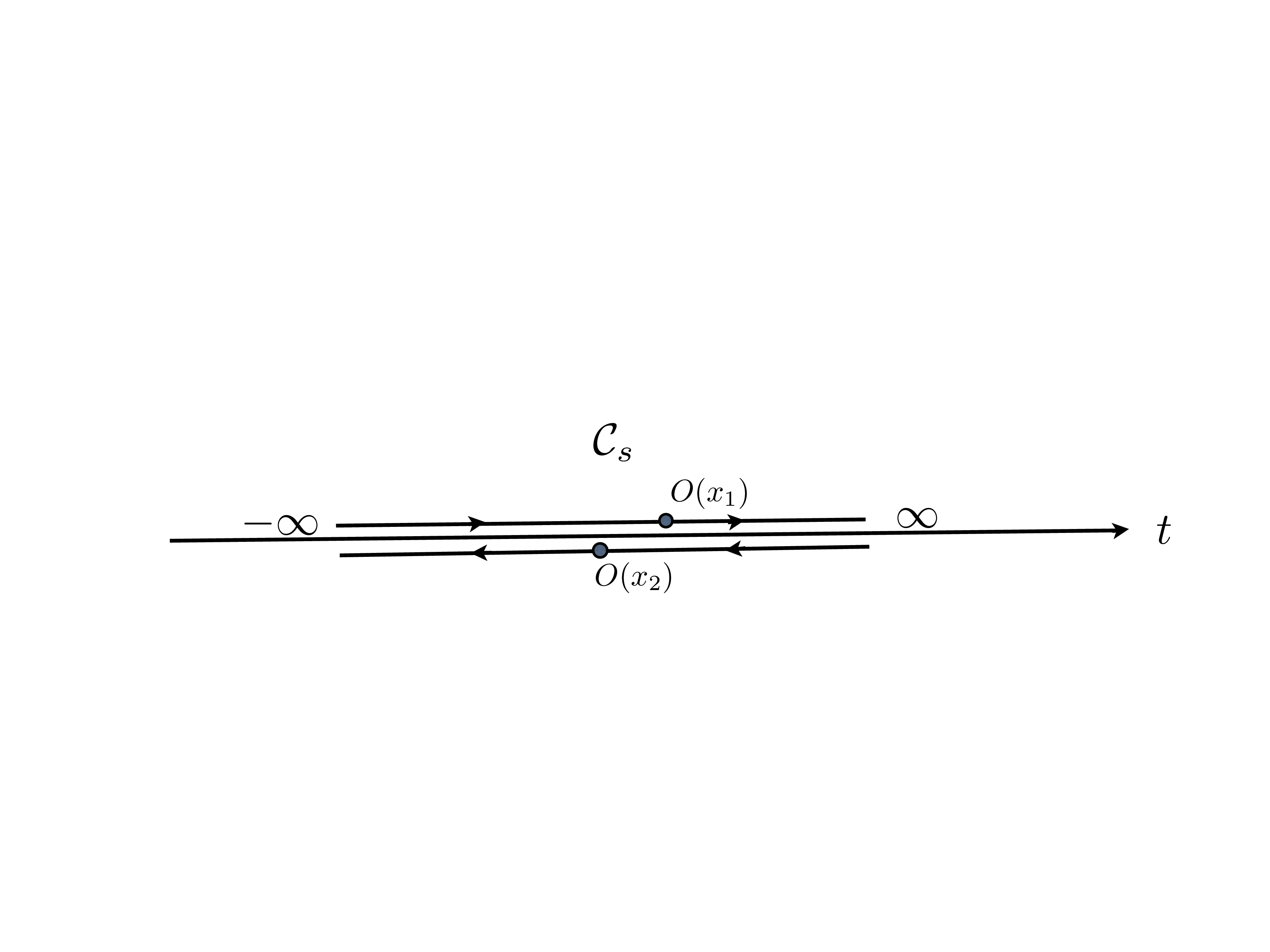}
\centering
\caption{The closed time Schwinger-Keldysh contour $\mathcal{C}_s$ is as above. The forward and backward directed parts of the contour have been displaced slightly above and below the real axis respectively just to distinguish them clearly. The operator insertions can be either in the forward or backward leg of the contour.}
\label{Keldyshcontour}
\end{figure}

To get complete information about the non-equilibrium Green's functions, we need to construct the partition function $W [J(x), K(x,y)]$ and consequently the effective action $\Gamma[\mathcal{O}(x), G(x,y)]$ over the so-called Schwinger-Keldysh closed real time contour $\mathcal{C}_s$, as shown in fig. \ref{Keldyshcontour}. 
This time contour travels from $-\infty$ to 
$\infty$ and then back from $\infty$ to $-\infty$, thus transversing the entire real line first forward and then backward. 

In fact, the full closed-time contour ordered Green's function $G_{\mathcal{C}_s}(x,y)$ can be written as a combination of the commutator and the anti-commutator, and hence as a sum over the spectral function and the statistical function as below 
\begin{eqnarray}\label{pe}
G_{\mathcal{C}_s}(x,y) = G_{\mathcal{K}}(x,y) - \frac{i}{2} \mathcal{A}(x,y), \text{sign}_{\mathcal{C}_s}(x^{0}-y^{0}),
\end{eqnarray} 
where $\text{sign}_{\mathcal{C}_s}(x^{0}-y^{0})$ is defined according to the contour-ordering over the full closed-time contour $\mathcal{C}_s$ \footnote{When both the operator insertions are on the upper contour we get the Feynman propagator, for instance. In general both the commutator and anti-commutator are needed to get the propagator over the full contour. Thus the contour is convenient for obtaining the independent dynamics of the non-equilibrium spectral and statistical functions.}.

The usefulness of the effective action $\Gamma[\mathcal{O}(x), G(x,y)]$ is realized only when we can reformulate it diagramatically. This has been done first in \cite{Cornwall} via a two-particle irreducible (2PI) effective action when the operator $O$ is an elementary field. A full discussion of this is outside the scope of the present paper. We do mention however, that though there is a diagrammatic formulation, this is non-perturbative as the lines of the diagrams are the full propagator $G(x,y)$. Therefore, the effective action $\Gamma[\mathcal{O}(x), G(x,y)]$ typically admits no perturbative expansion in a small parameter. 

Despite this drawback, in happy circumstances, we do get good approximations to experimental results or simulations, particularly when the coupling is weak and we can have a good intuition about the relevant diagrams to sum over in the effective action \cite{Rammer}. We need to resum over infinte diagrams to remove infrared divergences. There is no systematic understanding about how we can improve a given approximation while removing the infrared divergences in order to better the description. Nevertheless, this is so far the best field-theoretic tool at hand which gives a fundamental description of a physical process in an approximation that is uniform in time, i.e. which not only works at initial time but also far in the future. Furthermore this also reproduces quantum corrections to the Boltzmann equation.

\subsection{Redundancy of information in the effective action}
We will like to demonstrate here that the effective action $\Gamma [\mathcal{O}(x), G(x,y) ]$ formally does not contain any more information than the usual effective action $\Gamma[\mathcal{O}(x)]$. We recall $\Gamma [\mathcal{O}(x), G(x,y) ]$ is obtained from $Z[J(x), K(x,y)]$ by definition as in eqs. (\ref{fd}) and (\ref{fd1}), while $\Gamma[\mathcal{O}(x)]$ is obtained from the usual partition function $Z[J(x)]$, the generating functional of vacuum correlation functions. $\Gamma[\mathcal{O}(x)]$ is the usual one-particle irreducible (1PI) effective action when $O(x)$ is an elementary field. The effective action $\Gamma[O(x), G(x,y)]$ does not admit any perturbative expansion in a small parameter, but $\Gamma[O(x)]$ does. 

The effective action $\Gamma[\mathcal{O}(x), G(x,y)]$ is necessary to obtain a description which is a uniform approximation in time, but our current understanding of such approximations seem to be inadequate. It is not known how to have an approximation scheme which simultaneously takes care of unitarity in a controlled manner, is renormalizable by standard counterterms and satisfies all Ward identities (for recent progress see \cite{review1}). Therefore it is useful to have an argument for why in principle all the information is there in the usual effective action $\Gamma [\mathcal{O}(x)]$ for which such issues have been resolved. This will be useful for us later.

To show this, it is sufficient to prove that $W[J(x), K(x,y)]$ can be constructed from $W[J(x)]$, as $\Gamma[\mathcal{O}(x), G(x,y)]$ and $\Gamma[\mathcal{O}(x)]$ are obtained from $W[J(x), K(x,y)]$ and $W[J(x)]$ respectively via Legendre transforms. From (\ref{fd}) we readily observe that
\begin{equation}
\frac{\delta W[J,K]}{\delta K(x,y)} = \frac{1}{2}\Bigg(\frac{\delta W[J,K]}{\delta J(x)}\frac{\delta W[J,K]}{\delta J(y)} -\frac{\delta^2 W[J,K]}{\delta J(x) \delta J(y)} \Bigg),
\end{equation}
where we have used $\mathcal{O}(x) =\delta W[J,K] /\delta J(x)$ and $G(x,y)= \delta \mathcal{O}(x)/\delta J(y)$. Thus all functional $K-$ derivatives of $W[J,K]$ can be converted to functional $J-$ derivatives of $W[J,K]$. Therefore $W[J,K]$ can be constructed readily from $W[J, K = 0] = W[J]$. This is also expected because indeed a quantum field theory is uniquely defined once we know all connected vacuum correlation functions, hence $W[J]$.

\subsection{New relations between non-equilibrium Green's functions}

We will now investigate the general structure of non-equilibrium Green's functions of local bosonic operators in non-equilibrium states. Here the operators will be taken in the Heisenberg representation. 

In particular, we will establish some useful relations between the retarded propagator and the Feynman propagator, based on the assumption that the non-equilibrium parts of both admit systematic perturbative expansions.

The definition of the Feynman propagator is :
\begin{equation}
G_F(x_1, x_2) = -i \Big{\langle} T\left(O(x_1) O(x_2)\right)\Big{\rangle},
\end{equation}
where $T$ denotes time ordering.

We begin with the known relation (for proof see appendix A) \footnote{From now on we will denote the commutator as $\mathcal{A}(x_1, x_2)$, where it is implicitly implied that the inverse Wigner transform has been done on the spectral function $\mathcal{A}(k,x)$. Similarly we will denote $-i/2$ times the anti-commutator as $G_{\mathcal{K}}(x_1, x_2)$.}
\begin{equation}
G_F(x_1, x_2) = G_{\mathcal{K}}(x_1, x_2) - \frac{i}{2} \mathcal{A}(x_1, x_2)\,\text{sign}
(t_1 - t_2).
\end{equation}
Note this is just a special case of (\ref{pe}) when the operator insertion are in the forward part of the closed-time contour.

However,
\begin{eqnarray}
- i \mathcal{A}(x_1, x_2)\,\text{sign}
(t_1 - t_2) &=& - i \mathcal{A}(x_1, x_2)\,\Theta
(t_1 - t_2) +   i \mathcal{A}(x_1, x_2)\,\Theta
(t_2 - t_1)\nonumber\\
&=& G_R(x_1, x_2) + G_A (x_1, x_2).
\end{eqnarray}
Therefore,
\begin{equation}
G_F(x_1, x_2) = G_{\mathcal{K}}(x_1, x_2)
+ \frac{1}{2}\Big( G_R(x_1, x_2) + G_A(x_1, x_2) \Big),
\end{equation}
or in the Wigner transformed form 
\begin{equation}\label{feyngen}
G_F(k, x) = G_{\mathcal{K}}(k, x)
+ \text{Re}\, G_R(k, x),
\end{equation}
using $G_R*(k,x) = G_A(k,x)$ which has been proved in appendix A. 

We further note that $G_{\mathcal{K}}(k,x)$ is purely imaginary which follows from it's definition as proved in appendix A. Therefore, we conclude from (\ref{feyngen}) that
\begin{equation}\label{feynreim}
\text{Re} \,G_F(k,x) = \text{Re}\, G_R(k,x),
\quad \text{Im} \,G_F(k,x) = -iG_{\mathcal{K}}(k,x). 
\end{equation}

It is to be noted that unlike the retarded Green's function $G_R(k,x)$ or the advanced Green's function $G_A(k,x)$, the Feynman propagator is not analytic in $\omega$ either in the upper-half plane or the lower-half plane. This can be seen from the instance of the vacuum Feynman propagator itself in which positive frequency modes are propagated in the future and negative frequency modes are propagated in the past, i.e. the vacuum Feynman propagator 
is $\Theta(\omega) G_R(k) - \Theta(-\omega)G_A(k)$. The term $\Theta(\omega)G_R(k)$ implies lack of analyticity in $\omega$ in the lower-half plane as the positive frequency poles are shifted there with the standard $i\epsilon$ prescription, and similarly $\Theta(-\omega)G_A(k)$ implies there is no analyticity in the upper-half plane as the negative frequency poles are shifted there. 

Generally speaking, if a function is analytic in a variable in the upper half plane or lower half plane, with other variables held fixed, it can be fully constructed from it's imaginary part or real part. For instance as $\text{Im}\, G_R(k,x) = - (1/2) \mathcal{A}(k,x)$ (for proof see Appendix A), and as it is analytic in the upper-half plane in $\omega$, \footnote{Analyticity of $G_R(k,x)$ in lower-half $\omega$ plane is guaranteed by the $\Theta(t_1-t_2)$ term in $G_R(x_1, x_2)$.} the retarded Green's function can be constructed from the spectral function using
\begin{equation}
G_R (\omega, \mathbf{k}, \mathbf{x}, t) = \int_{-\infty}^{\infty} \frac{d\omega'}{2\pi}\frac{ \mathcal{A}(\omega', \mathbf{k}, \mathbf{x}, t)}{\omega -\omega' + i\epsilon}.
\end{equation}

In case of the Feynman propagator, it's real part $\text{Re}\, G_R(k,x)$ and it's imaginary part $G_{\mathcal{K}}(k,x)$ hold independent information in non-equilibrium states which cannot be kinematically constructed from the other because of non-analyticity in $\omega$ in the complex upper-half plane or the lower-half plane. 

We also note that as the Feynman propagator is symmetric, i.e. $G_F(x_1, x_2)= G_F(x_2, x_1)$, it follows that after Wigner transform that $G_F(k,x)= G_F(-k,x)$ as proved in appendix A.

The thermal Feynman propagator is a linear combination of the retarded or causal response  with weight $n_{\text{BE}}(\omega) + 1$ and advanced or anti-causal response with weight $-n_{\text{BE}}(\omega)$ for a given frequency, with $n_{\text{BE}}(\omega) = \frac{1}{e^{\frac{\omega}{T}}-1}$ being the Bose-Einstein distribution at temperature $T$. Thus, 
\begin{equation}\label{feyn1}
G_F(k) = \Big(n_{\text{BE}}(\omega)+ 1\Big)G_R(k) - n_{\text{BE}}(\omega)G_A(k).
\end{equation} 
As $T\rightarrow 0$,
\begin{equation}
n_{\text{BE}}(\omega) + 1 \rightarrow \Theta (\omega), \quad  -n_{\text{BE}}(\omega) \rightarrow \Theta (-\omega),
\end{equation}
so we recover the vacuum Feynman propagator in this limit.

It is not hard to see that (\ref{feyn1}) satisfies $G_F(k) = G_F(-k)$, because
\begin{equation}\label{inv}
n_{\text{BE}}(-\omega)+1= -
n_{\text{BE}}(\omega), \quad \text{or} \quad -n_{\text{BE}}(-\omega)= 
n_{\text{BE}}(\omega) + 1,
\end{equation}
and also $G_A(-\omega, -\mathbf{k})= G_R(\omega, \mathbf{k})$. 

We can rewrite the thermal Feynman propagator (\ref{feyn1}) as :
\begin{equation}
G_F(k) = \text{Re}\, G_R(k) + i \Big(2 n_{\text{BE}}(\omega) + 1\Big) \text{Im}\, G_R(k)
\end{equation} 
So, comparing (\ref{feyngen}) with the above we get
\begin{equation}
G_{\mathcal{K}}(k)= i \Big(2 n_{\text{BE}}(\omega) + 1\Big) \text{Im}\, G_R(k)
\end{equation}
However, $\text{Im}\, G_R(k) = - (1/2) \mathcal{A}(k)$. So,
\begin{equation}\label{thermalflucdis}
G_{\mathcal{K}}(k)= -i \Big( n_{\text{BE}}(\omega) + \frac{1}{2}\Big) \mathcal{A}(k).
\end{equation}
The above is the fluctuation-dissipation relation in thermal equilibrium. It is so-called because it implies the fluctuation-dissipation theorems \cite{Rammer}, like those relating the thermal electrical noise (current fluctuations) to the electric resistance.

We will now try to parameterize non-equilibrium fluctuation-dissipation relation. Using this, we will get a parameterization of non-equilibrium Green's functions.

Let us assume we are near equilibrium. Then the non-equilibrium state can be parametrized in terms of hydrodynamic and relaxational modes. This will allow us to do a perturbative expansion. Let us for instance linearize the state in terms of hydrodynamic fluctuations $\delta\mathbf{u}$ and $\delta T$. In this approximation, both the spectral and statistical functions will depend linearly on these variables. Thus the non-equilibrium fluctuation-dissipation relation will be linear in this approximation.

The exact nature of the dependence of the spectral and statistical functions on the hydrodynamic and non-hydrodynamic variables will be determined later in the holographic classical gravity approximation. As of now, we will take this dependence to be implicit.

It is also clear we can improve the linear approximation by including quadratic dependence of the spectral and statistical functions on the hydrodynamic and non-hydrodynamic variables which we will do systematically later. The general structure will be as follows :
\begin{eqnarray}
& G_{\mathcal{K}}(k,x) = G_{\mathcal{K}}^{\text{(eq)}}(k) + G_{\mathcal{K}}^{\text{(neq)}}(k,x), \quad \mathcal{A}(k,x) = \mathcal{A}^{\text{(eq)}}(k) +\mathcal{A}^{\text{(neq)}}(k,x), \nonumber\\
& G_{\mathcal{K}}^{\text{(neq)}}(k,x)= G_{\mathcal{K}}^{\text{(1)}}(k,x) + G_{\mathcal{K}}^{\text{(2)}}(k,x) + ... + G_{\mathcal{K}}^{\text{(n)}}(k,x) + ...  \, , &\nonumber\\
& \mathcal{A}^{\text{(neq)}}(k,x)= \mathcal{A}^{\text{(1)}}(k,x) + \mathcal{A}^{\text{(2)}}(k,x) + .... + \mathcal{A}^{\text{(n)}}(k,x) + ... \, .&
\end{eqnarray}
Above the superscript $(n)$ denotes the order of multi-linear dependence of the term on the non-equilibrium variables.

The non-equilibrium fluctuation dissipation relation can be parametrized as follows :
\begin{eqnarray}\label{fdneq}
G_{\mathcal{K}}^{\text{(eq)}}(k)&=& -i \Big( n_{\text{BE}}(\omega) + \frac{1}{2}\Big) \mathcal{A}^{\text{(eq)}}(k), \nonumber\\
G_{\mathcal{K}}^{\text{(1)}}(k,x) &=&  - i \int d^4 k_1   \,f^{(1,0)}(k, k_1, x)
\mathcal{A}^{\text{(eq)}}(k_1) -i\Big( n_{\text{BE}}(\omega) + \frac{1}{2}\Big) \mathcal{A}^{\text{(1)}}(k,x), ...\,.
\end{eqnarray}
Above $f^{(1,0)}(k, k_1, x)$ is real because each term in the expansion of $\mathcal{A}(k,x)$ is real while each term in the $G_{\mathcal{K}}(k,x)$ is purely imaginary. Also the Bose-Einstein distribution $n_{\text{BE}}(\omega)$ appearing above is always determined by the final equilibrium temperature. Furthermore as we have shown in appendix A, because $G_{\mathcal{K}}(x_1, x_2)$ is symmetric in $x_1$ and $x_2$, after Wigner transform it should satisfy $G_{\mathcal{K}}(k,x)= G_{\mathcal{K}}(-k, x)$. Similarly as shown in appendix A, $\mathcal{A}(k,x) = -\mathcal{A}(-k,x)$. Therefore, these imply that
\begin{eqnarray}\label{fprop}
&f^{(1,0)}(k,k_1,x) = f^{(1,0)}_{\text{S}}(k, k_1, x) +f^{(1,0)}_{\text{A}}(k,k_1,x), \quad \text{such that}&, \nonumber\\ 
&\text{$f^{(1,0)}_{\text{S}}(k, k_1, x)$ and $f^{(1,0)}_{\text{A}}(k, k_1, x)$ are real, and}&\nonumber\\
&f^{(1,0)}_{\text{S}}(k, k_1, x) = f^{(1,0)}_{\text{S}}(-k, k_1, x), \,  f^{(1,0)}_{\text{A}}(k, k_1, x) = - f^{(1,0)}_{\text{A}}(-k, -k_1, x). &
\end{eqnarray}

Similarly, at the next order
\begin{eqnarray}\label{fdneq2}
G_{\mathcal{K}}^{\text{(2)}}(k,x) &=&  -i\int d^4 k_1 d^4 k_2  \, f^{(2,0,0)}(k, k_1, k_2, x)
\mathcal{A}^{\text{(eq)}}(k_1)\mathcal{A}^{\text{(eq)}}(k_2) 
\nonumber\\&&
-i \int d^4 k_1 d^4 k_2 d^4 x_1\,  f^{(2,1,0)}(k, k_1, k_2, x, x_1)
\mathcal{A}^{\text{(1)}}(k_1, x_1)\mathcal{A}^{\text{(eq)}}(k_2)
\nonumber\\&&
- i \int d^4 k_1 d^4 x_1  \,f^{(2,1)}(k, k_1, x, x_1)
\mathcal{A}^{\text{(1)}}(k_1, x_1)
-i\Big( n_{\text{BE}}(\omega) + \frac{1}{2}\Big) \mathcal{A}^{\text{(2)}}(k,x), \, \text{etc.}
\end{eqnarray}
Again by definition $f^{(2,0,0)}(k, k_1, k_2, x)$, $f^{(2,1,0)}(k, k_1, k_2, x, x_1)$ and $f^{(2,1)}(k, k_1, x, x_1)$ have to be real and the Bose-Einstein distribution $n_{\text{BE}}(\omega)$ appearing above is always determined by the final equilibrium temperature.  Similarly the symmetry of the statistical function and anti-symmmetry of the spectral function in $x_1$ and $x_2$ prior to Wigner transform give further restrictions on $f^{(2,0,0)}(k, k_1, k_2, x)$, $f^{(2,1,0)}(k, k_1, k_2, x, x_1)$ and $f^{(2,1)}(k, k_1, x, x_1)$ as in (\ref{fprop}). The last two terms of (\ref{fdneq2}) are linear while the remaining terms are non-linear corrections to the equilibrium fluctuation-dissipation relation. Clearly non-linear corrections are expected because the Kadanoff-Baym equations which give equations of motion of the spectral and statistical functions are non-linear. We easily note as $G_{\mathcal{K}}^{(2)}$ and $\mathcal{A}^{(2)}$ are quadratic in non-equilibrium variables, while $G_{\mathcal{K}}^{(1)}$ and $\mathcal{A}^{(1)}$ are linear, $f^{(2,0,0)}$ depend quadratically while $f^{(2,1,0)}$ and $f^{(2,1)}$ depend linearly on non-equilibrium variables.

We will find the above parametrization of the non-equilibrium fluctuation-dissipation relation useful to determine it holographically. It can be also useful for experimental determination.

It then follows from eqs. (\ref{feynreim}) and (\ref{fdneq})  that the general structure of the non-equilibrium Feynman propagator for small perturbations away from equilibrium should be
\begin{eqnarray}\label{nfeyn}
G_F(k,x) &=& G^{\text{(eq)}}_F(k) + G^{\text{(neq)}}_F (k,x),  \quad G^{\text{(neq)}}_F (k,x) = G^{\text{(1)}}_F (k,x)+G^{\text{(2)}}_F (k,x) + .... +G^{\text{(n)}}_F (k,x)+ ... \, ,\nonumber\\
G^{\text{(eq)}}_F(k) &=& \Big(n_{\text{BE}}(\omega)+ 1\Big)G_R^{\text{(eq)}}(k) - n_{\text{BE}}(\omega)G_A^{\text{(eq)}}(k),
\nonumber\\ 
G^{\text{(1)}}_F(k,x) &=& iG^{\text{(1)}}_{\mathcal{K}}(k,x) + \text{Re}\, G_R^{\text{(1)}}(k,x),\nonumber\\
&=& \int d^4 k_1 \,f^{(1,0)}(k, k_1, x)   \Big( 
G_R^{\text{(eq)}}(k_1) - 
G_A^{\text{(eq)}}(k_1)\Big)\nonumber\\&&
+ \Big(n_{\text{BE}}(\omega)+ 1\Big)G_R^{\text{(1)}}(k,x) - n_{\text{BE}}(\omega)G_A^{\text{(1)}}(k,x).
\end{eqnarray}
We have also used $\mathcal{A}(k,x) = -2\text{Im}\, G_R(k,x)$ and $G_R*(k,x)= G_A(k,x)$ above. Similarly from (\ref{fdneq2}) we can derive that
\begin{eqnarray}\label{feynamp2}
G^{\text{(2)}}_F(k,x) &=& -\int d^4 k_1 d^4 k_2  \, f^{(2,0,0)}(k, k_1, k_2, x)
\Big(G_R^{\text{(eq)}}(k_1)-G_A^{\text{(eq)}}(k_1)\Big)\Big(G_R^{\text{(eq)}}(k_2)-G_A^{\text{(eq)}}(k_2)\Big)
\nonumber\\&&
-\int d^4 k_1 d^4 k_2 d^4 x_1\,  f^{(2,1,0)}(k, k_1, k_2, x, x_1)
\Big(G_R^{\text{(1)}}(k_1, x_1)-G_A^{\text{(1)}}(k_1, x_1)\Big)\Big(G_R^{\text{(eq)}}(k_2)-G_A^{\text{(eq)}}(k_2)\Big)
\nonumber\\&&
+\int d^4 k_1 d^4 x_1  \,f^{(2,1)}(k, k_1, x, x_1)
\Big(G_R^{\text{(1)}}(k_1, x_1)-G_A^{\text{(1)}}(k_1, x_1)\Big)\nonumber\\&&+ \Big(n_{\text{BE}}(\omega)+ 1\Big)G_R^{\text{(2)}}(k,x) - n_{\text{BE}}(\omega)G_A^{\text{(2)}}(k,x), \, \text{etc.}
\end{eqnarray}

We note that with the explicit form of fluctuation-dissipation relation (\ref{fdneq}), we can write any non-equilibrium Green's function as a functional of the non-equilibrium retarded Green's function. As the Wigner transformed advanced Green's function can be obtained from the Wigner transformed retarded Green's function simply by complex conjugation, an alternative representation is a weighted sum of the non-equilibrium retarded and advanced Green's function, where the weights are specified by the non-equilibrium fluctuation-dissipation relation.

Thus any non-equilibrium Green's function $G(k,x)$ should have the following structure in the linearized approximation :
\begin{eqnarray}\label{gnp}
G(k,x) &=& G^{\text{(eq)}}(k)+ G^{\text{(neq)}}(k,x), \quad G^{\text{(neq)}}(k,x) = G^{\text{(1)}}(k,x)+G^{\text{(2)}}(k,x)+ ...+G^{\text{(n)}}(k,x)+...\,,\nonumber\\
G^{\text{(eq)}}(k) &=& f_R^{\text{(eq)}}(\omega) G_R^{\text{(eq)}}(k) + f_A^{\text{(eq)}}(k)G_A^{\text{(eq)}}(k), \nonumber\\
G^{\text{(1)}}(k,x) &=&  \int d^4 k_1   \Big(f_R^{\text{(1,0)}}(k, k_1, x) 
G_R^{\text{(eq)}}(k_1) + f_A^{\text{(1,0)}}(k, k_1, x) 
G_A^{\text{(eq)}}(k_1)\Big)
\nonumber\\&& +f_R^{\text{(eq)}}(\omega) G_R^{\text{(1)}}(k,x) + f_A^{\text{(1)}}(k)G_A^{\text{(1)}}(k,x),
\end{eqnarray}
where $f_{R,A}^{\text{(eq)}}$ should be determined by the equilibrium fluctuation-dissipation relation and $f_{R,A}^{\text{(1,0)}}$ should be determined by $f^{(1,0)}$ in the non-equilibrium fluctuation-dissipation relation (\ref{fdneq}). Similarly,
\begin{eqnarray}\label{g2}
G^{\text{(2)}}(k,x) &=&  \int d^4 k_1 d^4 k_2  \Big(f_{RR}^{(2,0,0)}(k, k_1, k_2, x) 
G_R^{\text{(eq)}}(k_1)G_R^{\text{(eq)}}(k_2) \nonumber\\&&+ f_{RA}^{(2,0,0)}(k, k_1, k_2, x) 
G_R^{\text{(eq)}}(k_1)G_A^{\text{(eq)}}(k_2)+ f_{AR}^{(2,0,0)}(k, k_1, k_2, x) 
G_A^{\text{(eq)}}(k_1)G_R^{\text{(eq)}}(k_2)
\nonumber\\ && +f_{AA}^{(2,0,0)}(k, k_1, k_2, x) 
G_A^{\text{(eq)}}(k_1)G_A^{\text{(eq)}}(k_2)\Big)
\nonumber\\&& 
+\int d^4 k_1 d^4 k_2 d^4 x_1  \Big(f_{RR}^{\text{(2,1,0)}}(k, k_1, k_2, x, x_1) 
G_R^{\text{(1)}}(k_1,x_1)G_R^{\text{(eq)}}(k_2)\nonumber\\&& + f_{RA}^{\text{(2,1,0)}}(k, k_1, k_2, x, x_1) 
G_R^{\text{(1)}}(k_1,x_1)G_A^{\text{(eq)}}(k_2) + f_{AR}^{\text{(2,1,0)}}(k, k_1, k_2, x, x_1) 
G_A^{\text{(1)}}(k_1,x_1)G_R^{\text{(eq)}}(k_2)\nonumber\\&&
+f_{AA}^{\text{(2,1,0)}}(k, k_1, k_2, x, x_1) 
G_A^{\text{(1)}}(k_1,x_1)G_A^{\text{(eq)}}(k_2) \Big)\nonumber\\&&
+\int d^4 k_1 d^4 x_1 \Big(f_R^{\text{(2,1,0)}}(k, k_1, x,x_1) 
G_R^{\text{(1)}}(k_1, x_1) + f_A^{\text{(2,1,0)}}(k, k_1, x, x_1) 
G_A^{\text{(1)}}(k_1, x_1)\Big)\nonumber\\&&
+f_R^{\text{(eq)}}(\omega) G_R^{\text{(2)}}(k,x) + f_A^{\text{(eq)}}(k)G_A^{\text{(2)}}(k,x)\Big),\, \text{etc.}
\end{eqnarray}
Above $f^{(2,0,0)}_{RR}$, etc. should be determined $f^{(2,0,0)}$, etc. in (\ref{fdneq2}). We will find the above form of the non-equilibrium Green's functions useful for their holographic determination.

\section{Non-equilibrium geometries and the real scalar field}
In this section, we will first review the geometries which capture basic non-equilibrium processes like hydrodynamics and relaxation in the dual theory. Then we will study how the solution of a free massive real scalar field can be perturbatively obtained in these geometries. This real scalar field will be dual to a condensate, as for instance the chiral condensate of QCD. Finally we will determine which solutions of the bulk scalar field are regular at the horizon. We will show that the non-equilibrium corrections to the equilibrium solutions are unique, and determined by boundary conditions at the horizon which can be stated in a background independent manner, i.e. independently of the details of the dual non-equilibrium state, in the regime of validity of perturbation theory.

Most of this section is a review of the results of our previous works \cite{myself1, myself2, noneqspec}. The only new element that we introduce is a careful study of complex conjugation properties of various terms, leading to the background metric and solution of the bulk scalar field being real. 

\subsection{Holographic description of non-equilibrium states}
The solutions of vacuum Einstein's gravity with negative cosmological constant, which settle down to black holes with regular future horizons, holographically depict many fundamental non-equilibrium processes. Each such solution maps to a non-equilibrium state via gauge/gravity duality at strong coupling and large $N$. Each such non-equilibrium state can be characterized by the expectation value of the energy-momentum tensor alone, because this determines the dual solution in gravity which is a regular perturbation of anti-de Sitter black brane completely \cite{myself4}.

Conceptually, these states are similar to certain special solutions of the Boltzmann equation called conservative solutions, which are completely determined by the energy-momentum tensor \cite{myself1}. This is only a conceptual similarity because the Boltzmann equation is valid only in the weak coupling regime. Nevertheless, it is seen that one can systematically construct phenomenological equations for the energy-momentum tensor and any solution of these phenomenological equations can be lifted to a full solution of the Boltzmann equation. Thus such conservative solutions give a systematic approach to obtain phenomenology of irreversible processes in terms of energy-momentum tensor alone. At long times such solutions can be approximated by purely hydrodynamic solutions of Boltzmann equation, known as normal solutions \cite{Chapman} in literature. These normal solutions are thus special conservative solutions and allow us to compute linear and non-linear hydrodynamic transport coefficients at weak coupling and dilute densities, when the Boltzmann approximation is valid \footnote{Indeed such an approach is also valid for non-abelian gauge theories, like quantum chromo-dynamics (QCD) at temperature higher then the confining scale $\Lambda_{\text{QCD}}$ and low baryon densities \cite{Arnold}.}.

In the case of gravity, one can also systematically construct phenomenological equations involving the energy-momentum tensor alone, which when satisfied implies that the corresponding solution in gravity should have a regular future horizon \cite{myself1, myself2, myself3}. Just like in the case of the Boltzmann equation, these phenomenological equations can be expanded perturbatively in the derivative (hydrodynamic) and amplitude (non-hydrodynamic) expansion parameters as reviewed in appendix B. These equations have the most general structure valid in any field theory when the quantum corrections to the Boltzmann equation are considered. Also the values of phenomenological parameters are different at strong coupling.

At long time, all regular perturbations of anti-de Sitter black branes become purely hydrodynamic. In such cases, the metric can be constructed by the now well-known fluid/gravity correspondence \cite{fluidgravity1, Janik, fluidgravity2, Sayantani}.

This hydrodynamic limit is also a natural consequence of the phenomenological equations which admit purely hydrodynamic solutions \cite{myself1, myself2, myself3}. The phenomenological equations however also describe other non-equilibrium processes like decoherence and relaxation.

Another special class of solutions to these phenomenological equations describe homogeneous relaxation \cite{myself2}. In this case, it has been explicitly proved that the future horizon in corresponding solutions in gravity is indeed regular as outlined in appendix B.

In this paper, for illustrative purposes we will use two basic examples. The first example is that of the five-dimensional anti-de Sitter black brane perturbed by a hydrodynamic shear-mode.
The corresponding metric is \footnote{One can readily construct this metric using methods of \cite{myself4}.}:
\begin{eqnarray}\label{metric1}
ds^2 &=& \frac{l^2}{r^2} \frac{dr^2}{f\left(\frac{r r_0}{l^2}\right)} +   \frac{l^2}{r^2}\Big(-f\left(\frac{r r_0}{l^2}\right)dt^2 + dx^2 + dy^2 + dz^2 \Big)  \nonumber\\ &&-\frac{2l^2}{r^2}\Big(1-f\left(\frac{r r_0}{l^2}\right)\Big) \delta u_i (\mathbf{k}_{\text{(h)}}) e^{i \mathbf{k}_{\text{(h)}} \cdot \mathbf{x}}e^{ - \frac{\mathbf{k}_{\text{(h)}}^2}{4\pi T} t}dtdx^i  \nonumber\\
&& +\frac{2l^2}{r^2} \Big(-i \frac{l^2}{4r_0} \ k_{\text{(h)}i} \ \delta u_j (\mathbf{k}_{\text{(h)}})e^{i \mathbf{k}_{\text{(h)}} \cdot \mathbf{x}}e^{ - \frac{\mathbf{k}_{\text{(h)}}^2}{4\pi T} t}  \ h\left(\frac{r r_0}{l^2}\right)dx^i dx^j \Big) + O(\epsilon^2),
\end{eqnarray}
where,
\begin{eqnarray}\label{fh}
f(s) = 1 - s^4, \quad h(s) = -\text{ln}\, (1- s^4).
\end{eqnarray}
The first line in (\ref{metric1}) corresponds to the $AdS_5$ black brane at temperature $T = r_0/(\pi l^2)$, with $l$ being the asymptotic radius of curvature and $l^2/r_0$ being the radius of the horizon. The second and third lines describe the hydrodynamic perturbations. Here $\epsilon$ is the derivative expansion parameter which is $\mathbf{k}_{\text{(h)}}/T$.

The corresponding energy-momentum tensor is :
\begin{eqnarray}
t_{\mu\nu} &=& \text{diag}\, (\epsilon, p, p, p)- i \eta \Big( k_{\text{(h)}i} \delta\mathbf{u}(\mathbf{k}_{\text{(h)}})_j + k_{\text{(h)}j} \delta\mathbf{u}(\mathbf{k}_{\text{(h)}})_i\Big)e^{i \mathbf{k}_{\text{(h)}} \cdot \mathbf{x}}e^{ - \frac{\mathbf{k}_{\text{(h)}}^2}{4\pi T} t} + O(\epsilon^2),
\end{eqnarray}
where
\begin{equation}\label{em1}
\epsilon = 3p = \frac{3r_0^4}{2\kappa^2 l^5}, \quad \eta = \frac{r_0^3}{2\kappa^2 l^3}.
\end{equation}
Above $\epsilon$ and $p$ are the thermal energy-density and pressure, and $\eta$ is the shear-viscosity satisfying $\eta/s = 1/4\pi$ (applying standard thermodynamic relation to determine the entropy density $s$). Also $\kappa^2$ is given by $\kappa^2= (8\pi G_N)^{-1}$, where $G_N$ is the five-dimensional Newton's constant. 

The conservation of the energy-momentum tensor is required for (\ref{metric1}) to be a solution of vacuum Einstein's equations. In fact, this follows from the vector constraint of Einstein's equation. The conservation of energy-momentum tensor yields the Navier-Stokes' equation. The latter implies that the velocity perturbation $\delta\mathbf{u}(\mathbf{k}_{\text{(h)}})$ is transverse, i.e.
\begin{equation}
\mathbf{\delta u} (\mathbf{k}_{\text{(h)}})\cdot \mathbf{k}_{\text{(h)}} = 0.
\end{equation}
The above procedure can be extended non-linearly to yield higher derivative \cite{fluidgravity2, Sayantani} and non-linear corrections \cite{Sayantani} to the Navier-Stokes equations.

The second example will be that of homogeneous relaxation. The energy-momentum tensor for homogeneous relaxation takes the following form :
\begin{equation}\label{hem}
t_{\mu\nu} = \text{diag}\, (\epsilon, p, p, p) + \pi_{ij}^{\text{(nh)}}(t).
\end{equation}
It is easy to see that the above energy-momentum tensor is conserved for any $\pi_{ij}^{\text{(nh)}}(t)$. In case of a conformal field theory, we further require
\begin{equation}\label{tr}
\pi_{ij}^{\text{(nh)}}(t)\,\delta_{ij} = 0.
\end{equation} 
The above is also implied by the scalar constraint of Einstein's equation.

It is convenient to do a Fourier transform over time to define $\pi_{ij}^{\text{(nh)}}(\omega_{\text{(nh)}})$. The corresponding metric up to linear order in $\pi_{ij}^{\text{(nh)}}(\omega_{\text{(nh)}})$ is \cite{myself2}:
\begin{eqnarray}\label{metricnh}
ds^2 = \frac{l^2}{r^2} \frac{dr^2}{f\left(\frac{r r_0}{l^2}\right)} +   \frac{l^2}{r^2}\Big(-f\left(\frac{r r_0}{l^2}\right)dt^2 + dx^2 + dy^2  \Big)  +\frac{2l^2}{r^2} \Big(\pi_{ij}^{\text{(nh)}}(\omega_{\text{(nh)}})  \ \tilde{h}\left(\frac{r r_0}{l^2}, \omega_{\text{(nh)}} \right)dx^i dx^j \Big) + O(\delta^2),
\end{eqnarray}
with $\delta$ being the parameter of non-hydrodynamic amplitude expansion parameter - $\pi_{ij}^{\text{(nh)}}/p$. Furthermore, $\tilde{h}(s, \omega_{\text{(nh)}})$ follows the equation of motion :
\begin{eqnarray}
\frac{d^2 \tilde{h}(s,\omega_{\text{(nh)}})}{ds^2} - \frac{\Big(2 + (1 + 3\frac{r_*^4}{r_0^4})s^3 - 6\frac{r_*^4}{r_0^4}s^4\Big)}{sf(s)}\frac{d\tilde{h}(s,\omega_{\text{(nh)}})}{ds} + \frac{\omega_{\text{(nh)}}^2 l^4}{r_0^2}\Bigg(\frac{1}{f^2(s)} \Bigg)\tilde{h}(s,\omega_{\text{(nh)}})= 0,
\end{eqnarray}
such that
\begin{equation}
\tilde{h}(s, \omega_{\text{(nh)}}) = s^3 + O(s^4) \ \text{as $s\rightarrow 0$.}
\end{equation}
This uniquely defines $\tilde{h}$.

It can be shown that the (\ref{metricnh}) is regular at the horizon provided $\pi_{ij}^{\text{(nh)}}(\omega_{\text{(nh)}})$ has simple poles in $\omega_{\text{(nh)}}$ exactly at the location of the homogeneous quasinormal modes \cite{myself2}. These quasinormal frequencies have both real and imaginary parts, and unlike hydrodynamic modes both of these are large, i.e. $O(T)$. In case of the $AdS_5$ black brane these are known to be \cite{Starinets} :
\begin{eqnarray}\label{freqh}
\omega_{(n)\pm} &=& \pi T \ \Big[ \pm 1.2139 - 0. 7775 \ i \pm 2n (1\mp i)\Big], \ \text{for large n}.
\end{eqnarray}
This follows from the general phenomenological equations (see apendix B).

The above analysis can be readily extended non-linearly in the amplitude parameter - $\pi_{ij}^{\text{(nh)}}/p$, but we need to sum over all time derivatives at each order in the amplitude expansion to see manifest regularity (see appendix B). This is because the time derivatives of these modes are large, i.e. $O(T)$. This is usual for non-hydrodynamic relaxation modes even close to equilibrium.

The general phenomenological equations as discussed in appendix B have covariant form. In our examples we have written them in the laboratory frame, where the final equilibrium configuration is at rest. This is useful from point of view of comparing our results for Green's functions with experiments. 

The derivative and amplitude expansions in the laboratory frame is as follows. The derivative expansion will count powers of $\mathbf{k}_{\text{(b)}}/T$, where $\mathbf{k}_{\text{(b)}}$ will be the momentum of the background quasinormal mode. It will be useful also to consider the quasinormal modes off-shell (and impose the on-shell condition later). In that case, the derivative expansion will also count powers of $\omega_{\text{(h)}}/T$, with $\omega_{\text{(h)}}$ being real and the frequency of the hydrodynamic quasinormal modes without the dispersion relation imposed. For a generic quasinormal mode, the off-shell frequency will be denoted by $\omega_{\text{(b)}}$. The amplitude expansion will count powers of $\delta\mathbf{u}, \delta T/ T^{\text{(eq)}}$ and $\pi_{ij}^{\text{(nh)}}/p^{\text{(eq)}}$, with all powers of $\omega_{\text{(b)}}/T$ summed up at each order for non-hydrodynamic modes.

\subsection{The scalar field in non-equilibrium geometries}

Here we will study the a free real scalar field in the background non-equilibrium geometries described above. This real scalar field will be dual to an appropriate order parameter like the chiral condensate in the field theory.  The typical temperature of the non-equilibirum state will be above that of symmetry restoration. Thus the chiral condensate will have no expectation value. Furthermore the chemical potentials will be assumed to be small, so the flavor and baryon currents can be neglected. These conditions indeed are valid for the quark-gluon plasma produced by heavy ion collisions at Relativistic Heavy Ion Collider (RHIC) in Brookhaven and at A Large Ion Collider Experiment (ALICE) in CERN. 

Thus our non-equilibrium backgrounds can be modeled by the regular perturbations of $AdS$ black branes in vacuum Einstein's gravity. As we are interested in the two-point correlations in these non-equilibrium geometries, we can ignore the backreaction of the scalar field, which will contribute only to higher-point correlations.

It is convenient to Fourier transform the dependence on the boundary coordinates. 
We first need to do this for the background metric.

Let us take the example of the hydrodynamic shear-mode perturbation discussed before for concreteness.  As evident from (\ref{metric1}), to achieve this, we need to Fourier transform the velocity perturbation $\delta\mathbf{u}$. We have already done the Fourier transform in the boundary spatial coordinates, but we haven't done the Fourier transform of the time dependence yet.

\begin{figure}
 \begin{center}
\includegraphics[width = 0.3\textwidth]{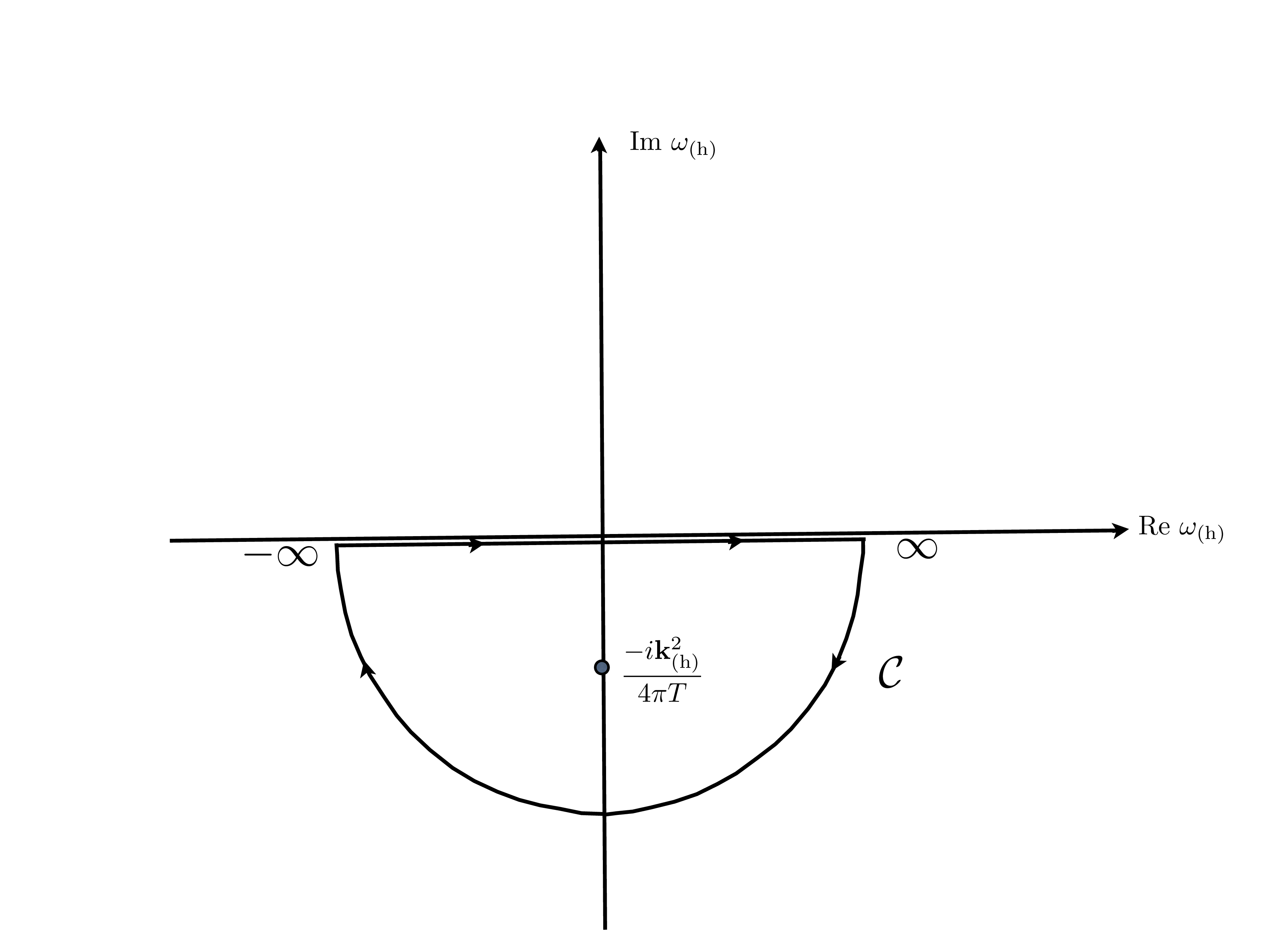}
\caption{Contour for integration $\mathcal{C}$ over $\omega_{\text{(h)}}$ which picks the contribution from the pole in the lower half plane}
\label{contour}
\end{center}
\end{figure}

The Fourier transform of $\delta\mathbf{u}(\mathbf{x},t)$ can be defined as \cite{noneqspec}:
\begin{equation}\label{uft}
\delta\mathbf{u}(\mathbf{k}_{\text{(h)}}, \omega_{\text{(h)}}) = - \Big(\frac{1}{2\pi i}\Big) \frac{\delta \mathbf{u}(\mathbf{k}_{\text{(h)}})}{\omega_{\text{(h)}} + i \frac{\mathbf{k}_{\text{(h)}}^2}{4\pi T}},
\end{equation}
such that $\delta \mathbf{u}^*(\mathbf{k}_{\text{(h)}}) = \delta \mathbf{u}(-\mathbf{k}_{\text{(h)}})$.  If we integrate over the $\omega_{\text{(h)}}$ over the real line and then closing in with the semicircle at infinity in the lower half plane as shown in fig. \ref{contour}, we find that
\begin{eqnarray}\label{uft1} \int_{\mathcal{C}} d\omega_{\text{(h)}}\,\, \delta\mathbf{u}(\mathbf{k}_{\text{(h)}}, \omega_{\text{(h)}}) \, e^{i\mathbf{k}_{\text{(h)}}\cdot \mathbf{x}}e^{-i\omega_{\text{(h)}}t} = 
\delta \mathbf{u}(\mathbf{k}_{\text{(h)}}) e^{i\mathbf{k}_{\text{(h)}}\cdot \mathbf{x}}e^{- \frac{\mathbf{k}_{\text{(h)}}^2}{4\pi T}t} .
\end{eqnarray}

Similarly in the case of the homogeneous perturbation given by (\ref{metricnh}) we can write $\pi_{ij}^{\text{(nh)}}(\omega_{\text{(nh)}})$ as
\begin{equation}\label{pift}
\pi_{ij}^{\text{(nh)}}(\omega_{\text{(nh)}})= - \sum_{n=0}^{\infty}\sum_{+,-}\Big(\frac{1}{2\pi i}\Big) \frac{a_{\text{(n)}ij}}{\omega_{\text{(nh)}} - \omega_{\text{(n)}\pm}},
\end{equation}
where $a_{\text{(n)}ij}$ are constants satisfying $a_{\text{(n)}ij}\delta_{ij}= 0$ and $\omega_{\text{(n)}\pm}$ satisfies (\ref{freqh}).

This can be generalized to any quasi-normal mode, which can also be represented as a solution of the linearized phenomenological equations. The dispersion relation of any quasi-normal mode can be written as :
\begin{eqnarray}\label{qd}
\omega_{\text{(b)}\pm}(\mathbf{k}_{\text{(b)}}) = \pm \text{Re}\,\omega_{\text{(b)}}(\mathbf{k}_{\text{(b)}}) + i  \text{Im}\,\omega_{\text{(b)}}(\mathbf{k}_{\text{(b)}}), \quad \text{Im}\,\omega_{\text{(b)}}(\mathbf{k}_{\text{(b)}}) &<& 0,
\end{eqnarray}
This gives rise to a complex pole in the Fourier transformed $\delta\mathbf{u}(\omega_{\text{(b)}}, \mathbf{k}_{\text{(b)}}), \delta T(\omega_{\text{(b)}}, \mathbf{k}_{\text{(b)}}), \pi_{ij}^{\text{(nh)}}(\omega_{\text{(b)}}, \mathbf{k}_{\text{(b)}})$ as below:
\begin{eqnarray}
\delta\mathbf{u}(\mathbf{k}_{\text{(b)}}, \omega_{\text{(b)}}) &=& - \Big(\frac{1}{2\pi i}\Big) \frac{\delta \mathbf{u}_\pm(\mathbf{k}_{\text{(b)}})}{\omega_{\text{(b)}} \pm \text{Re}\,\omega_{\text{(b)}}(\mathbf{k}_{\text{(b)}}) + i  \text{Im}\,\omega_{\text{(b)}}(\mathbf{k}_{\text{(b)}})}, \nonumber\\
\delta T(\mathbf{k}_{\text{(b)}}, \omega_{\text{(b)}}) &=& - \Big(\frac{1}{2\pi i}\Big) \frac{\delta T_\pm(\mathbf{k}_{\text{(b)}})}{\omega_{\text{(b)}} \pm \text{Re}\,\omega_{\text{(b)}}(\mathbf{k}_{\text{(b)}}) + i  \text{Im}\,\omega_{\text{(b)}}(\mathbf{k}_{\text{(b)}})}, \nonumber\\
\pi_{ij}^{\text{(nh)}}(\mathbf{k}_{\text{(b)}}, \omega_{\text{(b)}}) &=& - \Big(\frac{1}{2\pi i}\Big) \frac{a_{ij\pm}(\mathbf{k}_{\text{(b)}})}{\omega_{\text{(b)}} \pm \text{Re}\,\omega_{\text{(b)}}(\mathbf{k}_{\text{(b)}}) + i  \text{Im}\,\omega_{\text{(b)}}(\mathbf{k}_{\text{(b)}})}, \quad a_{ij\pm}(\mathbf{k}_{\text{(b)}})\delta_{ij}=0.
\end{eqnarray}
We will denote the frequency and momentum of the background perturbation as $\omega_{\text{(b)}}$ and $\mathbf{k}_{\text{(b)}}$ respectively to distinguish them clearly from those of the scalar field which will be denoted by $\omega$ and $\mathbf{k}$. Note both $\omega_{\text{(b)}}$ and $\omega$ are real by definition.

\begin{figure}
\begin{center}
\includegraphics[width = 0.3\textwidth]{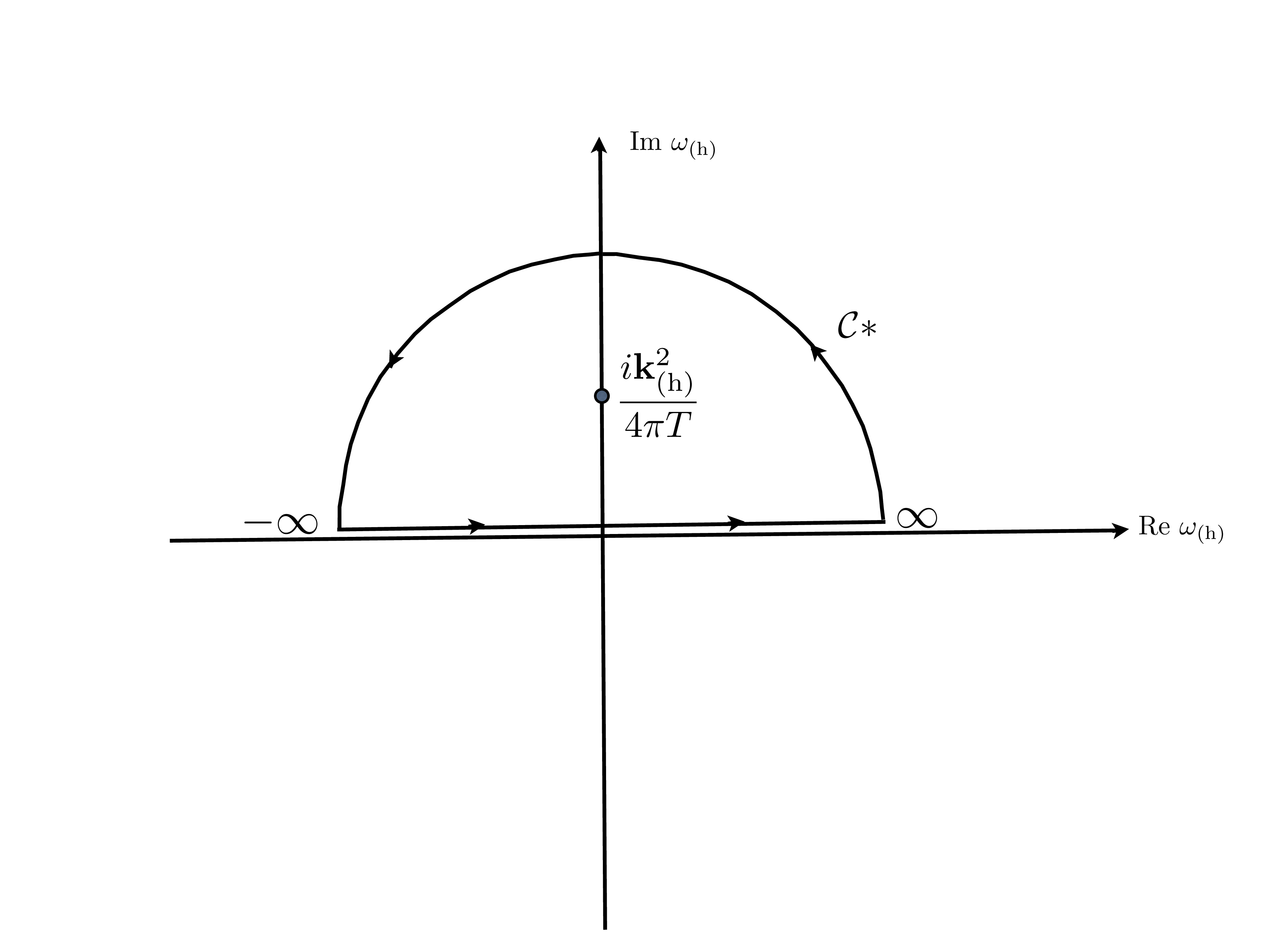} 
\caption{Contour for integration $\mathcal{C}*$ over $\omega_{\text{(h)}}$ which picks the contribution from the pole in the upper half plane}
\label{contourstar}
\end{center}
\end{figure}

The metric is real, and so are $\delta \mathbf{u}(\mathbf{x},t), \delta T(\mathbf{x},t)$ and $\pi_{ij}^{\text{(nh)}}(\mathbf{x},t)$. So we need to consider a pair of modes which are complex conjugates to each other. The complex conjugation involves a subtlety which has not been mentioned in our earlier work. It involves 
\begin{itemize}
\item reversing the sign of $\mathbf{k}_{\text{(b)}}$ and $\omega_{\text{(b)}}$, as for example we can see readily from (\ref{uft}) that
\begin{equation}
\delta\mathbf{u}(-\mathbf{k}_{\text{(h)}}, -\omega_{\text{(h)}}) =  \Big(\frac{1}{2\pi i}\Big) \frac{\delta \mathbf{u}(-\mathbf{k}_{\text{(h)}})}{\omega_{\text{(h)}} - i \frac{\mathbf{k}_{\text{(h)}}^2}{4\pi T}} =   \Big(\frac{1}{2\pi i}\Big) \frac{\delta \mathbf{u}*(\mathbf{k}_{\text{(h)}})}{\omega_{\text{(h)}} + i \frac{\mathbf{k}_{\text{(h)}}^2}{4\pi T}} = \delta\mathbf{u}*(\mathbf{k}_{\text{(h)}}, \omega_{\text{(h)}}), 
\end{equation}

\item the integration over $\omega_{(b)}$ which makes the mode on-shell should be done over the contour $\mathcal{C}*$ as shown in fig. \ref{contourstar} which goes over the real line and closes in the semicircle over the upper half plane instead of the lower half plane, as for example
\begin{eqnarray}
\int_{\mathcal{C}*} d\omega_{\text{(h)}}\,\, \delta\mathbf{u}(-\mathbf{k}_{\text{(h)}}, -\omega_{\text{(h)}}) \, e^{-i\mathbf{k}_{\text{(h)}}\cdot \mathbf{x}}e^{i\omega_{\text{(h)}}t} =
\delta \mathbf{u}(-\mathbf{k}_{\text{(h)}}) e^{-i\mathbf{k}_{\text{(h)}}\cdot \mathbf{x}}e^{- \frac{\mathbf{k}_{\text{(h)}}^2}{4\pi T}t}
 = \delta \mathbf{u}*(\mathbf{k}_{\text{(h)}}) e^{-i\mathbf{k}_{\text{(h)}}\cdot \mathbf{x}}e^{- \frac{\mathbf{k}_{\text{(h)}}^2}{4\pi T}t},
\end{eqnarray}
\end{itemize}
We note that the contour $\mathcal{C}*$ is a reflection of the original contour $\mathcal{C}$ over the real line. It is necessary to integrate the complex conjugates of (\ref{qd}) over $\mathcal{C}*$ so that the complex conjugate modes also decay in the future.  We readily see that we need to combine the two modes to produce a real perturbation, as for example
\begin{eqnarray}\label{sum2}
\int_{\mathcal{C}}d\omega_{\text{(b)}}\,\, \delta\mathbf{u}(\omega_{\text{(b)}},\mathbf{k}_{\text{(b)}}) e^{i\mathbf{k}_{\text{(b)}}\cdot\mathbf{x}}
e^{-i \omega_{\text{(b)}}t}+
\int_{\mathcal{C}*}d\omega_{\text{(b)}}
\,\,\delta\mathbf{u}(-\omega_{\text{(b)}},-\mathbf{k}_{\text{(b)}}) e^{-i\mathbf{k}_{\text{(b)}}\cdot\mathbf{x}}
e^{i \omega_{\text{(b)}}t}\nonumber\\
= \Big(\delta\mathbf{u}(\mathbf{k}_{\text{(b)}})
e^{i\mathbf{k}_{\text{(b)}}\cdot\mathbf{x}}
e^{-i\text{Re}\,\omega_{\text{(b)}}(\mathbf{k}_{\text{(b)}})t}
+\delta\mathbf{u}(-\mathbf{k}_{\text{(b)}})
e^{-i\mathbf{k}_{\text{(b)}}\cdot\mathbf{x}}
e^{i\text{Re}\,\omega_{\text{(b)}}(\mathbf{k}_{\text{(b)}})t}\Big)
e^{\text{Im}\,\omega_{\text{(b)}}(\mathbf{k}_{\text{(b)}})t},
\end{eqnarray}
for a general velocity perturbation $\delta\mathbf{u}(\mathbf{x},t)$.

The free scalar field satisfies the usual equation of motion in the non-equilibrium geometry which is
\begin{equation}
(\Box + m^2 )\phi(r, \mathbf{x}, t) = 0.
\end{equation}
The d'Alembertian $\Box$ however is that of the non-equilibrium geometry. 

Let us first consider the case when the perturbation of the geometry away from the $AdS$ black brane is linearized in $\delta\mathbf{u}, \delta T$ and $\pi_{ij}^{\text{(nh)}}$. Let these background perturbation be a sum of the specific mode given by the frequency $\omega_{\text{(b)}}$ and momentum $\mathbf{k}_{\text{(b)}}$ as in (\ref{qd}) and it's complex conjugate at frequency $-\omega_{\text{(b)}}$ and momentum $-\mathbf{k}_{\text{(b)}}$. 

The profile of the scalar field can be constructed as follows. Let us consider a specific mode of the scalar field at a frequency $\omega$ and momentum $\mathbf{k}$. Interacting with the quasinormal modes in the background, new modes at frequencies $\omega\pm\omega_{\text{(b)}}$ and momenta $\mathbf{k} \pm \mathbf{k}_{\text{(b)}}$ will be generated. So in the linearized non-equilibrium background the general solution of the scalar field will take the form :
\begin{eqnarray}\label{gp}
\Phi(\mathbf{x},t, r) &=& A(\omega, \mathbf{k})\Big(\Phi^{\text{(eq)}}(\omega, \mathbf{k}, r) e^{-i(\omega t- \mathbf{k}\cdot \mathbf{x})} + \int_{\mathcal{C}} d\omega_{\text{(b)}}\Phi^{\text{(neq)}}(\omega, \mathbf{k}, \omega_{\text{(b)}}, \mathbf{k}_{\text{(b)}}, r)e^{-i((\omega+\omega_{\text{(b)}}) t- (\mathbf{k}+\mathbf{k}_{\text{(b)}})\cdot \mathbf{x})} \nonumber\\
&& + \int_{\mathcal{C}*} d\omega_{\text{(b)}}\Phi^{\text{(neq)}}(\omega, \mathbf{k}, -\omega_{\text{(b)}}, -\mathbf{k}_{\text{(b)}}, r)e^{-i((\omega-\omega_{\text{(b)}}) t- (\mathbf{k}-\mathbf{k}_{\text{(b)}})\cdot \mathbf{x})}.
\end{eqnarray} 
We note the following points regarding the above equation :
\begin{itemize}
\item 
The non-equilibrium corrections given by $\Phi^{\text{(neq)}}(\omega, \mathbf{k}, \omega_{\text{(b)}}, \mathbf{k}_{\text{(b)}}, r)$ and $\Phi^{\text{(neq)}}(\omega, \mathbf{k}, -\omega_{\text{(b)}}, -\mathbf{k}_{\text{(b)}}, r)$ can be evaluated systematically in the derivative and amplitude expansions. 

\item Both $\Phi^{\text{(neq)}}(\omega, \mathbf{k}, \omega_{\text{(b)}}, \mathbf{k}_{\text{(b)}}, r)$ and $\Phi^{\text{(neq)}}(\omega, \mathbf{k}, -\omega_{\text{(b)}}, -\mathbf{k}_{\text{(b)}}, r)$ will be determined uniquely by the boundary conditions at the horizon.

\item We need to perform integrations over
$\omega_{\text{(b)}}$ using the contour prescriptions described above so that the background is on-shell.

\item The most general solution can be obtained as above by simply superposing over different equilibrium modes at $\omega$, $\mathbf{k}$ along with their unique non-equilibrium corrections, with an overall arbitrary coefficient $A(\omega, \mathbf{k})$.
\end{itemize}

As an explicit illustration let us examine the profile of the scalar field in the background metric (\ref{metric1}) with a hydrodynamic shear-wave perturbation.

The equation of motion of the equilibrium part $\Phi^{\text{(eq)}}(\omega,\mathbf{k})$ is
\begin{equation}
\Big(\Box^{ABB}_{\omega', \mathbf{k}'}+m^2\Big)\delta(\omega'-\omega)\delta^2(\mathbf{k}'-\mathbf{k})\Phi^{\text{(eq)}}(\omega, \mathbf{k}, r) = 0,
\end{equation}
where $\Box^{ABB}_{\omega, \mathbf{k}}$
is the Laplacian in the unperturbed AdS black brane metric given by :
\begin{eqnarray}\label{eqeom}
l^2\Box^{ABB}_{\omega, \mathbf{k}} = r^2 f\Big(\frac{rr_0}{l^2}\Big)\partial_r^2 +r\Big[-2 f\Big(\frac{rr_0}{l^2}\Big) +\frac{rr_0}{l^2}f'\Big(\frac{rr_0}{l^2}\Big)\Big] \partial_r 
+ r^2\Big[\frac{\omega^2}{f\Big(\frac{rr_0}{l^2}\Big)}-\mathbf{{k}^2}\Big].
\end{eqnarray}
Above, $f$ is the blackening function of the AdS Reissner-Nordstorm black brane which vanishes at the horizon located at $r = l^2/r_0$, and is as given in (\ref{fh}). 

The equations of motion of the non-equilibrium parts of the solution up to first order in the derivative expansion are :
\begin{eqnarray}\label{noneqeom}
\Big(\Box^{ABB}_{\omega', \mathbf{k}'}+m^2\Big)
\delta(\omega'-\omega-\omega_{\text{(h)}})
\delta^3(\mathbf{k}'-\mathbf{k}-
\mathbf{k}_{\text{(h)}})\Phi^{\text{(neq)}} (\omega, \mathbf{k},\omega_{\text{(h)}}, \mathbf{k}_{\text{(h)}},r) &=& V(\omega, \mathbf{k}, \omega_{\text{(h)}},\mathbf{k}_{\text{(h)}},r)\,\Phi^{\text{(eq)}}(\omega, \mathbf{k},r),\nonumber\\
\Big(\Box^{ABB}_{\omega', \mathbf{k}'}+m^2\Big)
\delta(\omega'-\omega+\omega_{\text{(h)}})
\delta^3(\mathbf{k}'-\mathbf{k}+
\mathbf{k}_{\text{(h)}})\Phi^{\text{(neq)}} (\omega, \mathbf{k},-\omega_{\text{(h)}}, -\mathbf{k}_{\text{(h)}},r) &=& V(\omega, \mathbf{k}, -\omega_{\text{(h)}},-\mathbf{k}_{\text{(h)}},r)\,\nonumber\\&& \Phi^{\text{(eq)}}(\omega, \mathbf{k},r),
\end{eqnarray}
with
\begin{eqnarray}\label{sterms}
V(\omega, \mathbf{k}, \omega_{\text{(h)}},\mathbf{k}_{\text{(h)}},r) &=& V_1(\omega, \mathbf{k}, \omega_{\text{(h)}},\mathbf{k}_{\text{(h)}},r)+V_2(\omega, \mathbf{k}, \omega_{\text{(h)}},\mathbf{k}_{\text{(h)}},r) + O(\epsilon^2), \nonumber\\
V_1(\omega, \mathbf{k}, \omega_{\text{(h)}},\mathbf{k}_{\text{(h)}},r) &=& \frac{2r^2}{l^2f\Big(\frac{rr_0}{l^2}\Big)}\,\omega\Big(1- f\Big(\frac{rr_0}{l^2}\Big)\Big)\delta \mathbf{u}(\omega_{\text{(h)}}, \mathbf{k}_{\text{(h)}})\cdot \mathbf{k}, \nonumber \\
V_2(\omega, \mathbf{k}, \omega_{\text{(h)}},\mathbf{k}_{\text{(h)}},r) &=&i \frac{r^2}{2r_0}  h\Big(\frac{rr_0}{l^2}\Big)k_ik_j k_{\text{(h)}i} \delta u_j(\omega_{\text{(h)}}, \mathbf{k}_{\text{(h)}}).
\end{eqnarray}
Above, $h(rr_0/l^2)$ gives the hydrodynamic correction to the background metric which is proportional to $k_{\text{(h)}i}\delta u_j + (i\leftrightarrow j)$, and is as given in (\ref{fh}). Also $V$ has been expanded up to first order in the derivative expansion only, the first term $V_1$ depicting the contribution  at the zeroth order in the derivative expansion, and $V_2$ depicting the contribution at the first order in the derivative expansion respectively. Both are linear in $\delta\mathbf{u}$ as we have treated the background in the quasi-normal mode approximation. Note the derivative expansion counts only the order of appearance of hydrodynamic frequency $\omega_{\text{(h)}}$ and $\mathbf{k}_{\text{(h)}}$ through derivatives of $\delta\mathbf{u}$, while dependence in $\omega$ and $\mathbf{k}$ should be treated exactly at each order. 

As indicated in (\ref{gp}), we need to integate over $\omega_{\text{(b)}}$ so that the background metric is on-shell. After this integration, 
the profile of the scalar field is :
\begin{eqnarray}\label{afterint}
\Phi(\mathbf{x},t, r) &=& \Phi^{\text{(eq)}}(\omega, \mathbf{k}, r) e^{-i(\omega t- \mathbf{k}\cdot \mathbf{x})} + 
\Big(\Phi^{\text{(neq)}}(\omega, \mathbf{k}, \mathbf{k}_{\text{(h)}}, r)e^{-i\omega t} 
e^{i (\mathbf{k}+\mathbf{k}_{\text{(h)}})\cdot \mathbf{x}}
\nonumber\\&&
+\Phi^{\text{(neq)}}(\omega, \mathbf{k}, -\mathbf{k}_{\text{(h)}}, r)e^{-i\omega t} 
e^{i (\mathbf{k}-\mathbf{k}_{\text{(h)}})\cdot \mathbf{x}}\Big)e^{-\frac{\mathbf{k}_{\text{(h)}}^2}{4\pi T}t}, 
\end{eqnarray}
Above, $\Phi^{\text{(neq)}}(\omega, \mathbf{k}, \mathbf{k}_{\text{(h)}}, r)$ and $\Phi^{\text{(neq)}}(\omega, \mathbf{k}, -\mathbf{k}_{\text{(h)}}, r)$ are
the residues of $\Phi^{\text{(neq)}}(\omega, \mathbf{k}, \omega_{\text{(h)}}, \mathbf{k}_{\text{(h)}}, r)$ and  
$\Phi^{\text{(neq)}}(\omega, \mathbf{k}, -\omega_{\text{(h)}}, -\mathbf{k}_{\text{(h)}}, r)$
at the poles $\omega_{\text{(h)}} = \mp i \mathbf{k}_{\text{(h)}}^2/(4\pi T)$ respectively. $\Phi^{\text{(neq)}}(\omega, \mathbf{k}, \omega_{\text{(h)}}, \mathbf{k}_{\text{(h)}}, r)$ and $\Phi^{\text{(neq)}}(\omega, \mathbf{k}, -\omega_{\text{(h)}},-\mathbf{k}_{\text{(h)}}, r)$ depend linearly on  $\delta\mathbf{u}(\omega_{\text{(h)}}, \mathbf{k}_{\text{(h)}})$ and $\delta\mathbf{u}(-\omega_{\text{(h)}}, -\mathbf{k}_{\text{(h)}})$ respectively through the $V$ terms in (\ref{noneqeom}). The contours $\mathcal{C}$ and $\mathcal{C}*$ pick up the residues of the poles in $\delta\mathbf{u}(\omega_{\text{(h)}}, \mathbf{k}_{\text{(h)}})$ and $\delta\mathbf{u}(-\omega_{\text{(h)}}, -\mathbf{k}_{\text{(h)}})$ in the upper and lower half planes respectively. 

It follows that $\Phi^{\text{(neq)}}\left(\omega, \mathbf{k},\mathbf{k}_{\text{(h)}}, r\right)$ takes the following form up to first order in the derivative expansion : 
\begin{equation}\label{phih1}
\Phi^{\text{(neq)}}(\omega,\mathbf{k},\mathbf{k}_{\text{(h)}},r) = \Phi^{\text{(neq)}}_1(\omega, \mathbf{k}, r) \   \delta\mathbf{u}( \mathbf{k}_{\text{(h)}})\cdot \mathbf{k} 
+ \Phi^{\text{(neq)}}_2(\omega, \mathbf{k}, r)\  k_i k_j k_{\text{(h)}i}\delta u_j( \mathbf{k}_{\text{(h)}}) + O(\epsilon^2).
\end{equation}
The similar expansion for $\Phi^{\text{(neq)}}(\omega,\mathbf{k},-\mathbf{k}_{\text{(h)}},r)$ is obtained simply by reversing the sign of $\mathbf{k}_{\text{(h)}}$ above.

The above procedure can be readily generalized when the background metric is expanded non-linearly in the perturbations parametrized by $\delta\mathbf{u}$, $\delta T$ and $\pi_{ij}^{\text{(nh)}}$

Let us focus on the case of the non-linear hydrodynamic background whose explicit forms can be found in the literature. Let us consider the Laplacian when we take into account quadratic dependence on two distinct velocity perturbations $\delta \mathbf{u}(k_{\text{(h)}})$ and $\delta \mathbf{u}(k_{\text{(h)}}')$ for instance, at a given order in the derivative expansion $m$ (i.e. at the $m$th order in the hydrodynamic momentum). The solution for $\Phi$ will receive a correction quadratic in the amplitude of velocity perturbation, and at $m$th order in the derivative expansion this takes the form
\begin{equation}
\Phi^{(2,m)}\Big(r, k, k_{\text{(h)}}, k_{\text{(h)}}'\Big)\, e^{i(k + k_{\text{(h)}} + k_{\text{(h)}}')\cdot x} + (k_{\text{(h)}}\rightarrow -k_{\text{(h)}}) + (k_{\text{(h)}}'\rightarrow -k_{\text{(h)}}') + (k_{\text{(h)}}\rightarrow -k_{\text{(h)}}, k_{\text{(h)}}'\rightarrow -k_{\text{(h)}}') .
\end{equation}
The radial dependence above can be determined the equation of motion :
\begin{equation}\label{noneqeomnl}
\Box^{ABB}_{k'} \delta^3 (k' -k -k_{\text{(h)}}-k_{\text{(h)}}' )
\Phi^{(2,m)}(r, k', k_{\text{(h)}}, k_{\text{(h)}}')= S^{(2,m)}(r, k,k_{\text{(h)}}, k_{\text{(h)}}'),
\end{equation}
where  $\Box^{ABB}_k$ is the Laplacian for a scalar with four-momentum $k$ in the unperturbed $AdS_5$ black brane and $S^{(2,m)}$ is the source term determined by the background perturbation. For $m=1$ i.e. at first order in the derivative expansion, the source term $S^{(2,1)}$ can contain terms like $(\mathbf{k}\cdot\delta\mathbf{u}(k_{\text{(h)}}))(\mathbf{k_{\text{(h)}}}\cdot\delta\mathbf{u}(k_{\text{(h)}}'))\Phi^{\text{(eq)}}$, $(\mathbf{k_{\text{(h)}}}\cdot\delta\mathbf{u}(k_{\text{(h)}}'))\Phi^{(1,0)}$ etc. Finally we need to integrate over $\omega_{\text{(h)}}$ and $\omega_{(h)}^\prime$ over $\mathcal{C}$ or $\mathcal{C}^\prime$ so that the background metric is taken on-shell.

Thus perturbatively the solution of the scalar field in an arbitrary background can be systematically expanded both in the derivative and amplitude expansions.

\subsection{The boundary conditions for regularity}

The behavior of the general solution $\Phi^{\text{(eq)}}(\omega, \mathbf{k}, r)$ in the equilibrium black brane background
can be split into one that is incoming at the horizon and another that is outgoing at the horizon. 
The incoming mode $\Phi^{\text{in}\text{(eq)}}(\omega, \mathbf{k}, r)$ can be defined uniquely by it's behavior near the horizon $r = \frac{l^2}{r_0}$ as below :
\begin{eqnarray}\label{solin}
\Phi^{\text{in(eq)}}(\omega, \mathbf{k}, r) &\approx& \Big(1- \frac{rr_0}{l^2}\Big)^{-i\frac{\omega}{4\pi T}} 
\end{eqnarray}
The outgoing mode  $\Phi^{\text{out(eq)}}(\omega, \mathbf{k})$ can be similarly defined uniquely by it's behavior near the horizon given by :
\begin{eqnarray}\label{solout}
\Phi^{\text{out(eq)}}(\omega, \mathbf{k}, r) &\approx& \Big(1- \frac{rr_0}{l^2}\Big)^{i\frac{\omega}{4\pi T}}
\end{eqnarray} 
Also, 
\begin{equation}\label{ccinout}
\Phi^{\text{in(eq)}}*(\omega, \mathbf{k}, r)=\Phi^{\text{in(eq)}}(-\omega, -\mathbf{k}, r) = \Phi^{\text{out(eq)}}(\omega, \mathbf{k}, r) = \Phi^{\text{out(eq)}}*(-\omega, -\mathbf{k}, r).
\end{equation}
In fact, the effect of reversing the sign of $\mathbf{k}$ is trivial as the solution can only depend on it's modulus due to the rotational symmetry of the black brane background. 

It is natural to include only the incoming solution when $\omega > 0$, because it respects the causal structure of the black brane which forbids anything coming out of the horizon classically.  We can also follow \cite{Horowitz} to argue that if we keep both incoming and outgoing modes, it will cause a singular back-reaction at the horizon. As we will be consider quasinormal mode perturbations of the background metric which are incoming at the horizon, we should exclude the outgoing solution of the scalar field to prevent singular backreaction at the horizon. Thus the general solution at the zeroth order is
\begin{equation}\label{solh0}
\Phi^{\text{(eq)}}(\omega, \mathbf{k}, r) = A^{\text{in(eq)}}(\omega, \mathbf{k})\Phi^{\text{in(eq)}}(\omega, \mathbf{k},r).
\end{equation}

Until the end of this subsection we will assume $\omega>0$. Later we will add a complex conjugate term necessary to make the full solution real. This complex conjugate will take care of the case $\omega<0$.

In order to study the regularity of the non-equilibrium solution let us first examine the approximation where the background metric perturbation is linearized in $\delta\mathbf{u}, \delta T$ and $\pi_{ij}^{\text{(nh)}}$. In particular let us study the case of the hydrodynamic shear wave perturbation.

It is easy to see from (\ref{noneqeom}) that the general form of the non-equilibrium part of the solution can be written as 
\begin{eqnarray}\label{solh1}
\Phi^{\text{(neq)}}(\omega,  \mathbf{k},\omega_{\text{(h)}},\mathbf{k}_{\text{(h)}}, r) &=& A^{\text{in(neq)}}\left(\omega, \mathbf{k},\omega_{\text{(h)}}, \mathbf{k}_{\text{(h)}}\right)\Phi^{\text{in(eq)}}(\omega +\omega_{\text{(h)}}, \mathbf{k} +\mathbf{k}_{\text{(h)}}, r) \nonumber\\&&+ A^{\text{out(neq)}}\left(\omega, \mathbf{k},\omega_{\text{(h)}}, \mathbf{k}_{\text{(h)}}\right)\Phi^{\text{in(eq)}}(\omega +\omega_{\text{(h)}}, \mathbf{k} +\mathbf{k}_{\text{(h)}}, r)\nonumber\\
&&+ A^{\text{in(eq)}}(\omega, \mathbf{k})\Phi^{\text{in(neq)}}(\omega, \mathbf{k}, \omega_{\text{(h)}}, \mathbf{k}_{\text{(h)}}, r) + (\omega_{\text{(h)}}\rightarrow -\omega_{\text{(h)}}, \mathbf{k}_{\text{(h)}}\rightarrow-\omega_{\text{(h)}}).
\end{eqnarray}
up to first order in the derivative expansion. The first two lines above indicate the homogeneous solutions of (\ref{noneqeom}) but now with $\omega$ replaced by $\omega+\omega_{\text{(h)}}$ and $\mathbf{k}$ replaced by $\mathbf{k} + \mathbf{k}_{\text{(h)}}$. The coefficients of these homogeneous solutions $A^{\text{in(neq)}}(\omega, \omega_{\text{(h)}}, \mathbf{k}, \mathbf{k}_{\text{(h)}})$ and $A^{\text{out(neq)}}(\omega, \omega_{\text{(h)}}, \mathbf{k}, \mathbf{k}_{\text{(h)}})$ respectively have to be linear in $\delta\mathbf{u}$ and have a consistent derivative expansion also. Their dependence on $\omega_{\text{(h)}}$ and $\mathbf{k}_{\text{(h)}}$ can be expanded systematically in terms of rotationally invariant scalars like $\delta\mathbf{u}\cdot \mathbf{k}$, $k_i k_j k_{\text{(h)}i}\delta u_j$, $\omega_{\text{(h)}}k_i k_j k_{\text{(h)}i}\delta u_j$, etc. Up to first order in the derivative expansion, only the first two scalars will apear. The coefficients of these scalars should be functions of $\omega$ and $\mathbf{k}$ only, as the depenedence on $\omega_{\text{(h)}}$ and $\mathbf{k}_{\text{(h)}}$ can be absorbed in coefficients of the scalars appearing at higher orders in the derivative expansion. Thus, up to first order in derivative expansion, we should have :
\begin{eqnarray}\label{coeffs}
A^{\text{in(neq)}}(\omega, \omega_{\text{(h)}},\mathbf{k}, \mathbf{k}_{\text{(h)}}) &=& A^{\text{in(neq)}}_1(\omega, \mathbf{k}) \   \delta\mathbf{u}(\omega_{\text{(h)}}, \mathbf{k}_{\text{(h)}})\cdot \mathbf{k} + A^{\text{in(neq)}}_2(\omega, \mathbf{k})\  k_i k_j k_{\text{(h)}i}\delta u_j(\omega_{\text{(h)}}, \mathbf{k}_{\text{(h)}}) + O(\epsilon^2),\nonumber\\
A^{\text{out(neq)}}(\omega, \omega_{\text{(h)}},\mathbf{k}, \mathbf{k}_{\text{(h)}}) &=& A^{\text{out(neq)}}_1(\omega, \mathbf{k}) \   \delta\mathbf{u}(\omega_{\text{(h)}}, \mathbf{k}_{\text{(h)}})\cdot \mathbf{k} + A^{\text{out(neq)}}_2(\omega, \mathbf{k})\  k_i k_j k_{\text{(h)}i}\delta u_j(\omega_{\text{(h)}}, \mathbf{k}_{\text{(h)}})+O(\epsilon^2).
\end{eqnarray}
We recall for the hydrodynamic shear mode $\delta\mathbf{u}\cdot\mathbf{k}_{\text{(h)}} =0$, so there are no more possible terms up to first order in $\mathbf{k}_{\text{(h)}}$. Above $ A^{\text{in(neq)}}_1(\omega, \mathbf{k})$, $ A^{\text{in(neq)}}_2(\omega, \mathbf{k})$, $ A^{\text{out(neq)}}_1(\omega, \mathbf{k})$ and $ A^{\text{out(neq)}}_2(\omega, \mathbf{k})$ are arbitrary.

The last line of (\ref{solh1}) denote the particular solution determined completely by the perturbation of the background (\ref{sterms}) and the zeroth order solution (\ref{solh0}) as appearing on the right hand side of (\ref{noneqeom}). There is no new constant appearing here. $\Phi^{\text{in(neq)}}(\omega, \mathbf{k}, \omega_{\text{(h)}}, \mathbf{k}_{\text{(h)}}, r)$ is the particular solution sourced by $\Phi^{\text{in(eq)}}(\omega, \mathbf{k}, r)$ and can be thought of as the non-equilibrium correction to the incoming mode. It's behavior at the horizon is given by :
\begin{eqnarray}\label{phiinneq}
\Phi^{\text{in(neq)}}(\omega, \mathbf{k}, \omega_{\text{(h)}}, \mathbf{k}_{\text{(h)}}, r)&\approx & i2\Bigg(\frac{\pi T l^2}{r_0}\Bigg)^2 \frac{\omega \delta \mathbf{u}(\omega, \mathbf{k}_{\text{(h)}})\cdot \mathbf{k}}{(2\omega+\omega_{\text{(h)}})\omega_{\text{(h)}}} \Bigg(1 - \frac{rr_0}{l^2}\Bigg)^{-i\frac{\omega}{4\pi T}}.
\end{eqnarray}
We observe this particular solution is simply proportional to the equilibrium incoming mode at the leading order and hence is regular at the horizon.

Just like in the case of equilibrium, we should discard the outgoing solution to prevent potentially harmful backreaction. Therefore, we should choose
\begin{equation}
A^{\text{out(neq)}}\left(\omega, \mathbf{k},\omega_{\text{(h)}}, \mathbf{k}_{\text{(h)}}\right) = 0, \quad\text{i.e.}\quad A^{\text{out(neq)}}_i\left(\omega, \mathbf{k}\right) = 0, \, \text{for i = 1,2, ..}\,.
\end{equation}

We recall that as in (\ref{gp}) we need to integrate over $\omega_{\text{(h)}}$ on the contour $\mathcal{C}$ given in fig. \ref{contour}. It is clear from the form of $A^{\text{in(neq)}}\left(\omega, \mathbf{k},\omega_{\text{(h)}}, \mathbf{k}_{\text{(h)}}\right) $ in (\ref{coeffs}) that the pole in the lower half plane of $\delta\mathbf{u}(\omega_{\text{(h)}},\mathbf{k}_{\text{(h)}})$ as given by (\ref{uft}) will produce a divergence at the horizon whose leading term is :
\begin{equation}\label{ld}
\Bigg(1- \frac{rr_0}{l^2}\Bigg)^{- i \frac{\omega}{4\pi T} - \frac{\mathbf{k}_{\text{(h)}}^2}{16\pi^2 T^2}}.
\end{equation}
This can be readily seen from the behavior of the incoming homogeneous solution $\Phi^{\text{in(eq)}}(\omega + \omega_{\text{(h)}},\mathbf{k}+ \mathbf{k}_{\text{(h)}}, r)$ near the horizon as given by (\ref{solin}), substituting $\omega_{\text{(h)}}$ with it's on-shell value. To get rid of this divergence, we also need to put 
\begin{equation}
A^{\text{in(neq)}}\left(\omega, \mathbf{k},\omega_{\text{(h)}}, \mathbf{k}_{\text{(h)}}\right) = 0, \quad\text{i.e.}\quad A^{\text{in(neq)}}_i\left(\omega, \mathbf{k}\right) = 0, \, \text{for i = 1,2, ..}\,.
\end{equation}

Similarly we can write the general solution for $\Phi^{\text{(neq)}}(\omega, \mathbf{k}, -\omega_{\text{(h)}}, -\mathbf{k}_{\text{(h)}}, r)$ in (\ref{noneqeom}), as a sum of  homogeneous
solutions given by $\Phi^{\text{in(eq)}}(\omega -\omega_{\text{(h)}}, \mathbf{k}-\mathbf{k}_{\text{(h)}}, r)$ and 
$\Phi^{\text{out(eq)}}(\omega - \omega_{\text{(h)}}, \mathbf{k}- \mathbf{k}_{\text{(h)}}, r)$ with coefficients 
$A^{\text{in(neq)}}(\omega, -\omega_{\text{(h)}},\mathbf{k}, -\mathbf{k}_{\text{(h)}})$ and 
$A^{\text{out(neq)}}(\omega, -\omega_{\text{(h)}},\mathbf{k}, -\mathbf{k}_{\text{(h)}})$ respectively. These coefficients are arbitrary but have a consistent derivative
expansion as in (\ref{coeffs}). The particular
solution is $A^{\text{in(eq)}}(\omega, \mathbf{k})\Phi^{\text{in(neq)}}(\omega, \mathbf{k}, -\omega_{\text{(h)}}, -\mathbf{k}_{\text{(h)}}, r)$ with coefficient
 $A^{\text{in(eq)}}(\omega, \mathbf{k})$ and $\Phi^{\text{out(neq)}}(\omega, \mathbf{k}, -\omega_{\text{(h)}}, -\mathbf{k}_{\text{(h)}}, r)$ which is completely determined by the equilibrium solution and is regular at the horizon.

Once again, we should put $A^{\text{out(neq)}}$ to zero to prevent harmful backreaction. We should also  $A^{\text{in(neq)}}$ to zero as the integration over contour $\mathcal{C}*$ (see fig. \ref{contourstar}) as in (\ref{gp}) will produce a leading divergence at the horizon of the form (\ref{ld}).

In a way these boundary condition for regularity is also a consequence of reality constraints (\ref{ainoutcc}) which relates $A^{\text{in(neq)}}$ to  $A^{\text{out(neq)}}$. We note as the background metric is real, for every background $(\omega_{\text{(b)}}, \mathbf{k}_{\text{(b)}})$ quasinormal mode, we need to include the complex conjugate  $(-\omega_{\text{(b)}}, -\mathbf{k}_{\text{(b)}})$ quasinormal mode. Thus (\ref{ainoutcc}) tells us if the coefficients of the outgoing homogeneous solutions are set to zero, so should be the coefficients of the ingoing homogeneous solutions.

The particular solutions which remain are regular at the horizon. Therefore regularity at the horizon uniquely determines the non-equilibrium correction to the profile of the scalar field in the unperturbed black brane background. The full solution is :
\begin{eqnarray}\label{gp1}
A(\omega, \mathbf{k})\Big(\Phi^{\text{(eq)}}(\omega, \mathbf{k}, r) e^{-i(\omega t- \mathbf{k}\cdot \mathbf{x})} + \int_{\mathcal{C}} d\omega_{\text{(h)}}\Phi^{\text{in(neq)}}(\omega, \mathbf{k}, \omega_{\text{(h)}}, \mathbf{k}_{\text{(h)}}, r)e^{-i((\omega+\omega_{\text{(h)}}) t- (\mathbf{k}+\mathbf{k}_{\text{(h)}})\cdot \mathbf{x})} \nonumber\\
+ \int_{\mathcal{C}*} d\omega_{\text{(h)}}\Phi^{\text{in(neq)}}(\omega, \mathbf{k}, -\omega_{\text{(h)}}, -\mathbf{k}_{\text{(h)}}, r)e^{-i((\omega-\omega_{\text{(h)}}) t- (\mathbf{k}-\mathbf{k}_{\text{(h)}})\cdot \mathbf{x})}\Big).
\end{eqnarray}

The above boundary conditions  apply to all quasinormal mode backgrounds. In each case, we need to put the coefficient of the outgoing mode at the horizon to zero to prevent harmful backreaction. Also the coefficient of the incoming mode should be put to zero, otherwise the $\omega_{\text{(b)}}$ integration will produce divergence at the horizon whose leading term will take the form :
\begin{equation}
\Bigg(1- \frac{rr_0}{l^2}\Bigg)^{- i \frac{\left(\omega + \, \text{Re}\,\omega_{\text{(b)}}(\mathbf{k}_{\text{(b)}})\right)}{4\pi T}}\Bigg(1- \frac{rr_0}{l^2}\Bigg)^{ \frac{\text{Im}\omega_{\text{(b)}}(\mathbf{k}_{\text{(b)}})}{4\pi T}}.
\end{equation}
The above is divergent because for any quasinormal mode $\text{Im}\,\omega_{\text{(b)}}(\mathbf{k}_{\text{(b)}})<0$.

The particular solution is regular at the horizons in all the instances checked so far. As for instance in case of the metric perturbed by the homogeneous relaxation mode in (\ref{metricnh}), the particular solution vanishes at the horizon due to factors like :
\begin{equation*}
\Big(1- \frac{rr_0}{l^2}\Big)^n \Bigg(\ln \Big(1- \frac{rr_0}{l^2}\Big)\Bigg)^m .
\end{equation*}
This particular solution gives the unique non-equilibrium correction to the equilibrium solution. The full solution as in (\ref{gp1}) has only one arbitrary constant which is $A(\omega, \mathbf{k})$.

The above boundary conditions are also valid when we take into account non-linear dependence of the background metric on $\delta\mathbf{u}, \delta T$ and $\pi_{ij}^\text{(nh)}$.

Let us consider the case of the scalar field in the non-linear hydrodynamic background. In particular, let us consider quadratic dependence on $\delta\mathbf{u}$ at $m$-th order in the derivative expansion. This non-equilibrium correction is $\Phi^{(2,m)}$ whose equation of motion is (\ref{noneqeomnl}). Clearly the general solution of $\Phi^{(2,m)}$ near the horizon can again be separated into two homogeneous pieces, the incoming and the outgoing modes, and a particular piece which has no arbitrary integration constant and is completely determined by the source term $S^{(2,m)}$. In order to prevent harful backreaction, we should put the coefficient of the outgoing mode to zero. Also as discussed before, the integration over the hydrodynamic frequencies $\omega_{\text{(h)}}$ and  $\omega_{\text{(h)}}'$ will produce a divergence at the horizon for the incoming mode, as for instance in the case above with dependence on two hydrodynamic shear wave background modes like :
\begin{equation}
\Big(1 - \frac{rr_0}{l^2}\Big)^{-i\frac{\omega}{4\pi T}- \frac{\mathbf{k}_{\text{(h)}}^2}{16 \pi^2 T^2}- \frac{\mathbf{k'}_{\text{(h)}}^2}{16 \pi^2 T^2} - ...}.
\end{equation}
Obviously the coefficient of the incoming mode has to depend on $\delta\mathbf{u}(\mathbf{k}_{\text{(h)}})$ and $\delta\mathbf{u}(\mathbf{k}_{\text{(h)}}^\prime)$ as required by the order in the perturbation expansion as in (\ref{coeffs}). The poles in $\omega_{\text{(h)}}$ and $\omega_{\text{(h)}}^\prime$ as in (\ref{uft}) produce the above divergent behavior after integration over contour $\mathcal{C}$ or $\mathcal{C}*$. In a consistent holographic set-up, the particular solution should be regular at the horizon.

Thus the full solution of the scalar field in the non-equilibrium background is undertermined only up to an overall constant $A(\omega, \mathbf{k})$. This is beacuse the equilibrium solution $\Phi^{\text{in}}(\omega, \mathbf{k})$ has
a unique non-equilibrium correction which is regular at the horizon.

We also note that for the full solution of $\Phi(r,\mathbf{x},t)$ to be real we need to superimpose also $\Phi^{\text{in}}(r,-\omega, -\mathbf{k})=\Phi^{\text{out}}(r, \omega, \mathbf{k})$ with $\Phi^{in}(r, \omega, \mathbf{k})$. Note the outgoing boundary condition is indeed natural for the negative frequency modes. As long as there is CPT invariance in the dual theory, the outgoing mode with negative frequency is the natural CPT conjugate of the incoming mode with positive frequency. 

We also note that the non-equilibrium correction to $\Phi^{\text{in}}(r,-\omega, -\mathbf{k})$ is also complex conjugate to the non-equilibrium correction to $\Phi^{\text{in}}(r,\omega,\mathbf{k})$. In particular the correction to $\Phi^{\text{in}}(r,-\omega, -\mathbf{k})$ due to the background quasinormal modes are complex conjugates to the correction to $\Phi^{\text{in}}(r,-\omega, -\mathbf{k})$ due to the complex conjugated quasinormal modes, i.e.
\begin{equation}
\Phi^{\text{in(neq)}}*(r, \omega, \mathbf{k}, \omega_{\text{(b)}}, \mathbf{k}_{\text{(b)}},\omega_{\text{(b)}}^\prime, \mathbf{k}_{\text{(b)}}^\prime, ...) = 
\Phi^{\text{in(neq)}}(r, -\omega, -\mathbf{k}, -\omega_{\text{(b)}}, -\mathbf{k}_{\text{(b)}},-\omega_{\text{(b)}}^\prime, -\mathbf{k}_{\text{(b)}}^\prime, ...) .
\end{equation}
As in the background perturbations for every $(\omega_{\text{(b)}}, \mathbf{k}_{\text{(b)}})$ mode there is the complex conjugate $(-\omega_{\text{(b)}}, -\mathbf{k}_{\text{(b)}})$ mode, the full non-equilibrium corrections of $\Phi^{in}(\omega, \mathbf{k})$ and $\Phi^{in}(-\omega,-\mathbf{k})$ are complex conjugates of each other. We can readily see that the complex conjugate is also regular at the horizon.

Similarly
\begin{equation}
\Phi^{\text{out(neq)}}*(r, \omega, \mathbf{k}, \omega_{\text{(b)}}, \mathbf{k}_{\text{(b)}},\omega_{\text{(b)}}^\prime, \mathbf{k}_{\text{(b)}}^\prime, ...) = 
\Phi^{\text{out(neq)}}(r, -\omega, -\mathbf{k}, -\omega_{\text{(b)}}, -\mathbf{k}_{\text{(b)}},-\omega_{\text{(b)}}^\prime, -\mathbf{k}_{\text{(b)}}^\prime, ...) .
\end{equation}

For the more general solution which is not necessarily regular at the horizon, we also need
\begin{eqnarray}\label{ainoutcc}
 A^{\text{in(neq)}}(\omega, \mathbf{k},  \omega_{\text{(b)}},\mathbf{k}_{\text{(b)}},\omega_{\text{(b)}}^\prime, \mathbf{k}_{\text{(b)}}^\prime, ... )  &=& A^{\text{out(neq)}}(\omega, \mathbf{k}, - \omega_{\text{(b)}},- \mathbf{k}_{\text{(b)}} , -\omega_{\text{(b)}}^\prime, -\mathbf{k}_{\text{(b)}}^\prime, ...).
\end{eqnarray}
for the solution to be real. We can readily see these imply specific relations between coefficients of the derivative expansion of both sides of the above equation. We can also derive similar requirements at higher orders in the amplitude expansion.

\section{Boundary conditions and non-equilibrium response functions}

In the previous section, we have studied the solution of a free scalar field in a non-equilibrium geometry. These geometries represents non-equilibrium states which are perturbatively connected to thermal equilibrium in the derivative and amplitude expansions. Utilizing these expansions, we will now build a framework in which we can determine the non-equilibrium corrections to the Green's functions holographically.

\subsection{On the physical significance of the boundary conditions at the horizon}

We have seen in the previous section that the non-equilibrium modification of the solution of the scalar field is determined uniquely by regularity. We required the non-equilibrium modes generated by the background quasi-normal modes to be regular at the horizon. It amounted to putting both the boundary conditions at the horizon, namely setting the coefficient of the incoming and outgoing homogeneous solutions to zero.

We may wonder what is the physical significance of putting both boundary conditions at the horizon in terms of the dual field theory. This has been partly explored in \cite{noneqspec}. 

Firstly, we indeed expect that the external source will receive non-equilibrium medium modifications just like the expectation value of the operator. This is because the external source as seen by the quasiparticles should be screened by the non-equilibrium collective excitations of the medium. Thus the Dirichlet boundary condition is the wrong thing to use for the non-equilibrium bulk modes as this would imply no such screening happens \cite{noneqspec}.

Secondly, the quasi-particle dispersion relation is modified by the non-equilibrium collective excitations of the medium. We know for instance the mass of the quasiparticles receive thermal contributions. Taking this thermal mass into account is necessary for curing infrared divergences in thermal field theory. In non-equilibrium field theory the effective mass is space-time dependent, which is natural given that the temperature is space-time dependent as well. Furthermore, the effective mass can depend as well on the velocity and shear-stress perturbations also. There is no systematic way to obtain the dependence of the effective mass as a function of the background temperature, velocity and shear-stress perturbations in non-equilibrium field theory.

Our holographic prescription gives a way to achieve this. The quasiparticle dispersion relation in equilibrum is obtained by solving for $\omega^{\text{(eq)}}(\mathbf{k})$ at a given $\mathbf{k}$, such that the external source satisfies $\mathcal{J}^{\text{(eq)}}(\omega^{\text{(eq)}}(\mathbf{k}), \mathbf{k})= 0$, the latter being determined by the incoming boundary condition at the horizon. This can be readily generalized to non-equilibrium to obtain the space-time dependent shifts of the quasi-particle dispersion relations as parametrized by the background velocity, temperature and shear-stress perturbations \cite{noneqspec}. This is reviewed in appendix C. 

The shift in the effective mass should come from resummation of infra-red divergences. As the infra-red physics is given by the dynamics of the horizon holographically, this indeed justifies why both the boundary conditions for the non-equilibrium modes should be applied at the horizon. Thus we determine the non-equilibrium modification of the source uniquely and hence the spacetime shifts of the quasiparticle dispersion relations. 

Building on this intuition, we can claim that the dependence of the non-equilibrium contributions of the Green's functions on the temperature, velocity and shear-stress perturbations characterizing the collective excitations should also come from resumming infra-red divergences. Hence, we can build on the hypothesis that the boundary conditions which will determine the non-equilibrium contributions to the propagators should be applied at the horizon. 

Thus we will examine general boundary conditions at the horizon which need not lead to a regular solution in the bulk. In the next subsection, we will study how the non-equilibrium corrections to the external source and the expectation value of the operator are parametrized by these boundary conditions. Then in the following subsection, we will use linear response theory and field theoretic consistency to determine the relevant boundary conditions for a given non-equilibrium Green's function.

We will find eventually that regularity at the horizon determines all the non-equilibrium Green's functions in the holographic classical gravity approximation. In the final subsection we will return to  non-equilibrium field theory to intuitively understand this.

\subsection{Non-equilibrium modifications of external source and expectation value}

For the moment, let us assume we have specified a pair of boundary conditions at the horizon which need not give a solution which is regular at the horizon. This means we have made specific choices of $A^{\text{in(neq)}}(\omega, \mathbf{k}, \omega_{\text{(b)}}, \mathbf{k}_{\text{(b)}})$ and $A^{\text{in(neq)}}(\omega, \mathbf{k}, \omega_{\text{(b)}}, \mathbf{k}_{\text{(b)}})$, the coefficients of the two homogeneous solutions. Any such choice has to be consistent with derivative and amplitude expansions. As for instance in (\ref{coeffs}), this amounts to making arbitrary choices for $A^{\text{in(neq)}}_i(\omega, \mathbf{k})$ and $A^{\text{out(neq)}}_i(\omega, \mathbf{k})$.

After such choices are made, the entire solution is determined uniquely. We can readily determine the normalizable and non-normalizable modes from the solution by following it's behavior at the boundary. The full non-normalizable mode, which is sum of the equilibrium and non-equilibrium parts, is identified with the external source coupling to the dual operator. Let us denote this as $\mathcal{J}(x)$. The full normalizable mode is identified with the expectation value of the dual operator in the dual state, namely $\mathcal{O}(x)$.

Thus the equilibrium part of the source and expectation value of the operator can be obtained from the behavior of $\Phi^{\text{(eq)}}(k,r)$ near $r=0$, as below :
\begin{equation}
\Phi^{\text{(eq)}}(k,r) \approx \mathcal{J}^{\text{(eq)}}(k)r^{4-\Delta} + \mathcal{O}^{\text{(eq)}}(k)r^{\Delta}.
\end{equation}
Above $\Delta$ is the anomalous dimension of the dual operator and is related to the mass of the bulk scalar field by
\begin{equation}
\Delta = 2 + \sqrt{4+m^2 l^2}.
\end{equation}
The non-equilibrium corrections to the source and the expectation value of the operator in the dual non-equilibrium state can be obtained from the behavior of $\Phi^{\text{(neq)}}(k, k_{\text{(b)}}, r)$ and $\Phi^{\text{(neq)}}(k, -k_{\text{(b)}}, r)$ near $r=0$,
\begin{equation}
\Phi^{\text{(neq)}}(k,\pm k_{\text{(b)}}, r) \approx \mathcal{J}^{\text{(neq)}}(k, \pm k_{\text{(b)}})r^{4-\Delta} + \mathcal{O}^{\text{(neq)}}(k, \pm k_{\text{(b)}})r^{\Delta}.
\end{equation}
In case of the hydrodynamic shear mode background, for example, the non-equilibrium source and expectation value takes the form as can be seen from (\ref{solh1}):
\begin{eqnarray}
\mathcal{(J,O)}^{\text{(neq)}}(\omega,  \mathbf{k},\pm\omega_{\text{(h)}},\pm\mathbf{k}_{\text{(h)}}) &=& A^{\text{in(neq)}}\left(\omega, \mathbf{k},\pm\omega_{\text{(h)}}, \pm\mathbf{k}_{\text{(h)}}\right)\mathcal{(J,O)}^{\text{in(eq)}}(\omega \pm\omega_{\text{(h)}}, \mathbf{k} \pm\mathbf{k}_{\text{(h)}}) \nonumber\\&&+ A^{\text{out(neq)}}\left(\omega, \mathbf{k},\pm\omega_{\text{(h)}}, \pm\mathbf{k}_{\text{(h)}}\right)\mathcal{(J,O)}^{\text{in(eq)}}(\omega \pm\omega_{\text{(h)}}, \mathbf{k} \pm\mathbf{k}_{\text{(h)}})\nonumber\\
&&+ A^{\text{in(eq)}}(\omega, \mathbf{k})\mathcal{(J,O)}^{\text{in(neq)}}(\omega, \mathbf{k}, \pm\omega_{\text{(h)}}, \pm\mathbf{k}_{\text{(h)}}).
\end{eqnarray}
Above $\mathcal{(J,O)}^{\text{in(eq), out(eq)}}(\omega \pm\omega_{\text{(h)}}, \mathbf{k} \pm\mathbf{k}_{\text{(h)}})$ are obtained from asymptotic expansions of $\Phi^{\text{in(eq), out(eq)}}(\omega \pm\omega_{\text{(h)}}, \mathbf{k} \pm\mathbf{k}_{\text{(h)}},r)$. Similarly $\mathcal{(J,O)}^{\text{in(neq)}}(\omega, \mathbf{k}, \pm\omega_{\text{(h)}}, \pm\mathbf{k}_{\text{(h)}})$ are obtained from the asymptotic expansions of $\Phi^{\text{in(neq)}}(\omega, \mathbf{k}, \pm\omega_{\text{(h)}}, \pm\mathbf{k}_{\text{(h)}},r)$. We note each of $\Phi^{\text{in(eq), out(eq)}}(\omega \pm\omega_{\text{(h)}}, \mathbf{k} \pm\mathbf{k}_{\text{(h)}},r)$ and $\Phi^{\text{in(neq)}}(\omega, \mathbf{k}, \pm\omega_{\text{(h)}}, \pm\mathbf{k}_{\text{(h)}},r)$ are solutions of the radial equation, therefore have the same asymptotic expansions.

Furthermore, using the specific form of the coefficients $A^{\text{in(neq), out(neq)}}$ as given in (\ref{coeffs}), it is easy to see that the non-equilibrium modifications of the external source and the expectation value take the form :
\begin{eqnarray}\label{jo1}
\mathcal{(J,O)}^{\text{(neq)}}(\omega, \pm\omega_{\text{(h)}},\mathbf{k}, \pm\mathbf{k}_{\text{(h)}}) &=& \mathcal{(J,O)}^{\text{(neq)}}_1(\omega, \mathbf{k}) \   \delta\mathbf{u}(\pm\omega_{\text{(h)}}, \pm\mathbf{k}_{\text{(h)}})\cdot \mathbf{k} \nonumber\\
&& \pm \mathcal{(J,O)}^{\text{(neq)}}_2(\omega, \mathbf{k})\  k_i k_j k_{\text{(h)}i}\delta u_j(\pm\omega_{\text{(h)}}, \pm\mathbf{k}_{\text{(h)}})+ O(\epsilon^2),
\end{eqnarray}
with
\begin{eqnarray}\label{jo2}
\mathcal{(J,O)}^{\text{(neq)}}_i(\omega, \mathbf{k}) = A^{\text{in(neq)}}_i(\omega, \mathbf{k})\mathcal{(J,O)}^{\text{in(eq)}}
(\omega,\mathbf{k}) + A^{\text{out(neq)}}_i(\omega, \mathbf{k})\mathcal{(J,O)}^{\text{out(eq)}}
(\omega,\mathbf{k}) + \mathcal{(J,O)}^{\text{in(neq)}}_i(\omega, \mathbf{k}).
\end{eqnarray}
In this form it is explicitly clear how the non-equilibrium modifications of the source and the expectation value depend on the boundary conditions at the horizon, namely in the choice of $A^{\text{in(neq), out{(neq)}}}_i (\omega,\mathbf{k})$.  

The full space-time dependence of the source $\mathcal{J}(x)$ and the expectation value of the operator $\mathcal{O}(x)$ thus take the form :
\begin{eqnarray}
\mathcal{(J,O)}(\omega, \mathbf{k}, \mathbf{x},t) &=& \mathcal{(J,O)}^{\text{(eq)}}(\omega,\mathbf{k})e^{i(\mathbf{k}\cdot\mathbf{x}-\omega t)} +\int_{\mathcal{C}}\, d\omega_{\text{(b)}}  \mathcal{(J,O)}^{\text{(neq)}}(\omega,\mathbf{k}, \omega_{\text{(b)}}, \mathbf{k}_{\text{(b)}})e^{i((\mathbf{k}+
\mathbf{k}_{\text{(b)}})\cdot\mathbf{x}-(\omega
+\omega_{\text{(b)}})t)} \nonumber\\
&& + \int_{\mathcal{C}*}\, d\omega_{\text{(b)}}  \mathcal{(J,O)}^{\text{(neq)}}(\omega,\mathbf{k}, -\omega_{\text{(b)}}, -\mathbf{k}_{\text{(b)}})e^{i((\mathbf{k}-
\mathbf{k}_{\text{(b)}})\cdot\mathbf{x}-(\omega
-\omega_{\text{(b)}})t)}.
\end{eqnarray}

After the integrations over $\omega_{\text{(b)}}$ required to take the background on-shell are done as above, we are left with the residues at the poles of the quasi-normal modes. Therefore,
\begin{eqnarray}\label{jog}
\mathcal{(J,O)}(\omega,\mathbf{k},\mathbf{x},t) &=&\mathcal{(J,O)}^{\text{(eq)}}(\omega,\mathbf{k})e^{i(\mathbf{k}\cdot\mathbf{x}-\omega t)} +  \mathcal{(J,O)}^{\text{(neq)}}(\omega,\mathbf{k},  \mathbf{k}_{\text{(b)}})e^{i((\mathbf{k}+
\mathbf{k}_{\text{(b)}})\cdot\mathbf{x}))}
e^{-i(\omega+ \text{Re}\,\omega_{\text{(b)}}(\mathbf{k}_{\text{(b)}})) t}e^{-\text{Im}\,\omega_{\text{(b)}}(\mathbf{k}_{\text{(b)}})t} \nonumber\\
&& +\mathcal{(J,O)}^{\text{(neq)}}(\omega,\mathbf{k},  -\mathbf{k}_{\text{(b)}})e^{i((\mathbf{k}-
\mathbf{k}_{\text{(b)}})\cdot\mathbf{x}))}
e^{-i(\omega- \text{Re}\,\omega_{\text{(b)}}(\mathbf{k}_{\text{(b)}})) t}e^{-\text{Im}\,\omega_{\text{(b)}}(\mathbf{k}_{\text{(b)}})t} ,
\end{eqnarray}
Thus in the case of the hydrodynamic shear-mode background we get
\begin{eqnarray}
\mathcal{(J,O)}(\omega, \mathbf{k},\mathbf{x},t) &=& \mathcal{(J,O)}^{\text{(eq)}}(\omega,\mathbf{k})e^{i(\mathbf{k}\cdot\mathbf{x}-\omega t)} +  \mathcal{(J,O)}^{\text{(neq)}}(\omega,\mathbf{k},  \mathbf{k}_{\text{(h)}})e^{i((\mathbf{k}+
\mathbf{k}_{\text{(h)}})\cdot\mathbf{x}))}
e^{-i\omega t}e^{-\frac{\mathbf{k}_{\text{(h)}}^{2}}{4\pi T}t} \nonumber\\
&& +\mathcal{(J,O)}^{\text{(neq)}}(\omega,\mathbf{k},  -\mathbf{k}_{\text{(h)}})e^{i((\mathbf{k}-
\mathbf{k}_{\text{(h)}})\cdot\mathbf{x}))}
e^{-i\omega t}e^{-\frac{\mathbf{k}_{\text{(h)}}^{2}}{4\pi T}t},
\end{eqnarray}
where
\begin{eqnarray}\label{jo3}
\mathcal{(J,O)}^{\text{(neq)}}(\omega,\mathbf{k}, \pm\mathbf{k}_{\text{(h)}}) &=& \mathcal{(J,O)}^{\text{(neq)}}_1(\omega, \mathbf{k}) \   \delta\mathbf{u}( \pm\mathbf{k}_{\text{(h)}})\cdot \mathbf{k} \nonumber\\
&& \pm \mathcal{(J,O)}^{\text{(neq)}}_2(\omega, \mathbf{k})\  k_i k_j k_{\text{(h)}i}\delta u_j( \pm\mathbf{k}_{\text{(h)}})+ O(\epsilon^2),
\end{eqnarray}
with $ \mathcal{(J,O)}^{\text{(neq)}}_i$ given by (\ref{jo2}). Comparing (\ref{jo1}) and (\ref{jo3}), we see that we have extracted the residue of $\delta\mathbf{u}(\pm\omega_{\text{(h)}},\pm\mathbf{k}_{\text{(h)}})$ as given in (\ref{uft}) at the poles $\omega_{\text{(h)}} = \pm i\mathbf{k}_{\text{(h)}}^2/(4\pi T)$. In general $\mathcal{(J,O)}^{\text{(neq)}}(\omega, \mathbf{k}, \pm\mathbf{k}_{\text{(b)}})$ will take the form
\begin{eqnarray}\label{gso}
\mathcal{(J,O)}^{\text{(neq)}}(\omega, \mathbf{k}, \pm \mathbf{k}_{\text{(b)}}) &=& A^{\text{in(neq)}}(\omega, \mathbf{k}, \pm\mathbf{k}_{\text{(b)}}) \mathcal{(J,O)}^{\text{in(eq)}}(\omega, \mathbf{k})+ A^{\text{out(neq)}}(\omega, \mathbf{k}, \pm\mathbf{k}_{\text{(b)}}) \mathcal{(J,O)}^{\text{out(eq)}}(\omega, \mathbf{k})\nonumber\\ &&+ \mathcal{(J,O)}^{\text{in(neq)}}(\omega, \mathbf{k}, \pm\mathbf{k}_{\text{(b)}}),
\end{eqnarray}
with $A^{\text{in(neq), out(neq)}}$ and $\mathcal{(J,O)}^{\text{in(neq)}}$ having  consistent derivative and amplitude expansions. As for instance $A^{\text{in(neq)}}(\omega, \mathbf{k}, \mathbf{k}_{\text{(b)}}) = A^{\text{in(neq)}}_3(\omega, \mathbf{k})\pi_{ij}(\mathbf{k}_{\text{(b)}})k_i k_j + ...\,$.

The above discussion is readily generalized by including non-linearities in the dynamics of the non-equilibrium perturbations of the background. As for instance, if we take into account the quadratic dependence on two hydrodynamic shear modes $\delta\mathbf{u}(\mathbf{k}_{\text{(h)}})$ and  $\delta\mathbf{u}(\mathbf{k}_{\text{(h)}}^\prime)$, the corrections will take the form,
\begin{eqnarray}\label{jognl}
(\mathcal{J,O})^{(neq)}(\omega, \mathbf{k}, \mathbf{k}_{\text{(h)}}, \mathbf{k}_{\text{(h)}}^\prime)
e^{i(\mathbf{k}_{\text{(h)}}
+\mathbf{k}_{\text{(h)}}^\prime)
\cdot\mathbf{x}}
e^{-i\omega t}
e^{-\Big(\frac{\mathbf{k}_{\text{(h)}}^2
+\mathbf{k}_{\text{(h)}}^{\prime 2}}{4\pi T}\Big)t} + (\mathbf{k}_{\text{(h)}}\rightarrow -\mathbf{k}_{\text{(h)}})+ (\mathbf{k}_{\text{(h)}}^\prime\rightarrow -\mathbf{k}_{\text{(h)}}^\prime),
\end{eqnarray}
after doing integrations over $\omega_{\text{(h)}}$ and $\omega_{\text{(h)}}^\prime$. Furthermore,
\begin{equation}
(\mathcal{J,O})^{(neq)}(\omega, \mathbf{k}, \mathbf{k}_{\text{(h)}}, \mathbf{k}_{\text{(h)}}^\prime) = (\mathcal{J,O})^{(neq)}_3(\omega, \mathbf{k})\,\left(\delta\mathbf{u}( \mathbf{k}_{\text{(h)}})\cdot \mathbf{k}\right)\left(\delta\mathbf{u}( \mathbf{k}_{\text{(h)}}^\prime)\cdot \mathbf{k}_{\text{(h)}}\right) + ....\, ,
\end{equation}
at the first order in the derivative expansion. All these terms are fixed uniquely in terms of the boundary conditions at the horizon.

The general form of the non-equilibrium corrections to source and expectation value as in (\ref{gso}) after including non-linear dynamics of the background involves
\begin{eqnarray}\label{gsonl}
\mathcal{(J,O)}^{\text{(neq)}}(\omega, \mathbf{k}, \pm \mathbf{k}_{\text{(b)}1},..., \pm \mathbf{k}_{\text{(b)}n}) &=& A^{\text{in(neq)}}(\omega, \mathbf{k},\pm \mathbf{k}_{\text{(b)}1},..., \pm \mathbf{k}_{\text{(b)}n}) \mathcal{(J,O)}^{\text{in(eq)}}(\omega, \mathbf{k})
\nonumber\\&&+ A^{\text{out(neq)}}(\omega, \mathbf{k}, \pm \mathbf{k}_{\text{(b)}1},..., \pm \mathbf{k}_{\text{(b)}n}) \mathcal{(J,O)}^{\text{out(eq)}}(\omega, \mathbf{k})\nonumber\\ &&+ \mathcal{(J,O)}^{\text{in(neq)}}(\omega, \mathbf{k},\pm \mathbf{k}_{\text{(b)}1},..., \pm \mathbf{k}_{\text{(b)}n}),
\end{eqnarray}
which will appear with a spatio temporal factor 
\begin{equation*}
e^{i(\pm \mathbf{k}_{\text{(b)}1}\pm ... \pm\mathbf{k}_{\text{(b)}n})\cdot\mathbf{x}}
e^{-i\left(\pm\text{Re}\,\omega_{(b)1}(\mathbf{k}_{\text{(b)}1}) \pm ...\pm \text{Re}\,\omega_{(b)n}(\mathbf{k}_{\text{(b)}n})\right)t}e^{\left(\pm\text{Im}\,\omega_{(b)1}(\mathbf{k}_{\text{(b)}1}) \pm ...\pm \text{Im}\,\omega_{(b)n}(\mathbf{k}_{\text{(b)}n})\right)t}.
\end{equation*}

\subsection{Mapping linear response to non-equilibrium Green's functions}

There is a basic linear response to an external perturbation in any physical system - the causal response. This causal response gives the retarded Green's function.  

Holographically, the equilibrium causal response maps to the incoming boundary condition at the horizon \cite{incoming}. This is expected to be the case because it is the only boundary condition that is consistent with the causal structure of the black hole which forbids anything propagating out of the horizon classically.

We have seen that the incoming equilibrium solution of the bulk scalar field has a unique non-equilibrium correction which is regular at the horizon and is free of potentially harmful back-reaction at the future horizon. We need regularity at the horizon to preserve the global causal structure of the black brane. We recall that this non-equilibrium solution which is regular at the horizon requires very specific boundary conditions at the horizon. These boundary conditions have been determined even after including non-linearities in the dynamics of the perturbations of the background quasi-normal modes.

Furthermore, this non-equilibrium solution which is regular at the future horizon uniquely determines the non-equilibrium modifications to the external source and expectation value of the dual operator. It has been proposed that it determines the non-equilibrium corrections to the causal response, and therefore the non-equilibrium retarded Green's function \cite{noneqspec}.

The retarded Green's function in any state is given by the causal response in that state.  Thus the holographic non-equilibrium retarded Green's function is :
\begin{equation}\label{rdef}
G_R(\mathbf{x}_1, t_1; \mathbf{x}_2 t_2)= \int d\omega d^3k \frac{\mathcal{O}(\omega, \mathbf{k},\mathbf{x}_1, t_1)}{\mathcal{J}(\omega, \mathbf{k}, \mathbf{x}_2,t_2)}.
\end{equation}

Let us go back to the linearized approximation when the non-equilibrium perturbation in the background is given by a single quasi-normal mode or a single relaxation mode in the dual theory. Then the full spatio-temporal form of $\mathcal{J}$ and $\mathcal{O}$ above is as in (\ref{jog}).The boundary conditions required for regularity at the horizon are that the coefficients of the homogeneous solutions $A^{\text{in(neq), out(neq)}}$ should be set to zero, so that the non-equilibrium corrections to external source/ expectation value are given fully by $\mathcal{(J,O)}^{\text{in(neq)}}(\omega, \mathbf{k}, \pm\mathbf{k}_{\text{(b)}})$ respectively.

One can see that after doing the Wigner transform (for intermediate steps see appendix D) the non-equilibrium retarded Green's function takes the following form \cite{noneqspec}:
\begin{eqnarray}\label{retpw}
G_R(\omega, \mathbf{k}, \mathbf{x}, t) &=& \int d\omega_1 \int d^3 k_1 \, G_R^{\text{(eq)}}(\omega_1, \mathbf{k}_1)\Bigg[\delta(\omega- \omega_1)\delta^3 (\mathbf{k} -\mathbf{k}_1) \nonumber\\&&- \frac{1}{2\pi i}
\Bigg(\mathcal{M}(\omega_1, \mathbf{k}_1, \mathbf{k}_{\text{(b)}})\delta^3\Big(\mathbf{k} - \mathbf{k}_1 -\frac{\mathbf{k}_{\text{(b)}}}{2}\Big)\frac{1}{\Big(\omega - \omega _1 -\frac{\text{Re}\,\omega_{\text{(b)}}(\mathbf{k}_{(b)})}{2}- i \frac{\text{Im}\,\omega_{\text{(b)}}(\mathbf{k}_{\text{(b)}})}{2}\Big)}\nonumber\\
&&\qquad\qquad\qquad\qquad-\mathcal{N}(\omega_1, \mathbf{k}_1, \mathbf{k}_{\text{(b)}})\delta^3\Big(\mathbf{k} - \mathbf{k}_1 +\frac{\mathbf{k}_{\text{(b)}}}{2}\Big)\frac{1}{ \Big(\omega - \omega _1 + \frac{\text{Re}\,\omega_{\text{(b)}}(\mathbf{k}_{(b)})}{2} + i \frac{\text{Im}\,\omega_{\text{(b)}}(\mathbf{k}_{\text{(b)}}))}{2}\Big)}\Bigg)\nonumber\\&&\qquad\qquad\qquad\qquad
\nonumber\\
&&
\qquad\qquad\qquad\qquad
e^{i \mathbf{k}_{\text{(b)}}\cdot\mathbf{x}}e^{- i\text{Re}\,\omega_{\text{(b)}}(\mathbf{k}_{(b)})t} e^{\text{Im}\,\omega_{\text{(b)}}(\mathbf{k}_{\text{(b)}})t}
\nonumber\\&&
+\frac{1}{2\pi i}
\Bigg(\mathcal{M}(\omega_1, \mathbf{k}_1, -\mathbf{k}_{\text{(b)}})\delta^3\Big(\mathbf{k} - \mathbf{k}_1 +\frac{\mathbf{k}_{\text{(b)}}}{2}\Big)\frac{1}{\Big(\omega - \omega _1 +\frac{\text{Re}\,\omega_{\text{(b)}}(\mathbf{k}_{(b)})}{2}+ i \frac{\text{Im}\,\omega_{\text{(b)}}(\mathbf{k}_{\text{(b)}})}{2}\Big)}\nonumber\\&&\qquad\qquad\qquad\qquad
-\mathcal{N}(\omega_1, \mathbf{k}_1, -\mathbf{k}_{\text{(b)}})\delta^3\Big(\mathbf{k} - \mathbf{k}_1 -\frac{\mathbf{k}_{\text{(b)}}}{2}\Big)\frac{1}{\Big(\omega - \omega _1 -\frac{\text{Re}\,\omega_{\text{(b)}}(\mathbf{k}_{(b)})}{2}- i \frac{\text{Im}\,\omega_{\text{(b)}}(\mathbf{k}_{\text{(b)}})}{2}\Big)}\Bigg)\nonumber\\
&&
\qquad\qquad\qquad\qquad
e^{-i \mathbf{k}_{\text{(b)}}\cdot\mathbf{x}}
e^{i\text{Re}\,\omega_{\text{(b)}}(\mathbf{k}_{(b)})t} e^{ \text{Im}\,\omega_{\text{(b)}}(\mathbf{k}_{\text{(b)}})t}\Bigg],
\end{eqnarray}
with 
\begin{equation}
G_R^{\text{(eq)}}(\omega, \mathbf{k}) = \frac{\mathcal{O}^{\text{(eq)}}(\omega, \mathbf{k})}{\mathcal{J}^{\text{(eq)}}(\omega, \mathbf{k})}
\end{equation}
being the equilibrium retarded Green's function and
\begin{equation}
\mathcal{M}(\omega_1, \mathbf{k}_1, \pm\mathbf{k}_{\text{(b)}}) =\frac{ \mathcal{O}^{\text{in(neq)}}(\omega_1, \mathbf{k}_1, \pm\mathbf{k}_{\text{(b)}})}{\mathcal{O}^{\text{(eq)}}(\omega_1, \mathbf{k}_1)}, \quad \mathcal{N}(\omega_1, \mathbf{k}_1, \mathbf{k}_{\text{(b)}})=\frac{ \mathcal{J}^{\text{in(neq)}}(\omega_1, \mathbf{k}_1, \pm\mathbf{k}_{\text{(b)}})}{\mathcal{J}^{\text{(eq)}}(\omega_1, \mathbf{k}_1)}.
\end{equation}
The complex conjugate mode corrections included above needed for complete field-theoretic consistency (see below) were not included in our earlier work \cite{noneqspec}.

The crucial elements to note are :
\begin{itemize}
\item The non-equilibrium contributions involve two pieces, spatio-temporally comoving with the background quasi-normal modes at momenta $\pm \mathbf{k}_{\text{(b)}}$ respectively.
\item The non-equilibrium contributions to the equilibrium retarded propagator at frquency $\omega$ and momentum $\mathbf{k}$ involves a convolution with support at momenta $\mathbf{k} \pm \mathbf{k}_{\text{(b)}}/2$ and poles at frequencies $\omega\pm\text{Re}\,\omega_{\text{(b)}}(\mathbf{k}_{\text{(b)}})/2\pm i\text{Im}\,\omega_{\text{(b)}}(\mathbf{k}_{\text{(b)}})/2$.
\end{itemize}
These features reflect the classical gravity approximation. We will see how these generalize beyond the quasi-normal mode approximation of the background.

Also note that as the non-equilibrium correction to the equilibrium solution which is regular at the future horizon is unique, the overall solution is undetermined up to an overall multiplicative constant $A(\omega, \mathbf{k})$ as in (\ref{gp}). This multiplicative constant $A(\omega, \mathbf{k})$ cancels between the numerator and denominator in (\ref{rdef}). Therefore the non-equilibrium retarded Green's function is uniquely determined by this holographic prescription.

The advanced Green's function should be naturally related to the advanced response. Thus,
\begin{equation}\label{addef}
G_A(\mathbf{x}_1, t_1; \mathbf{x}_2, t_2)  = \int d\omega d^3k \frac{\mathcal{O}(\omega, \mathbf{k},\mathbf{x}_2, t_2)}{\mathcal{J}(\omega, \mathbf{k}, \mathbf{x}_1,t_1)}.
\end{equation}
The above also follows from the definitions that $G_A(\mathbf{x}_1, t_1; \mathbf{x}_2, t_2) = G_R(\mathbf{x}_2, t_2; \mathbf{x}_1, t_1)$ (for proof of the latter see appendix A). For the holographic definition of advanced response, both $\mathcal{O}$ and $\mathcal{J}$ above must be evaluated in the same non-equilibrium solution as in (\ref{rdef}) in the case of the retarded propagator, i.e. in the solution involving the unique correction to the equilibrium incoming mode which is regular at the future horizon. Therefore, $\mathcal{O}$ and $\mathcal{J}$ take the form as in (\ref{jog}) with $\mathcal{(J,O)}^{\text{(neq)}}=\mathcal{(J,O)}^{\text{in(neq)}}$.

From appendix A, it follows that after the Wigner transform $G_A(\omega, \mathbf{k}, \mathbf{x}, t) = G_R(-\omega, 
-\mathbf{k}, \mathbf{x}, t)$. Therefore $G_A(\omega, \mathbf{k}, \mathbf{x}, t)$ is given by (\ref{retpw}) with $(\omega,\mathbf{k})$ replaced by $(-\omega, -\mathbf{k})$. 

Also $G_R*(k,x) = G_A(k,x)$ as proved in appendix A. This is also satisfied as can be seen from (\ref{retpw}) using $G_A(k,x) = G_R(-k,x)$. To do this one first reverses the sign of the integrated variables $\omega_1$ and $\mathbf{k}_1$ and then uses results from section III.C which imply that $\mathcal{(J,O)}^{\text{(eq)}}*(k)  =\mathcal{(J,O)}^{\text{(eq)}}(-k)$ and $\mathcal{(J,O)}^{\text{in(neq)}}*(k, \mathbf{k}_{\text{(b)}}) =\mathcal{(J,O)}^{\text{in(neq)}}(-k, -\mathbf{k}_{\text{(b)}})$. Thus our holographic prescriptions for the non-equilibrium retarded and advanced propagators pass important field-theoretic conssitency tests.

At the beginning of this section, we have already made the hypothesis that non-equilibrium Green's functions in the holographic classical gravity approximation should be determined by appropriate boundary conditions at the horizon. These boundary conditions should be determined by the consistency of the map with field-theoretic requirements.

In order to map non-equilibrium Green's function to boundary conditions at the horizon, we start from the gravity side. Consider an arbitrary solution of the bulk scalar field in the non-equilibrium geometry which is consistent with the derivative/amplitude expansions but not necessarily regular at the horizon. This means that the space-time profile of the source and expectation value takes the form (\ref{jog}) with the non-equilibrium parts taking the general form (\ref{gso}). We should make a choice of both $A^{\text{in(neq)}}(\omega, \mathbf{k}, \mathbf{k}_{\text{(b)}})$ and $A^{\text{out(neq)}}(\omega, \mathbf{k}, \mathbf{k}_{\text{(b)}})$ to specify the solution uniquely. This amounts to making choices of coefficients of $A^{\text{in(neq)}}_i(\omega, \mathbf{k})$ and $A^{\text{out(neq)}}_i(\omega, \mathbf{k})$ of terms like $\delta\mathbf{u}(\mathbf{k}_{\text{(b)}})\cdot\mathbf{k}$, $\pi_{ij}^{\text{(nh)}}(\mathbf{k}_{\text{(b)}})k_i k_j$ in their derivative/amplitude expansions as in (\ref{coeffs}).

The most well-defined response functions on the gravity side are the causal and advanced response functions. Let us consider the following response function evaluated in an arbitrary non-equilibrium solution :
\begin{eqnarray}\label{grav}
G(x_1,x_2) &=& \int d\omega d^3 k \Bigg(f_R^{\text{(eq)}}(\omega)\frac{O(\omega,\mathbf{k}, \mathbf{x}_1, t_1)}{J(\omega,\mathbf{k}, \mathbf{x}_2, t_2)}
 +f_A^{\text{(eq)}}(\omega)\frac{O(\omega,\mathbf{k}, \mathbf{x}_2, t_2)}{J(\omega,\mathbf{k}, \mathbf{x}_1, t_1)}\Bigg).
\end{eqnarray}

Physically the boundary conditions given by choice of $A^{\text{in(neq)}}(\omega, \mathbf{k}, \mathbf{k}_{\text{(b)}})$ and $A^{\text{out(neq)}}(\omega, \mathbf{k}, \mathbf{k}_{\text{(b)}})$ can be interpreted as follows. The full response function above at equilibrium is the sum of the causal response with weight $f_R^{\text{(eq)}}(\omega)$ and the advanced response with weight $f_A^{\text{(eq)}}(\omega)$. The boundary conditions $A^{\text{in(neq)}}(\omega, \mathbf{k}, \mathbf{k}_{\text{(b)}})$ and $A^{\text{out(neq)}}(\omega, \mathbf{k}, \mathbf{k}_{\text{(b)}})$ thus shift the weight of the non-equilibrium causal response and the non-equilibrium advanced response in the total  weighted sum, because individually they are given by the boundary conditions $A^{\text{in(neq)}}(\omega, \mathbf{k}, \mathbf{k}_{\text{(b)}})$ and $A^{\text{out(neq)}}(\omega, \mathbf{k}, \mathbf{k}_{\text{(b)}})$ set to zero.

It is then easy to repeat the steps which lead from (\ref{rdef}) to (\ref{retpw}) as explicitly shown in appendix D. We find that (\ref{grav}) reproduces our general parametrization of any non-equilibrium Green's function as a weighted sum of the retarded and advanced Green's functions (\ref{gnp}), which is valid at the leading order in amplitude expansion, with the identification
\begin{eqnarray}\label{fra}
f^{(1,0)}_{R}(\omega, \mathbf{k}, \omega_1, \mathbf{k}_1, \mathbf{x} , t) &=&-\frac{f_{R}^{\text{(eq)}}(\omega)}{2\pi i}\Bigg( \tilde{\mathcal{M}}\left(\omega_1, \mathbf{k}_1, \mathbf{k}_{\text{(b)}}\right)\delta^3\Big(\mathbf{k} - \mathbf{k}_1 -\frac{\mathbf{k}_{\text{(b)}}}{2}\Big)\frac{1}{\Big(\omega - \omega _1 -\frac{\text{Re}\,\omega_{\text{(b)}}(\mathbf{k}_{(b)})}{2}- i \frac{\text{Im}\,\omega_{\text{(b)}}(\mathbf{k}_{\text{(b)}})}{2}\Big)}\nonumber\\
&&+\tilde{\mathcal{N}}\left(\omega_1, \mathbf{k}_1, \mathbf{k}_{\text{(b)}}\right)\delta^3\Big(\mathbf{k} - \mathbf{k}_1 +\frac{\mathbf{k}_{\text{(b)}}}{2}\Big)\frac{1}{ \Big(\omega - \omega _1 + \frac{\text{Re}\,\omega_{\text{(b)}}(\mathbf{k}_{(b)})}{2} + i \frac{\text{Im}\,\omega_{\text{(b)}}(\mathbf{k}_{\text{(b)}}))}{2}\Big)}\Bigg)\nonumber\\&&\qquad\qquad\qquad\qquad
\nonumber\\
&&
\qquad\qquad\qquad\qquad
e^{i \mathbf{k}_{\text{(b)}}\cdot\mathbf{x}}e^{- i\text{Re}\,\omega_{\text{(b)}}(\mathbf{k}_{(b)})t} e^{\text{Im}\,\omega_{\text{(b)}}(\mathbf{k}_{\text{(b)}})t}
\nonumber\\&&
+\frac{f_{R}^{\text{(eq)}}(\omega)}{2\pi i}
\Bigg(\tilde{\mathcal{M}}\left(\omega_1, \mathbf{k}_1, -\mathbf{k}_{\text{(b)}}\right)\delta^3\Big(\mathbf{k} - \mathbf{k}_1 +\frac{\mathbf{k}_{\text{(b)}}}{2}\Big)\frac{1}{\Big(\omega - \omega _1 +\frac{\text{Re}\,\omega_{\text{(b)}}(\mathbf{k}_{(b)})}{2}+ i \frac{\text{Im}\,\omega_{\text{(b)}}(\mathbf{k}_{\text{(b)}})}{2}\Big)}\nonumber\\&&
+\tilde{\mathcal{N}}\left(\omega_1, \mathbf{k}_1, -\mathbf{k}_{\text{(b)}}\right)\delta^3\Big(\mathbf{k} - \mathbf{k}_1 -\frac{\mathbf{k}_{\text{(b)}}}{2}\Big)\frac{1}{\Big(\omega - \omega _1 -\frac{\text{Re}\,\omega_{\text{(b)}}(\mathbf{k}_{(b)})}{2}- i \frac{\text{Im}\,\omega_{\text{(b)}}(\mathbf{k}_{\text{(b)}})}{2}\Big)}\Bigg)\nonumber\\
&&
\qquad\qquad\qquad\qquad
e^{-i \mathbf{k}_{\text{(b)}}\cdot\mathbf{x}}
e^{i\text{Re}\,\omega_{\text{(b)}}(\mathbf{k}_{(b)})t} e^{ \text{Im}\,\omega_{\text{(b)}}(\mathbf{k}_{\text{(b)}})t},
\end{eqnarray}
where
\begin{eqnarray}\label{mn}
\tilde{\mathcal{M}}(\omega_1, \mathbf{k}_1, \pm\mathbf{k}_{\text{(b)}}) &=&\frac{ \mathcal{O}^{\text{(neq)}}(\omega_1, \mathbf{k}_1, \pm\mathbf{k}_{\text{(b)}})-\mathcal{O}^{\text{in(neq)}}(\omega_1, \mathbf{k}_1, \pm\mathbf{k}_{\text{(b)}})}{\mathcal{O}^{\text{(eq)}}(\omega_1, \mathbf{k}_1)}\nonumber\\
&=& \frac{ A^{\text{in(neq)}}(\omega_1, \mathbf{k}_1, \pm\mathbf{k}_{\text{(b)}}) \mathcal{O}^{\text{in(eq)}}(\omega_1, \mathbf{k}_1)+ A^{\text{out(neq)}}(\omega_1, \mathbf{k}_1, \pm\mathbf{k}_{\text{(b)}}) \mathcal{O}^{\text{out(eq)}}(\omega_1, \mathbf{k}_1)}{\mathcal{O}^{\text{(eq)}}(\omega_1, \mathbf{k}_1)}, \nonumber\\
\tilde{\mathcal{N}}(\omega_1, \mathbf{k}_1, \pm\mathbf{k}_{\text{(b)}}) &=& \frac{\mathcal{J}^{\text{(neq)}}(\omega_1, \mathbf{k}_1, \pm\mathbf{k}_{\text{(b)}})- \mathcal{J}^{\text{in(neq)}}(\omega_1, \mathbf{k}_1, \pm\mathbf{k}_{\text{(b)}})}{\mathcal{J}^{\text{(eq)}}(\omega_1, \mathbf{k}_1)} \nonumber\\
&=& \frac{ A^{\text{in(neq)}}(\omega_1, \mathbf{k}_1, \pm\mathbf{k}_{\text{(b)}}) \mathcal{J}^{\text{in(eq)}}(\omega_1, \mathbf{k}_1)+ A^{\text{out(neq)}}(\omega_1, \mathbf{k}_1, \pm\mathbf{k}_{\text{(b)}}) \mathcal{J}^{\text{out(eq)}}(\omega_1, \mathbf{k}_1)}{\mathcal{J}^{\text{(eq)}}(\omega_1, \mathbf{k}_1)}.
\end{eqnarray}
Also $f^{(1,0)}_A(k,k_1, x)$ is given by the above equations (\ref{fra}) and (\ref{mn}) with $f_R^{\text{(eq)}}(\omega)$ replaced by $f_A^{\text{(eq)}}(\omega)$ and $(\omega, \mathbf{k})$ replaced by $(-\omega, -\mathbf{k})$.

Thus there is a map between between parameters $f^{(1,0)}_{R,A}$ in the leading order parametrization (\ref{gnp}) of non-equilibrium Green's functions required for consistent pertubation expansions and leading order boundary conditions $A^{\text{in(neq), out(neq)}}$ for the non-equilibrium modes set at the horizon. However, $f^{(1,0)}_{R,A}$ should also satisfy other field-theoretic requirements, aside from producing consistent derivative and amplitude expansions.

Let us work this out explicitly for the non-equilibrium Feynman propagator. In this case we should set in the gravity response function (\ref{grav}) $f_R^{\text{(eq)}}(\omega)= n_{\text{BE}}(\omega)+ 1$ and $f_A^{\text{(eq)}}(\omega) = f_R^{\text{(eq)}}(-\omega)= -n_{\text{BE}}(\omega)$ to reproduce the equilibrium Feynman propagator.

The holographic non-equilibrium Feynman propagator $G_F(x_1, x_2)$ is therefore given by 
\begin{eqnarray}\label{holfeyn}
G_F(x_1,x_2) &=& \int d\omega d^3 k \Bigg(\Big(n_{\text{BE}}(\omega)+ 1\Big)\frac{O(\omega,\mathbf{k}, \mathbf{x}_1, t_1)}{J(\omega,\mathbf{k}, \mathbf{x}_2, t_2)}
 -n_{\text{BE}}(\omega)\frac{O(\omega,\mathbf{k}, \mathbf{x}_2, t_2)}{J(\omega,\mathbf{k}, \mathbf{x}_1, t_1)}\Bigg),
\end{eqnarray}
with both $\mathcal{J}$ and $\mathcal{O}$ evaluated in a non-equilibrium solution with boundary conditions given by $A^{\text{in(neq)}}(\omega, \mathbf{k}, \mathbf{k}_{\text{(b)}})$ and $A^{\text{out(neq)}}(\omega, \mathbf{k}, \mathbf{k}_{\text{(b)}})$, both of which we should determine now using field theoretic consistency conditions. 

Note the holographic Feynman propagator (\ref{holfeyn}) is symmetric in $x_1$ and $x_2$ as required. This can be seen by exchanging $x_1$ and $x_2$ while reversing the signs of the integrated variables $\omega$ and $\mathbf{k}$.

Let us now compare the holographic form of the non-equilibrium Feynman propagator (\ref{holfeyn}) with the parametrization (\ref{nfeyn}) that yields consistent perturbative expansions obtained earlier.  Our comparison yields that
\begin{eqnarray}
f^{(1,0)}(k, k_1, x) &=& -i\int_{-\infty}^\infty \frac{d\omega^{\prime}}{2\pi}P\Bigg(\frac{f_R^{(1,0)}(k, \omega^{\prime}, \mathbf{k}_1, x)+f_A^{(1,0)}(k, \omega^{\prime}, \mathbf{k}_1, x)}{\omega^{\prime}-\omega_1}\Bigg)
\nonumber\\
&& -\frac{1}{2}\Big(f_R^{(1,0)}(k,k_1,x)-f_A^{(1,0)}(k,k_1,x)\Big),
\end{eqnarray}
where $f_{R}^{(1,0)}(k,k_1,x)$ is given by (\ref{fra}) with $f_R^{\text{(eq)}}(\omega)= n_{\text{BE}}(\omega)+ 1$ and $f_{A}^{(1,0)}(k,k_1x)$ is also given by (\ref{fra}) with $f_R^{\text{(eq)}}(\omega)$ replaced by $f_A^{\text{(eq)}}(\omega)=f_R^{\text{(eq)}}(-\omega) = -n_{\text{BE}}(\omega)$. Also $P$ denotes the principal value.

For the case of the Feynman propagator, using (\ref{ainoutcc}) and $f_A^{\text{(eq)}}(\omega)=f_R^{\text{(eq)}}(-\omega)$ we find that
\begin{eqnarray}
f_R^{(1,0)}(k, k_1, x) = f_A^{(1,0)}(-k, -k_1, x).
\end{eqnarray}
We can now compare with the requirements (\ref{fprop}) to obtain
\begin{eqnarray}
f^{(1,0)}_{\text{S}}(k,k_1,x) &=& 0,
\nonumber\\
f^{(1,0)}_{\text{A}} (k,k_1,x)&=& f^{(1,0)}(k, k_1,x).
\end{eqnarray}
We recall that the above implies that the constraint of symmetry in $x_1$ and $x_2$ is satisfied for the holographic $G_F(x_1, x_2)$, which is also evident in the defining form (\ref{holfeyn}).

There are two crucial field theoretic requirements in (\ref{fprop}) which should be additionally satisfied in the above identifcations. These are that $f^{(1,0)}_{\text{S}}(k,k_1,x)$ and $f^{(1,0)}_{\text{A}}(k,k_1,x)$ should be real. One can readily see from (\ref{fra}) that it is impossible to satisfy these for all $x$ unless both $f_R^{(1,0)}(k, k_1,x)$ and $f_A^{(1,0)}(k, k_1, x)$ both vanish identically. The latter is possible only when $A^{\text{in(neq)}}$ and $A^{\text{out(neq)}}$ are both set to zero giving the solution which is regular at the horizon. This implies there are no non-equilibrium shifts in the weights of the causal and advanced responses in (\ref{holfeyn}).

Thus we have proved when the quasinormal approximation of the background is valid,
\begin{equation}\label{holffin}
G_F(k, x) = \Big(n_{\text{BE}}(\omega)+1\Big)G_R(k,x) - n_{\text{BE}}(\omega) G_A(k,x).
\end{equation}
This simply follows from \ref{holfeyn} using boundary conditions where $A^{\text{in(neq)}}$ and $A^{\text{out(neq)}}$ are set to zero in 
(\ref{holfeyn}) and then doing Wigner transform.

We can now go beyond the quasinormal mode approximation of the background. This means taking into account non-linearities in the dynamics of the hydrodynamic and relaxational modes constituting the non-equilibrium state.

We note that (\ref{rdef}) still gives the holographic retarded Green's function because it follows from linear response theory - the expectation value and source are obtained from the full non-equilibrium solution which depends non-linearly on the non-equilibrium perturbations $\delta\mathbf{u}, \delta T$ and $\pi_{ij}^{\text{(nh)}}$ and is regular at the future horizon. We have discussed how we obtain this full non-equilibrium solution in section III.C - namely we apply the same boundary conditions which cut down both the homogeneous incoming and outgoing solutions at each order in the amplitude perturbation expansion.

The retarded response function depends non-linearly on each term in the perturbation expansion though. Let us denote $(n)$ as the order in the amplitude expansion, i.e. it counts the total number of pertubation parameters $\delta\mathbf{u}, \delta T$ and $\pi_{ij}^{\text{(nh)}}$ as in section III.C. Then the retarded Green's function up to $n$-th order in the amplitude expansion is obtained from
\begin{equation*}
\frac{\mathcal{O}^{\text{(eq)}}+\mathcal{O}^{(1)}+
...+\mathcal{O}^{(n)}}{\mathcal{J}^{\text{(eq)}}+\mathcal{J}^{(1)}+
...+\mathcal{J}^{(n)}}.
\end{equation*}
At the $n$-th order we not only get $\mathcal{O}^{(n)}/\mathcal{J}^{(n)}$ but also $(\mathcal{O}^{(1)}/\mathcal{J}^{(1)})^n$. Above we have suppressed the derivative expansion, but it clear we can do a double expansion in the derivatives and amplitudes of the background perturbation close to equilibrium.

Similarly the advanced Green's function is given by (\ref{addef}), but the expectation value and the source are obtained from the full non-equilibrium solution which is regular at the horizon. One can readily check that the full holographic retarded and advanced propagators also satisfy the relation $G_R*(k,x) = G_A(k,x)$ simply repeating the same steps discussed earlier to all orders in the perturbation expansion.

One can now map the holographic linear response as given by (\ref{grav}) to a given non-equilibrium propagator. In (\ref{grav}) we need to use an appropriate boundary conditions at the horizon instead of those fixed by regularity at each order in the amplitude expansion. These boundary conditions give the non-equilibrium shifts of the weight of the retarded and advanced Green's functions which consitute the given non-equilibrium propagator.

It is also easy to see that, at the second order in the amplitude expansion, we will obtain the same structure (\ref{g2}) as we argued from field theory from the holographic response (\ref{grav}). Namely $f^{(2,0,0)}_{RR, RA, AR, AA}(k, k_1, k_2, x)$ will depend on the boundary conditions on the solutions $\Phi^{(2,n,m)}$, where $n$ and $m$ denote the order of the derivative expansion acting on each of the two amplitude expansions. These are reflected by the boundary conditions given by $A^{(2,n,m)\text{in(neq)}}(k,  \pm k_{\text{(b)}1}, \pm k_{\text{(b)}2})$ and $A^{(2,n,m)\text{out(neq)}}(k,  \pm k_{\text{(b)}1}, \pm k_{\text{(b)}2})$. We note that the reality constraints (\ref{ainoutcc}) tells us there are exactly four independent boundary conditions for the four independent parameters  $f^{(2,0,0)}_{RR, RA, AR, AA}(k, k_1, k_2, x)$. Without writing the detailed expression for $f^{(2,0,0)}_{RR, RA, AR, AA}(k, k_1, k_2, x)$ we can readily see that taking all orders in derivative expansion into account,
\begin{itemize}
\item the dependence on $\mathbf{k}$ is given by a sum of four terms which get supported by $\delta^3 (\mathbf{k} -\mathbf{k}_1- \mathbf{k}_2 \pm \mathbf{k}_{\text{(b)}1} \pm \mathbf{k}_{\text{(b)}2} )$ respectively,
\item as function of $\omega$, these terms have poles in $\omega - \omega_1 -\omega_2 \pm \omega(k_{\text{(b)}1}) \pm\omega(k_{\text{(b)}2})$ respectively, where $\omega(k_{\text{(b)}1})$ and $\omega(k_{\text{(b)}2})$ are the complex dispersion relations of the two quasinormal modes appearing at the quadratic order,
\item the space-time dependences are given by $e^{i(\pm\mathbf{k}_{\text{(b)}1}\pm\mathbf{k}_{\text{(b)}2})
\cdot \mathbf{x})}e^{-i(\pm\omega(\mathbf{k}_{\text{(b)}1})\pm\omega(
\mathbf{k}_{\text{(b)}2}))t}$ respectively.
\end{itemize}
One can similarly work out that $f^{(2,1,0)}(k,k_1, k_2, x, x_1)$ are dertermined by $A^{(1)\text{in(neq), out(neq)}}(k, k_{\text{(b)}})$, the boundary conditions appearing at the linear order in the amplitude expansion. 

We can now specialize to the holographic Feynman propagator as given by (\ref{holfeyn}). We can see that we reproduce the  general form (\ref{feynamp2}) at the second order in the amplitude expansion. We get $f^{(2,0,0)}(k,k_1, k_2, x)$ similarly from the boundary conditions at the quadratic order and satisfy the properties listed above. Once again these satisfy the symmetry requirements but not reality constraints at all $\mathbf{x}$ and $t$ unless both $A^{\text{in(neq)}}(k, k_{\text{(b)}1}, k_{\text{(b)}2})$ and $A^{\text{out(neq)}}(k, k_{\text{(b)}1}, k_{\text{(b)}2})$ are set to zero, thus the solution is taken to be exactly the same as given by regularity at the horizon. Similarly one can show that $f^{(2,1,0)}(k, k_1, k_2, x, x_1)$ in (\ref{feynamp2}) should also vanish. This result generalizes to higher orders in the amplitude expansion.

Thus we show that the holographic Feynman propagator is given by (\ref{holffin}) to all orders in the amplitude expansion. In other words only the solution which is regular at the horizon contributes in the linear response (\ref{holfeyn}) holographically in determining the non-equilibrium Feynman propagator.

Given that both the non-equilibrium retarded Green's function and Feynman propagator are determined by the regular solution, all non-equilibrium propagators are also determined as well by the regular solution in the gravitational response (\ref{grav}). Thus the boundary conditions which determine the non-equilibrium corrections to the Green's functions are determined by regularity alone.

\subsection{On why regularity at the horizon is sufficient}

Our main result is that the regularity at the horizon and linear response theory together determine all the non-equilibrum Green's functions holographically. We can try to understand the physical significance of this result in terms of the dual field theory. 

Clearly the non-equilibrium dynamics of the expectation value of the operator, as for instance the energy-momentum tensor, is determined by regularity at the horizon. Indeed this is how hydrodynamic transport coefficients are determined. In field theory this is given by the effective action $\Gamma[\mathcal{O}]$ as discussed in section II.

We recall from section II.A that the non-equilibrium evolution of the propagators is given by the effective action $\Gamma[\mathcal{O}, G]$ evaluated over the Schwinger-Keldysh contour. In section II.A we also showed that the effective action $\Gamma[\mathcal{O}, G]$ can be obtain from $\Gamma[\mathcal{O}]$ applying functional identities. Thus, we should require no new information to determine the dynamics of non-equilibrium Green's functions other than those required to determine the non-equilibrium evolution of the expectation value of the operator.

Holographically, this should imply that, just as regularity at the horizon is a sufficient principle to determine the non-equilibrium evolution of the expectation value of the operator, it should be a sufficient principle to determine the non-equilibrium Green's functions also. So the results of section II.A give a field theoretic justification of the holographic arguments advanced here.

\section{The holographic non-equilibrium fluctuation-dissipation relation}

In the previous section we have proved that the holographic non-equilibrium Feynman propagator is given by (\ref{holffin}). This result is true in non-equilibrium states close to thermal equilibrium, or more precisely in states perturbatively connected to equilibrium in derivative and amplitude expansions. This result holds in the absence of any external forces. Also this result is valid even when the non-linearities in the dynamics of hydrodynamic and non-hydrodynamic variables characterizing the non-equilibrium state are taken into account.

From the holographic Feynman propagator (\ref{holffin}), one can readily obtain the non-equilibrium statistical function using (\ref{feynreim}). Thus we obtain the holographic non-equilibrium fluctuation-dissipation relation :
\begin{equation}\label{holflucdis}
G_{\mathcal{K}}(k,x) =i \Big(2 n_{\text{BE}}(\omega)+1\Big) \text{Im}\, G_R(k,x) = -i\Big(n_{\text{BE}}(\omega)+\frac{1}{2}\Big)
\mathcal{A}(k,x).
\end{equation}
We recall again that the Bose-Einstein distribution $n_{\text{BE}}(\omega)$ above is determined by the temperature of final thermal equilibrium. Thus at long times we recover the usual equilibrium fluctuation-dissipation relation.

We may wonder if this result is in contradiction with locality and causality of the dual field theory, because it depends on the temperature of thermal equilibrium to be attained in the future. This is not the case because of the following reasons :
\begin{itemize}
\item The fluctuation-dissipation relation (\ref{holflucdis}) relates the Wigner transformed statistical function $G_{\mathcal{K}}(k,x)$ and $\mathcal{A}(k,x)$. The Wigner transform involves Fourier transform in the relative coordinate $x_1 - x_2$ both in $G_{\mathcal{K}}(x_1, x_2)$ and $\mathcal{A}(x_1, x_2)$. The Fourier transform receives contributions from all values of the relative coordinate $x_1 - x_2$, thus both $G_{\mathcal{K}}(k,x)$ and $\mathcal{A}(k,x)$ are non-local in space and time by construction.
\item The result (\ref{holflucdis}) is true strictly in the absence of external forces. In such situations, no energy is being pumped in externally, therefore the final equilibrium temperature is fixed by the total energy of the initial state. So there is no teleological adjustment involved as it may naively appear to be the case.
\end{itemize}

Crucially $G_{\mathcal{K}}(k,x)$ and $\mathcal{A}(k,x)$ behave as local objects in space-time typically at weak coupling. In this case, the Kadanoff-Baym equations governing their evolution reduces approximately to Boltzmann equation which is local in space and time. At strong coupling we do not expect this to be the case, therefore we have no reason to expect that $G_{\mathcal{K}}(k,x)$ and $\mathcal{A}(k,x)$ will be related by local data. What we find is that the total conserved energy in the laboratory frame is sufficient to relate them - we do not need any more details of the non-equilibrium state, provided the field theory can be described holographically and the classical gravity approximation with a minimally coupled bulk scalar is valid. Individually  $G_{\mathcal{K}}(k,x)$ and $\mathcal{A}(k,x)$ though, carries detailed information of the relaxational modes of the system and their non-linear dynamics.

We also note that in the late equilibrium state where our derivative and amplitude expansions are valid, the background hydrodynamic and non-hydrodynamic variables satisfy generic phenomenological equations which can be derived from classical gravity (for details see appendix B). These phenomenological equations, just like Navier-Stokes equation hold irrespctively of initial conditions. The dual geometries have regular future horizons. Nevertheless, we can expect that not all solutions of these generic equations can be lifted to solutions in the field theory. Only a class of dual non-equilibrium geometries will also be regular in the past, and these will be genuine solutions in field theory. However, regularity in the past cannot be analyzed in perturbative derivative and amplitude expansions. Thus there will indeed be hidden constrants in the initial conditions for $\delta\mathbf{u}$, $\delta T$ and $\pi_{ij}^{\text{(nh)}}$ coming from the requirement of regularity in the past. Nevertheless, these do not affect our main conclusions, at least for the large class of non-equilibrium states which admit permit perturbative derivative and amplitude expansions in the future.

We may also ask how this result may generalize to higher point non-equilibrium Green's functions. To address this, we have to study the backreaction of the scalar field on the background geometry with an arbitrary external source at the boundary. Regarding the generalization we make the following conjectures :
\begin{itemize}
\item In the classical gravity approximation, the fully causal higher point Green's function $\Theta (t_n - t_{n-1})....\Theta(t_2 - t_1)\Big{\langle}[O(x_n), [O(x_{n-1}),......[O(x_2),O(x_1)]......]]\Big{\rangle}$ can be determined from the regular solution alone.
\item The other higher point Green's functions can be obtained from the causal higher-point function using their general field-theoretic properties and the consistency of their allowed forms with the classical gravity approximation.
\end{itemize}
The first point above simply says that a regular future horizon determines the full causal structure and hence the causal Green's functions. The second point above generalizes our derivation of the holographic Feynman propagator. This conjecture says that we do not need to construct any partition function to obtain non-equilibrium correlation functions - general dynamical principles should suffice to determine them as in field theory. In future work, we would like to test or prove this conjecture.

\section{Do our results survive stringy corrections to classical gravity?}

We will find here that we cannot certainly say that our results will survive the stringy corrections which typically give $1/\sqrt{\lambda}$ corrections in the strong coupling limit as a function of the dimensionless coupling $\lambda$. In gauge theories, $\lambda$ is the 't Hooft coupling.

Nevertheless all of our conclusions generalize even if there is stringy corrections in gravity dynamics (involving higher orders in curvatures and their derivatives), provided the scalar field dual to the operator is minimally coupled to gravity. The latter is certainly not guaranteed, but could be true in supersymmetric contexts. 

Firstly, we will show that our presciptions which give solutions which are regular at the future horizon in the non-equilibrium geometry, indeed remain valid as long as the bulk scalar field is minimally coupled to gravity.

We observe that the structure of the homogeneous incoming solution near the horizon can be determined from geometric optics, if the bulk scalar field is minimally coupled to gravity. We can certainly construct an appropriate function of $r$ which we denote as $r_*(r)$ such that the incoming radial null geodesic at the horizon is :
\begin{equation*}
v = t - r_*(r).
\end{equation*}
Clearly $ r_*(r)$ has to increase monotonically as $r$ moves towards the horizon because of blue-shifting. The homogeneous incoming wave solution in the thermal background by a quasi-normal mode perturbation will always behave like :
\begin{equation*}
\approx e^{-i(\omega+\omega_{\text{(b)}})v}
\end{equation*}
near the horizon at the leading order, as the geometrical optics approximation is always good at the horizon due to the blue-shifting. Therefore, as long as thermal geometry is stable, the integration over $\omega_{(b)}$ which we need to do to put the metric on-shell, will produce a divergent factor :
\begin{equation*}
e^{-\text{Im}\,\omega_{\text{(b)}}(\mathbf{k}_{\text{(b)}}) r*(r)}.
\end{equation*}
If the thermal background is stable, all it's quasi-normal fluctuatons should satisfy $\text{Im}\,\omega_{\text{(b)}}(\mathbf{k}_{\text{(b)}}) < 0$. Therefore the homogeneous incoming solution will always be divergent at the horizon. 

On the other hand the argument that the outgoing modes will cause divergent backreaction at the horizon is based on analyticity in $\omega$ - this remains true as long as the thermal backgound is stable.

In any consistent classical gravity, the black brane dual to the thermal background should be stable, thus we need to cut down both the homogeneous incoming and outgoing modes for regularity at the horizon.

The chain of arguments determining non-equilibrium propagators, similarly will also carry through all the way to the non-equilibrium fluctuation-dissipation relation (\ref{holflucdis}). This will be true provided the black brane do not support quasi-normal modes where both the momentum and frequency could be imaginary, or where the frequency/momentum is imaginary while the momentum/frequency vanishes respectively. One can check, if this happens, the reality arguments which cut down other boundary conditions to determine non-equilibirum propagators other than those required by regularity, will no longer hold. 

In fact, a quasinormal mode with imaginary momentum at zero frequency, will signify the existence of hair. Also, a quasinormal mode with imaginary frequency at zero momentum can be interpreted as a thermal tensor-like Goldstone mode. The first situation is unlikely as usually black branes do not have hair. The second situation is unlikely as usally in field theories we do not have spontaneous breaking of Lorentz symmetry.  

Even if the bulk scalar remains minimally coupled taking stringy corrections into account, we still need some care. Though the equations of motion may admit systematic perturbative expansion in $1/\sqrt{\lambda}$, all solutions which thermalize to black branes need not admit such expansion. Also we cannot take the limit $\lambda \rightarrow 0$ smoothly, as in this limit curvature diverges asymptotically. Therefore, we still need to be on the strong coupling side, for our results to generalize.

We may conclude that the nonequilibrium fluctuation-dissipation relation (\ref{holflucdis}) requires not only the validity of the classical gravity approximation, but also the classical theory of gravity to be nearly Einstein's theory, and the bulk scalar field to be minimally coupled to gravity.

\section{Concluding remarks}

The key points of this paper may be summarized as follows. 
\begin{itemize}
\item The holographic non-equilibrium spectral function carries systematic information about the relaxation modes of the system and their collective non-linear dynamics.
\item The holographic prescription for obtaining the non-equilibrium spectral function is independent of the non-equilibrium state i.e. the gravitational background and is obtained from requiring regularity at the horizon at strong coupling and large $N$.
\item This holographic prescription should also hold in any theory of classical gravity which has a stable thermal background as a solution as long as the bulk scalar dual to the bosonic operator is minimally coupled to gravity.
\item It is possible to map the parametrization of non-equilibrium Green's functions required for their consistent derivative/amplitude expansions to linear response functions in gravity using arbitrary boundary conditions for the non-equilibrium modes.
\item Field-theoretic consistency determines the holographic non-equilibrium Feynman propagator.  
\item From the holographic non-equilibrium Feynman propagator, we obtain the non-equilibrium fluctuation-dissipation relation.
\item The holographic non-equilibrium fluctuation-dissipation relation holds universally for any non-equilibrium state at strong coupling in the classical gravity approximation and thus provides a potentially strong test of the applicability of holographic duality.
\end{itemize}
Except for the first two points, all these have been derived in this paper. 

We may now discuss how can the holographic scenario be tested in RHIC or ALICE. Our discussion will assume that the plasma produced by the heavy-ion collision undergo three stages of evolution - (i) the initial phase (ii) a strongly coupled nearly conformal phase, and (iii) the final hadron gas phase. The classical gravity approximation studied here can only apply to the middle phase of evolution. This picture is supported by lattice studies as in \cite{Gavai} where it was found that QCD is inded close to a strongly coupled fixed point at a scale of about 175 MeV (which is nearly the temperature of the quark-gluon plasma produced at RHIC/ALICE) for small chemical potentials.  This is also consistent with the observed incredibly short time of thermalization ($\approx$ 1 fm/c) \cite{Adams, Florkowski}.

We note that in the early stage of 
the ultra-relativistic collisions we may use perturbative QCD which gives a quantum kinetic theory involving quarks and gluons. In this case, we may use models like \cite{Geiger}. We may also use color glass condensate models where the two particle irreducible (2PI) effective action technique has been used to obtain growth of  fluctuations and correlations of various operators \cite{Hatta}. In the late stage of expansion of the fireball, we can use the Ultra-relativistic 
Quantum Molecular Dynamics (UrQMD) simulations which models the quantum kinetics of the hadron gas \cite{Bleicher}. The holographic fluctuation and correlation of the chiral condensate needs to be matched with the quantum kinetics of the early and late phase of the expansion to obtain a complete picture that can be tested experimentally.

In the holographic phase, we may use the gravitational backgrounds provided by Einstein's gravity with anti de-Sitter boundary conditions to calculate the non-equilibrium spectral function. The non-equilibrium spectral function gives us a way to determine the collective non-linear dynamics of the relaxation modes and thus check if the phenomenological equations obtained from gravity really describe the space-time evolution of the fireball in the strongly coupled phase. The latter can also be tested independently by detailed measurement of transverse momentum spectra and elliptic flow coefficient of the hadron gas. A mutual confirmation will be a strong test for holography.

The non-equilibrium spectral function combined with the non-equilibrium fluctuation-dissipation relation gives sufficient data to match with correlations of emission of pions and their resonances in the later stage of expansion, as pions and their resonances have the same quantum numbers as the chiral condensate. Thus we can build a more sophisticated theory of pion interferometry to reconstruct the expansion of the fireball and test the holographic non-equilibrium fluctuation-dissipation relation. As the fluctuation-dissipation relation holds universally for any non-equilibrium state in the classical gravity approximation, it's validation will indeed be a very strong test for holography.

In fact, it has been observed that statistical models using pure thermal estimates work very well for predicting the spectrum of hadrons produced from the fireball with the temperature being set to the value at which the fireball thermalizes \cite{Braun}, \cite{Adams}. This is surprising given that the phase transition does not occur adiabatically. This may already hint that the holographic non-equilibrium fluctuation dissipation relation is valid to a good accuracy.

The statistical function of the chiral condensate carries direct information about the production of pions and their resonances. Thus a time-resolved study of emission of pions and their resonances will allow us to determine the non-equilibrium statistical function once we know how to match it's evolution to the early and final phases.

A more refined time-resolved study of correlations of emitted pions and their resonances will allow determination of the non-equilibrium spectral function. This also needs to be matched with the early and late stages of evolution. Having thus determined the non-equilibrium statistical and spectral functions individually, we can check the non-equilibrium fluctuation-dissipation relation.

The matching can be done by noting the following that in the holographic phase, the non-equilibrium state, and also the non-equilibrium spectral and statistical functions, can be parametrized by a few non-equilibrium variables like the hydryodynamic variables and the shear-stress tensor. These variables represent the expectation value of a handful of operators like the energy-momentum tensor. Thus the matching can be realized by tracking the evolution of these operators. Some interesting progress has already been made in this direction \cite{Gardim}.

It thus looks possible that the holographic non-equilibrium fluctuation-dissipation relation can be tested in the future in heavy-ion collisions. This may also lead to an accurate experimental determination of the temperature at which the plasma thermalizes.

It will be also interesting to understand how the non-equilibrium fluctuation-dissipation relation be measured in order parameter correlations in quantum critical systems.

\begin{acknowledgments}

The author would like to thank Giuseppe Policastro, Umut Gursoy, Kostas Skenderis and Jan de Boer for useful discussions. The author would like to thank Giuseppe Policastro for comments on the manuscript. The author thanks Costas Bachas, Atish Dabholkar and Marios Petropoulos for encouragement in pursuing holographic non-equilibrium physics. The author thanks Souvik Banerjee for checking some equations and also for comments on the manuscript. The author would also like to thank CERN Theory Division, Albert Einstein Institute, Golm and ITF, University of Amsterdam for opportunities to present this work prior to publication. The research of the author is presently supported by the grant number ANR-07-CEXC-006 of L'Agence Nationale de La Recherche. The paper has been rewritten in it's present form when the author has been visiting CERN Theory Division.

\end{acknowledgments}

\appendix
\section{Useful identities}
\textbf{Consequences of symmetry and transposition}
\newline\newline
Consider a Hermitean scalar operator $O(x)$. The retarded Green's function for this operator is defined as 
\begin{equation}\label{retd}
G_R(x_1, x_2) = - i \Theta(t_1 - t_2)\Big{\langle}\Big[O(x_1), O(x_2)\Big]\Big{\rangle}.
\end{equation}
The advanced Green's function is defined as
\begin{equation}\label{advd}
G_A(x_1, x_2) = i \Theta(t_2 - t_1)\Big{\langle}\Big[O(x_1), O(x_2)\Big]\Big{\rangle}.
\end{equation}
It follows from the definitions that
\begin{equation}\label{sym}
G_A(x_1, x_2) = G_R(x_2, x_1).
\end{equation}
We may also see that 
\begin{equation}\label{ccgr}
G_R*(x_1, x_2) = G_R(x_1, x_2), \quad G_A*(x_1, x_2) = G_A(x_1, x_2).
\end{equation}

Let us define the center-of-mass coordinate as $x = (x_1+x_2)/2$ and the relative coordinate $r = x_1 - x_2$. The Wigner transformed retarded and advanced Green's function are :
\begin{equation}
G_{R, A}(k, x) = \int d^4 r\, e^{ik\cdot r} G_{R,A}(x,r).
\end{equation}
It follows from (\ref{sym}) that $G_A(x, r)= G_R(x, -r)$. We thus see that
\begin{eqnarray}\label{id1}
G_{R}(-k, x) &=& \int d^4 r\, e^{-ik\cdot r} G_{R}(x,r) \nonumber\\
&=&\int d^4 r\, e^{ik\cdot r} G_{R}(x,-r)\nonumber\\
&=&\int d^4 r\, e^{ik\cdot r} G_{A}(x,r)\nonumber\\
&=& G_A(k, x).
\end{eqnarray}
In the second line above we have changed variables from $r$ to $-r$. 

Similarly, (\ref{ccgr}) implies that $G_R*(x,r)= G_R(x, r)$ and $G_A*(x,r)= G_A(x, r)$. It follows then
\begin{eqnarray}\label{id2}
G_{R}*(k, x) &=& \int d^4 r\, e^{-ik\cdot r} G_{R}*(x,r) \nonumber\\
&=&\int d^4 r\, e^{-ik\cdot r} G_{R}(x, r)\nonumber\\
&=& G_A(k, x).
\end{eqnarray}
In the second line above, we have used $G_R*(x,r)= G_R(x, r)$. Observing that the second line coincides with the first equality in (\ref{id1}), our final conclusion follows.
We can similarly prove that $G_A*(k, x)= G_R(k,x)$.

We now turn to the Feynman propagator $G_F(x_1, x_2)$ which is symmetric. Therefore, $G_F(x, r) = G_F(x, -r)$. Therefore it follows that the Wigner transform of the Feynman propagator should satisfy :
\begin{eqnarray}\label{id3}
G_{F}(-k, x) &=& \int d^4 r\, e^{-ik\cdot r} G_{F}(x,r) \nonumber\\
&=&\int d^4 r\, e^{ik\cdot r} G_{F}(x,-r)\nonumber\\
&=&\int d^4 r\, e^{ik\cdot r} G_{F}(x,r)\nonumber\\
&=& G_F(k, x).
\end{eqnarray}
In the second line above, we have changed 
variables from $r$ to $-r$, and in the third line we have used $G_F(x, r) = G_F(x, -r)$. 

Similarly it follows from symmetry in $x_1$ and $x_2$ prior to Wigner transform that $G_{\mathcal{K}}(k, x) = G_{\mathcal{K}}(-k, x)$.

Using (\ref{id1}), we also note that $\text{Re}\, G_R(k,x) = (1/2)(G_R(k, x)+ G_A(k,x)) = (1/2)(G_A(-k, x)+ G_R(-k,x)) = \text{Re}\, G_R(-k,x)$.
\newline\newline
\textbf{The statistical function is purely imaginary}
\newline\newline
According to (\ref{statb}), the inverse Wigner transform of the statistical function is given by the anti-commutator as below :
\begin{equation}
G_{\mathcal{K}}(x, r) = -\frac{i}{2}  \Big{\langle}\Big{\{ }O\Big(x + \frac{r}{2}\Big), O\Big(x- \frac{r}{2}\Big)\Big{\} } \Big{\rangle} .
\end{equation}
Clearly,
\begin{equation}
G_{\mathcal{K}}(x,r) = G_{\mathcal{K}}(x,-r), \quad G_{\mathcal{K}}*(x,r) = -G_{\mathcal{K}}(x,r).
\end{equation}
Therefore,
\begin{eqnarray}\label{id4}
G_{\mathcal{K}}*(k, x) &=& \int d^4 r\, e^{-ik\cdot r} G_{\mathcal{K}}*(x,r) \nonumber\\
&=&-\int d^4 r\, e^{-ik\cdot r} G_{\mathcal{K}}(x, r)\nonumber\\
&=& -\int d^4 r\, e^{ik\cdot r} G_{\mathcal{K}}(x, -r)\nonumber\\
&=&-\int d^4 r\, e^{ik\cdot r} G_{\mathcal{K}}(x, r)\nonumber\\
&=& -G_{\mathcal{K}}(k,x)
\end{eqnarray}
In the second line above, we have used $G_{\mathcal{K}}*(x,r) = -G_{\mathcal{K}}(x,r)$, in the third line we have changed variables from $r$ to $-r$, and in the fourth line, we have used $G_{\mathcal{K}}(x,r) = G_{\mathcal{K}}(x,-r)$.

Clearly, $G_{\mathcal{K}}*(k,x) = -G_{\mathcal{K}}(k,x)$ implies that $G_{\mathcal{K}}(k,x)$ is purely imaginary.
\newline\newline
\textbf{The spectral function and the retarded Green's function}
\newline\newline
According to (\ref{specb}), the inverse Wigner transform of the spectral function is given by the commutator as below :
\begin{equation}
\mathcal{A}(x, r) =  \Big{\langle}\Big[ O\Big(x + \frac{r}{2}\Big), O\Big(x- \frac{r}{2}\Big)\Big]  \Big{\rangle} .
\end{equation}
It follows from (\ref{retd}) and(\ref{advd}) that
\begin{equation}
\mathcal{A}(x,r) = i \Big(G_R(x,r) - G_A(x,r)\Big) = - 2\text{Im}\,G_R(x,r) ,
\end{equation}
using $\Theta (t_1 - t_2) + \Theta(t_2 - t_1) = 1$.
Clearly then after Wigner transform,
\begin{equation}
\mathcal{A}(k,x) = i \Big(G_R(k,x) - G_A(k, x)\Big) = - 2\,\text{Im}\,G_R(k,x).
\end{equation}
Using (\ref{id1}), we readily see from the above that
\begin{equation}
\mathcal{A}(k,x)= -\mathcal{A}(-k,x).
\end{equation}
\newline\newline
\textbf{The Feynman propagator as a sum of the statistical and spectral functions}
\newline\newline
The definition of the Feynman propagator is :
\begin{equation}
G_F(x_1, x_2) = -i \Big{\langle} T\left(O(x_1) O(x_2)\right)\Big{\rangle},
\end{equation}
where $T$ denotes time ordering.

It follows from the definition of the Feynman propagator that
\begin{eqnarray}
G_F(x_1, x_2) &=& - i \Big{\langle} O(x_1) O(x_2) \Big{\rangle} \ \text{if $t_1 > t_2$} 
\nonumber\\
&=& - \frac{i}{2} \Big{\langle} O(x_1) O(x_2) + O(x_2) O(x_1) \Big{\rangle} - \frac{i}{2} \Big{\langle} O(x_1) O(x_2) - O(x_2) O(x_1) \Big{\rangle} \ \text{if $t_1 > t_2$} \nonumber\\
&=& G_{\mathcal{K}}(x_1, x_2) - \frac{i}{2} \mathcal{A}(x_1, x_2)\ \text{if $t_1 > t_2$}.
\end{eqnarray}
Similarly,
\begin{eqnarray}
G_F(x_1, x_2) &=& - i \Big{\langle} O(x_2) O(x_1) \Big{\rangle} \ \text{if $t_2 > t_1$} 
\nonumber\\
&=& - \frac{i}{2} \Big{\langle} O(x_1) O(x_2) + O(x_2) O(x_1) \Big{\rangle} + \frac{i}{2} \Big{\langle} O(x_1) O(x_2) - O(x_2) O(x_1) \Big{\rangle} \ \text{if $t_2 > t_1$} \nonumber\\
&=& G_{\mathcal{K}}(x_1, x_2) + \frac{i}{2} \mathcal{A}(x_1, x_2)\ \text{if $t_2 > t_1$}.
\end{eqnarray}
Combining these, we obtain
\begin{equation}
G_F(x_1, x_2) = G_{\mathcal{K}}(x_1, x_2) - \frac{i}{2} \mathcal{A}(x_1, x_2)\,\text{sign}
(t_1 - t_2).
\end{equation}

\section{The general phenomenological equations for irreversible processes}

The phenomenological equations generalizing hydrodynamics which can be obtained from gravity can be argued to be as follows. We can do a kinematic Landau-Lifshitz decomposition of the energy-momentum tensor as below :
\begin{equation}\label{ll}
t_{\mu\nu} = \epsilon\, u_\mu u_\nu + p\, 
P_{\mu\nu} + \pi_{\mu\nu},
\end{equation} 
with $u^\mu$ being a time-like vector of unit norm (so $u_\mu u^\nu = -1$), and $P_{\mu\nu}$ being the projection tensor in the spatial plane orthogonal to $u^\mu$ given by $P_{\mu\nu} = u_\mu u_\nu + \eta_{\mu\nu}$. 

We can define a local temperature using the equation of state $\epsilon(T)$ locally. We can also set $p$ by using the local equation of state $p(T)$. Then $\pi_{\mu\nu}$ denotes the local non-equilirbium part of the energy-momentum tensor. 

Also $u^\mu$ is defined to be the local velocity of energy transport, therefore
\begin{equation}\label{fpi}
u^\mu \pi_{\mu\nu}= 0.
\end{equation}
Thus $\pi_{\mu\nu}$ has six independent components only. Together with $u^\mu$ and $T$, it gives ten independent variables to parametrize ten independent components of $t_{\mu\nu}$. 

Furthermore, conformal invariance requires 
\begin{equation}
\epsilon = 3p, \quad \pi_{\mu\nu}\eta^{\mu\nu} = 0.
\end{equation}
This reduces the number of variables to nine, cutting down the independent components of $\pi_{\mu\nu}$ from six to five.

To proceed further, we need to decompose $\pi_{\mu\nu}$ into two parts, a purely hydrodynamic part $\pi_{\mu\nu}^{\text{(h)}}$ and a non-hydordynamic part $\pi_{\mu\nu}^{\text{(nh)}}$ which is dynamically independent of the hydrodynamic variables though they do couple to them \cite{myself1}, \cite{myself2}. Thus,
\begin{equation}\label{split}
\pi_{\mu\nu} = \pi_{\mu\nu}^{\text{(h)}} + \pi_{\mu\nu}^{\text{(nh)}}.
\end{equation}
The fluid/gravity correspondence tells us that at strong coupling, purely hydrodynamic states exist in the dual field theory in the derivative expansion. Thus we can consistently set the non-hydrodynamic part $\pi_{\mu\nu}^{\text{(nh)}}$ to zero. 

The general form of the purely hydrodynamic part $\pi_{\mu\nu}^{\text{(h)}}$ can be readily obtained order by order in the derivative expansion by constructing the conformally covariant tensors which are algebraic functionals of the derivatives of the hydrodynamic variables. The derivative expansion parameter is typical length scale of variation with respect to the thermal wavelength. The transport coefficients are determined via regularity at the future horizon \cite{Sayantani}. Thus
\begin{equation}
\pi_{\mu\nu}^{\text{(h)}} = -\eta P_\mu^{\phantom{\mu}\rho}
P_\nu^{\phantom{\mu}\sigma}
\Big(\partial_\rho u_\sigma +\partial_\sigma u_\rho - \frac{2}{3}P_{\rho\sigma}(\partial\cdot u)\Big) + O(\epsilon^2).
\end{equation}
Putting $\pi_{\mu\nu}^{\text{(nh)}} = 0$, plugging $\pi_{\mu\nu} = \pi_{\mu\nu}^{\text{(h)}}$ in the form of $t_{\mu\nu}$ in (\ref{ll}), and demanding $\partial^\mu t_{\mu\nu}= 0$ results in the hydrodynamic equations of motion.

Energy and momentum are always conserved and this follows directly from the constraints of Einstein's equations. This implies that 
\begin{equation}\label{consv}
\partial^\mu t_{\mu\nu}= \partial^\mu \Big(  \epsilon(T)\, u_\mu u_\nu + p(T)\, 
P_{\mu\nu} + \pi_{\mu\nu}^{\text{(h)}}+ \pi_{\mu\nu}^{\text{(nh)}}\Big)= 0.
\end{equation}
The above are sufficient to determine the evolution of $u^\mu$ and $T$, but not that of the non-hydrodynamic shear-stress tensor $\pi_{\mu\nu}^{\text{(nh)}}$. We require five more equations to determine the latter. These should be provided by the regularity of the horizon in gravity as in the case of homogeneous relaxation. We need a separate expansion parameter - the amplitude parameter $\pi_{ij}^{\text{(nh)}}/p$. As in the case of homogeneous relaxation, though the spatial derivates of $\pi_{ij}^{\text{(nh)}}$ can be expected to be small near equilibrium, the time-derivatives will be $O(T)$. The Lorentz and Weyl covariant generalization of the time-derivative is 
$\mathcal{D} = (u\cdot\partial)+ ...$.

The equation of motion for $\pi_{ij}^{\text{(nh)}}$ should be such that (i) it can be consistently put to zero (ii) it is Lorentz and Weyl covariant and (iii) all time-derivatives are summed over at each order in the amplitude expansion. Requiring these, it follows that this equation should take the form \cite{myself2}, \cite{myself3}:
\begin{eqnarray}\label{pheneq}
\left(\displaystyle\sum\limits_{n=0}^{\infty}D_{R}^{(1,n)}
\mathcal{D}^n\right)\pi_{\mu \nu}^{\text{\text{\text{(nh)}}}} &=&
\frac{1 }{2}
\displaystyle\sum\limits_{n=0}^{\infty}
\Bigg(\Big(\lambda_1^{(n)}\mathcal{D}^n
\pi_{\mu}^{\text{(nh)}\alpha}\Big)
\partial_{(\alpha}u_{\nu)}+
\Big(\lambda_1^{(n)}\mathcal{D}^n
\pi_{\nu}^{\text{(nh)}\alpha}\Big)\partial_{(\alpha}u_{\mu)}- \frac{2}{3}P_{\mu
\nu} \Big(\lambda_1^{(n)}\mathcal{D}^n
\pi_{\alpha\beta}^{\text{(nh)}}\Big)\partial^\alpha u^\beta
\nonumber\\
&&+
\Big(\lambda_2^{(n)}\mathcal{D}^n
\pi_{\mu}^{\text{(nh)}\alpha}\Big)\partial_{[\alpha}u_{\nu]}
+\Big(\lambda_2^{(n)}\mathcal{D}^n
\pi_{\nu}^{\text{(nh)}\alpha}\Big)\partial_{[\alpha}u_{\mu]}\Bigg) \nonumber\\&
&- \displaystyle\sum\limits_{n=0}^{\infty}\displaystyle\sum\limits_{m=0}^{n}D_R^{(2,n,m)}\Bigg[\mathcal{D}^m
\pi_{\mu}^{\text{(nh)}\alpha}\mathcal{D}^n
\pi_{\alpha\nu}^{\text{\text{\text{(nh)}}}}-\frac{1}{3}P_{\mu\nu} \ \mathcal{D}^{m}
\pi_{\alpha\beta}^{\text{(nh)}}\mathcal{D}^n \pi^{\text{(nh)}\alpha\beta}\Bigg]\nonumber\\
&&+ O(\epsilon^2\delta, \epsilon\delta^2 ,\delta^3)
\end{eqnarray}
up to given orders in the derivative/amplitude expansion. Above $(...)$ and $[...]$ denotes symmetrization and anti-symmetrization of enclosed indices respectively. 

Let us see how we can derive these equations from gravity for the special case of homogeneous relaxation. In this case, the flow is at rest, so we can go to an inertial frame where $u^\mu = (1, 0, 0,0)$. Also the temperature $T$ is constant in space and time. Furthermore (\ref{fpi}) implies that the non-zero components of $\pi_{\mu\nu}$ are the spatial components $\pi_{ij}$. The conservation of energy and momentum implies that $\pi_{ij}(t)$ is an arbitrary function of time. Nevertheless, regularity at the horizon can be expected to be guaranteed only when $\pi_{ij}(t)$ follows a definite equation of motion.

One can construct the dual metric in such cases perturbatively in the so-called amplitude expansion $\pi_{ij}^{\text{(nh)}}(t)/p$. Omitting the details, the metric up to second order in the amplitude expansion, takes the form below \cite{myself2}:
\begin{eqnarray}\label{metrichr}
ds^2 &=& \frac{l^2}{r^2}\frac{dr^2}{f\Big(\frac{r r_0}{l^2}\Big)}+ \frac{l^2}{r^2}\Bigg(- f\Big(\frac{r r_0}{l^2}\Big) dt^2 + d\mathbf{x}^2 \nonumber\\
&&+ \sum_{n=1}^{\infty}r_0^{4+n}f^{(1,n)} \Big(\frac{r r_0}{l^2}\Big) \Big(\frac{d}{dt}\Big)^n\pi_{ij}^{\text{(nh)}}dx^i dx^j\nonumber \\
&& +\sum_{n=1}^{\infty}\sum_{m=1}^{n}r_0^{8+n+m}f_1^{(2,n, m)} \Big(\frac{r r_0}{l^2}\Big) \Big(\frac{d}{dt}\Big)^n\pi_{kl}^{\text{(nh)}}\Big(\frac{d}{dt}\Big)^m\pi_{kl}^{\text{(nh)}}\, dt^2 \nonumber\\
&&+\sum_{n=1}^{\infty}\sum_{m=1}^{n}r_0^{8+n+m}f_2^{(2,n, m)} \Big(\frac{r r_0}{l^2}\Big) \Big(\frac{d}{dt}\Big)^n\pi_{kl}^{\text{(nh)}}\Big(\frac{d}{dt}\Big)^m\pi_{kl}^{\text{(nh)}}\, d\mathbf{x}^2 \nonumber \\
&&+\sum_{n=1}^{\infty}\sum_{m=1}^{n}r_0^{8+n+m}f_3^{(2,n, m)} \Big(\frac{r r_0}{l^2}\Big) \Bigg(\Big(\frac{d}{dt}\Big)^n\pi_{ik}^{\text{(nh)}}\Big(\frac{d}{dt}\Big)^m\pi_{kj}^{\text{(nh)}} + (i\leftrightarrow j)- \frac{2}{3}\delta_{ij}\Big(\frac{d}{dt}\Big)^n\pi_{kl}^{\text{(nh)}}\Big(\frac{d}{dt}\Big)^m\pi_{kl}^{\text{(nh)}}\Bigg)\, dx^i dx^j \nonumber\\ && + O(\delta^3)\Bigg).
\end{eqnarray} 
We note that at each order in the amplitude expansion we have summed over all time-derivatives of $\pi_{ij}^{\text{(nh)}}(t)$. Thus at the first order in amplitude expansion, we get an infinite number of radial functions $f^{(1,n)}$ correponding to the $n$-th time-derivative. These can be determined uniquely by requiring $f^{(1,1)}(s) = s^4 + O(s^8)$ and $f^{(1,n)}(s) = O(s^{4+n})$ for $n>1$. In these Schwarzchild coordinates it turns out that $f^{(1,n)}$ vanishes for odd $n$. Similarly at the second order in the amplitude expansion we get infinite number of radial functions $f^{(2,n,m)}_1$, $f^{(2,n,m)}_2$ and $f^{(2,n,m)}_3$ all of which can be determined uniquely from their boundary behavior $f^{(2,n,m)}_i=O(s^{8+n+m})$ for $i=1,2,3$. Also in these Schwarzchild coordinates $f^{(2,n,m)}_i$ vanishes when $n+m$ is odd.

The regularity analysis of the metric (\ref{metrichr}) is subtle and involves it's translation to Eddington-Finkelstein coordinates. One finds that the metric is regular, if up to second order in the amplitude expansion $\pi_{ij}^{\text{(nh)}}(t)$ satisfies the following equation of motion \cite{myself2} :
\begin{eqnarray}\label{hreq}
\sum_{n=0}^{\infty} D^{(1,n)}\Big(\frac{d}{dt}\Big)^n \pi_{ij}^{\text{(nh)}} + \sum_{n=0}^{\infty}\sum_{m=0}^{n} D^{(2,m,n)} \Bigg(\Big(\frac{d}{dt}\Big)^n \pi_{ik}^{\text{(nh)}} \Big(\frac{d}{dt}\Big)^m \pi_{kj}^{\text{(nh)}} - \frac{1}{3}\delta_{ij}\Big(\frac{d}{dt}\Big)^n \pi_{kl}^{\text{(nh)}} \Big(\frac{d}{dt}\Big)^m \pi_{kl}^{\text{(nh)}}\Bigg) \nonumber\\+ O(\delta^3)= 0 .
\end{eqnarray}
In order to see regularity of the metric at future horizon, order by order in the amplitude expansion, it is necessary to sum over all time-derivatives at each order. One can determine the non-dydrodynamic phenomenological coefficients $D^{(1,n)}$ and $D^{(2,n,m)}$ in terms of complicated recursion relations. The first few terms are :
\begin{equation}\label{ds}
D_R^{(1,0)}= -\pi T, \ \  D_R^{(1,1)}=-\Big(\frac{\pi}{2} - \frac{1}{4} \ \ln 2\Big) (\pi T)^2, \ \ \text{etc.;} \quad D_R^{(2,0,0)} = \frac{1}{2(\pi T)^4}, \ \text{etc.}
\end{equation}
That it is necessary to sum over all time-derivatives at each order in the amplitude expansion to obtain regularity at the horizon, can be understood from the first order in the amplitude expansion itself. It can be shown that the series $\sum D_R^{(1,n)}(-i\omega)^n$ has simple zeroes at the location of the discrete homogenous quasi-normal modes (with $\mathbf{k}= 0$) in the complex lower half plane. The location of such quasi-normal modes are approximately given by \cite{Starinets}
\begin{equation}
\omega_{\text{(n)}} \approx \pi T\Big[ \pm 1.2139  - 0.775 \, i \pm 2n (1\mp i)\Big] .
\end{equation}
We observe that both the real and imaginary parts are of $O(T)$ signifying that the time-derivatives of $\pi_{ij}^{\text{(nh)}}$ are of $O(T)$. Thus, we do need to sum over all time-derivatives as we cannot do a small frequency expansion even in the linearized approximation.

The equation (\ref{pheneq}) reduces to (\ref{hreq}) in the special case when $T$ and $u^\mu$ are constant is space and time, and furthermore we go to the laboratory frame where $u^\mu = (1,0, 0,0)$. Thus $D_R^{(1,n)}$ and $D_R^{(2,n,m)}$ should be as given in (\ref{ds}). The coefficients $\lambda_1^{(n)}$ and $\lambda_2^{(n)}$ denote coupling of $\pi_{\mu\nu}^{\text{(nh)}}$ and it's local time-derivatives to hydrodynamic variables. To determine these, one needs to construct metrics with regular future horizons for configuration where $u^\mu$ and $T$ vary spatially and temporally, while $\pi_{\mu\nu}^{\text{(nh)}}$ is non-zero.

\section{Non-equilibrium shifts in quasiparticle dispersion relations}

Holographically the vanishing of the source gives the quasi-particle poles (which can be so broad that we may not call them quasi-particles). At equilibrium $\mathcal{J}^{\text{in(eq)}}(\omega, \mathbf{k})$ vanishes only when $\omega$ takes certain discrete values at a given $\mathbf{k}$. We may select a branch of quasi-particle pole given by:
\begin{equation}
\omega = \omega^{\text{(eq)}}(\mathbf{k})\quad \text{such that}\quad \mathcal{J}^{\text{in(eq)}}(\omega^{\text{(eq)}}(\mathbf{k}), \mathbf{k})= 0.
\end{equation}
These correspond to the poles which can be complex generally.
The non-equilibrium modification depends on space and time, and takes the form
\begin{equation}
\omega = \omega^{\text{(eq)}}(\mathbf{k}) + \delta\omega(\mathbf{k}, \mathbf{x}, t).
\end{equation}
The non-equilibrium shift in the pole $\delta\omega(\mathbf{k}, \mathbf{x}, t)$ can be obtained by solving $\mathcal{J}(\mathbf{x},t)= 0$ perturbatively \cite{noneqspec}. This amounts to solving the linear complex equation :
\begin{eqnarray}\label{nshift}
 \delta\omega(\mathbf{k}, \mathbf{x}, t) \, \partial_{\omega}\mathcal{J}^{\text{in(eq)}}
\Big(\omega&=& \omega^{\text{(eq)}}(\mathbf{k}), \mathbf{k}\Big) = -\mathcal{J}^{\text{in(neq)}}\Big(\omega=\omega^{\text{(eq)}}(\mathbf{k}) ,\mathbf{k},  \mathbf{k}_{\text{(b)}}\Big)
e^{- i \text{Re}\,\omega_{\text{(b)}}(\mathbf{k}_{\text{(b)}})t}e^{\text{Im}\,\omega_{\text{(b)}}(\mathbf{k}_{\text{(b)}})t}
e^{i\mathbf{k}_{\text{(b)}}
\cdot\mathbf{x}}\nonumber\\&&
-\mathcal{J}^{\text{in(neq)}}\Big(\omega=\omega^{\text{(eq)}}(\mathbf{k}),\mathbf{k},  -\mathbf{k}_{\text{(b)}}\Big)e^{i \text{Re}\,\omega_{\text{(b)}}(\mathbf{k}_{\text{(b)}}))t}e^{\text{Im}\,\omega_{\text{(b)}}(\mathbf{k}_{\text{(b)}})t}
e^{-i\mathbf{k}_{\text{(b)}}
\cdot\mathbf{x}}.
\end{eqnarray}
The non-equilibrium parts of the source, or equivalently the right hand side of the above is determined completely by the equilibrium source by our prescription at the horizon. This is precisely what we expect in the dual field theory. The non-equilibrium shift in the pole should be determined solely by the infrared behavior. For example, if the temperature fluctuates, so should the thermal mass. This gives a good justification of our prescrition. Our general expression (\ref{nshift}) shows that the pole of the resummed non-equilibrium propagator can depend also on the velocity and shear-stress perturbations also.

It is evident from (\ref{nshift}) that $\delta\omega(\mathbf{k}, \mathbf{x}, t)$ will be comoving with the relaxational mode constituting the non-equilibrium state (or the dual quasinormal mode) as below :
\begin{eqnarray}
 \delta\omega(\mathbf{k}, \mathbf{x}, t) \, &=&\delta\omega(\mathbf{k}, \mathbf{k}_{\text{(b)}})
e^{- i \text{Re}\,\omega_{\text{(b)}}(\mathbf{k}_{\text{(b)}})t}e^{\text{Im}\,\omega_{\text{(b)}}(\mathbf{k}_{\text{(b)}})t}
e^{i\mathbf{k}_{\text{(b)}}
\cdot\mathbf{x}}\nonumber\\&&
+\delta\omega(\mathbf{k}, -\mathbf{k}_{\text{(b)}})e^{i \text{Re}\,\omega_{\text{(b)}}(\mathbf{k}_{\text{(b)}}))t}e^{\text{Im}\,\omega_{\text{(b)}}(\mathbf{k}_{\text{(b)}})t}
e^{-i\mathbf{k}_{\text{(b)}}
\cdot\mathbf{x}}.
\end{eqnarray}
Furthermore, $\delta\omega(\mathbf{k}, \mathbf{k}_{\text{(b)}})$ will have a consistent expansion in the derivative/amplitude expansion which follows from that of $\mathcal{J}^{\text{in(neq)}}(\omega, \mathbf{k}, \mathbf{k}_{\text{(b)}})$. For example, in the case of the hydrodynamic shear wave background $\delta\omega(\mathbf{k}, \mathbf{k}_{\text{(h)}})$ takes the form :
\begin{equation}
\delta\omega(\mathbf{k}, \mathbf{k}_{\text{(h)}})= \delta\omega_1(\mathbf{k})\delta\mathbf{u}
\cdot\mathbf{k} + \delta\omega_2(\mathbf{k})k_i k_j k_{\text{(h)}i}\delta u_j
\end{equation}
up to first order in the derivative expansion. We thus explicitly see in the above example how our holographic prescription gives the non-equilibrium shift in the quasi-particle pole parametrized in terms of the background velocity perturbation.

\section{Wigner transform of the holographic retarded Green's function}
Writing down explicitly we get,
\begin{eqnarray}\label{retp1}
G_{R}(\mathbf{x}_1, t_1, \mathbf{x}_2, t_2) &=&\int d\omega d^3 k\, \frac{ \mathcal{O}(\omega, \mathbf{k},\mathbf{x}_1, t_1)}{\mathcal{J}(\omega, \mathbf{k},\mathbf{x}_2,t_2)} = \int d\omega
\int d^3 k 
\\\nonumber&&   \Bigg(\mathcal{O}^{\text{(eq)}}(\omega,\mathbf{k})e^{-i\omega t_1}e^{i\mathbf{k}\cdot \mathbf{x}_1} + \mathcal{O}^{\text{in(neq)}}(\omega,\mathbf{k},  \mathbf{k}_{\text{(b)}})e^{- i (\omega + \text{Re}\,\omega_{\text{(b)}}(\mathbf{k}_{\text{(b)}}))t_1}e^{\text{Im}\,\omega_{\text{(b)}}(\mathbf{k}_{\text{(b)}})t_1}
e^{i(\mathbf{k}+\mathbf{k}_{\text{(b)}}
)\cdot\mathbf{x}_1}\\\nonumber&&+\mathcal{O}^{\text{in(neq)}}(\omega,\mathbf{k},  -\mathbf{k}_{\text{(b)}})e^{- i (\omega -\text{Re}\,\omega_{\text{(b)}}(\mathbf{k}_{\text{(b)}}))t_1}e^{\text{Im}\,\omega_{\text{(b)}}(\mathbf{k}_{\text{(b)}})t_1}
e^{i(\mathbf{k}-\mathbf{k}_{\text{(b)}}
)\cdot\mathbf{x}_1}\Bigg)/\Bigg(\mathcal{J}^{\text{(eq)}}(\omega,\mathbf{k})e^{-i\omega t_2}e^{i\mathbf{k}\cdot \mathbf{x}_2} \\\nonumber&&+ \mathcal{J}^{\text{in(neq)}}(\omega,\mathbf{k},  \mathbf{k}_{\text{(b)}})e^{- i (\omega + \text{Re}\,\omega_{\text{(b)}}(\mathbf{k}_{\text{(b)}}))t_2}e^{\text{Im}\,\omega_{\text{(b)}}(\mathbf{k}_{\text{(b)}})t_2}
e^{i(\mathbf{k}+\mathbf{k}_{\text{(b)}}
)\cdot\mathbf{x}_2}\\\nonumber&& +
\mathcal{J}^{\text{in(neq)}}(\omega,\mathbf{k},  -\mathbf{k}_{\text{(b)}})e^{- i (\omega -\text{Re}\,\omega_{\text{(b)}}(\mathbf{k}_{\text{(b)}}))t_2}e^{\text{Im}\,\omega_{\text{(b)}}(\mathbf{k}_{\text{(b)}})t_2}
e^{i(\mathbf{k}-\mathbf{k}_{\text{(b)}}
)\cdot\mathbf{x}_2}\Bigg).
\end{eqnarray}
The above can be approximated in the derivative/amplitude expansion as below :
\begin{eqnarray}\label{retp2}
G_{R}(\mathbf{x}_1, t_1, \mathbf{x}_2, t_2)&=& \int d\omega \int d^3k  \ \ e^{-i\omega(t_1 -t_2)}e^{i\mathbf{k}\cdot(\mathbf{x}_1 -\mathbf{x}_2)}\frac{\mathcal{O}^{\text{(eq)}}(\omega, \mathbf{k})}{\mathcal{J}^{\text{(eq)}}(\omega, \mathbf{k})}\nonumber\\&& \Bigg(1 + \frac{ \mathcal{O}^{\text{in(neq)}}(\omega, \mathbf{k}, \mathbf{k}_{\text{(b)}})}{\mathcal{O}^{\text{(eq)}}(\omega, \mathbf{k})}e^{i\mathbf{k}_{\text{(b)}}\cdot\mathbf{x}_1}
e^{-i \text{Re}\,\omega_{\text{(b)}}(\mathbf{k}_{\text{(b)}})t_1}e^{ \text{Im}\,\omega_{\text{(b)}}(\mathbf{k}_{\text{(b)}})t_1}
\nonumber\\&&+\frac{\mathcal{O}^{\text{in(neq)}}(\omega, \mathbf{k}, -\mathbf{k}_{\text{(b)}})}{\mathcal{O}^{\text{(eq)}}(\omega, \mathbf{k})}e^{-i\mathbf{k}_{\text{(b)}}\cdot\mathbf{x}_1}
e^{i \text{Re}\,\omega_{\text{(b)}}(\mathbf{k}_{\text{(b)}})t_1}e^{ \text{Im}\,\omega_{\text{(b)}}(\mathbf{k}_{\text{(b)}})t_1}
\nonumber\\&&
-\frac{ \mathcal{J}^{\text{in(neq)}}(\omega, \mathbf{k},\mathbf{k}_{\text{(b)}})}{\mathcal{J}^{\text{(eq)}}(\omega, \mathbf{k})}e^{i\mathbf{k}_{\text{(b)}}\cdot\mathbf{x}_2}
e^{-i \text{Re}\,\omega_{\text{(b)}}(\mathbf{k}_{\text{(b)}})t_2}e^{ \text{Im}\,\omega_{\text{(b)}}(\mathbf{k}_{\text{(b)}})t_2}\nonumber\\&&
-\frac{ \mathcal{J}^{\text{in(neq)}}(\omega, \mathbf{k},-\mathbf{k}_{\text{(b)}})}{\mathcal{J}^{\text{(eq)}}(\omega, \mathbf{k})}e^{-i\mathbf{k}_{\text{(b)}}\cdot\mathbf{x}_2}
e^{i \text{Re}\,\omega_{\text{(b)}}(\mathbf{k}_{\text{(b)}})t_2}e^{ \text{Im}\,\omega_{\text{(b)}}(\mathbf{k}_{\text{(b)}})t_2}\Bigg).
\end{eqnarray}
It is easy to take the Wigner transform of the above to obtain (\ref{retpw}).

\end{document}